\documentstyle[epsf,floats,aps,prd,eqsecnum]{revtex}
\def\be{\begin{equation}}
\def\ee{\end{equation}}
\def\bea{\begin{eqnarray}}
\def\nn{\nonumber}
\def\eea{\end{eqnarray}}
\def\beq{\begin{equation}}
\def\eeq{\end{equation}}
\def\beqn{\begin{eqnarray}}
\def\eeqn{\end{eqnarray}}
\def\Gperp{{G_L(\sigma)}_{,x_\perp x_\perp}}
\def\Gz{{G_L(\sigma)}_{,zz}}

\def\Gi{{G_L(\sigma)}_{,i}}

\def\Gperpdiv{G_{L,x_\perp x_\perp}^{\rm div}}
\def\Gperpfin{G_{L,x_\perp x_\perp}^{\rm fin}}
\def\Gzdiv{G_{L,zz}^{\rm div}}
\def\Gzfin{G_{L,zz}^{\rm fin}}
\def\Gidiv{G_{L,i}^{\rm div}}
\def\Gifin{G_{L,i}^{\rm fin}}

\begin{document}
\title{Recent Advances in Stochastic Gravity: Theory and Issues}
\author{B. L. Hu 
    \thanks{Electronic address: {\tt hub@physics.umd.edu}} }
    \address{Department of Physics, University of Maryland,
             College Park, Maryland 20742-4111, U.S.A.}
\author{E. Verdaguer
    \thanks{Electronic address: {\tt verdague@ffn.ub.es,
              enric@physics.umd.edu}}} 
    \address{Departament de F\'{\i}sica Fonamental,
             Universitat de Barcelona,\\
             Av.~Diagonal 647, 08028 Barcelona, Spain}
\date{Submitted Oct. 19, 2001 -- Erice Lectures May 2001. \\
To appear in {\it Advances in the Interplay between
Quantum and Gravity Physics} edited by V. De Sabbata, Kluwer, 2002}

\maketitle

\begin{abstract}

Whereas semiclassical gravity is based on 
the semiclassical Einstein equation with sources given by 
the expectation value of the stress-energy tensor of quantum fields,
stochastic semiclassical gravity is based 
on the Einstein-Langevin equation, which has in addition sources due to 
the noise kernel. 
The noise kernel is the vacuum expectation value of
the  (operator-valued) stress-energy bi-tensor 
which describes the fluctuations of quantum matter fields in
curved spacetimes. 
We  show how a consistent stochastic semiclassical
theory can be formulated as a perturbative
generalization of semiclassical gravity which describes the back
reaction of the lowest order stress-energy fluctuations.
The original approach \cite{Physica} used
in the early investigations leading to the establishment of this field
\cite{ELE} (for a review emphasizing ideas, see \cite{stogra}) was based on
quantum open system concepts 
(where the metric field acts as the ``system'' 
of interest and the matter fields as part of its ``environment'') 
and the influence functional method.
Here, following Refs. \cite{MV0,MV1} we first give an axiomatic
derivation of the Einstein-Langevin equations and then
show how they can also be derived by the original method
based on the influence functional. As a first application we 
solve these equations following Ref. \cite{MV2},
and compute the two-point correlation
functions for the linearized Einstein tensor and for the metric
perturbations in a Minkowski background. We then turn to 
the important issue of the validity of semiclassical gravity
by examining the criteria based on the ratio of the variance of 
fluctuations  of the stress-energy  
tensor of a quantum field to its mean. We show  a 
calculation of these quantities performed
in Ref. \cite{HP0,PH1} for a  massless scalar
field in the Minkowski and the Casimir vacua  as a function of
an intrinsic scale 
defined by introducing a smeared field or by point separation.
Contrary to prior claims,
the ratio of variance to mean-squared being of the order unity does not 
necessarily imply the failure of semiclassical gravity. 
Expressions for the variance to mean-squared ratio  as a
function of the ratio of the intrinsic to extrinsic scale (defined by the
separation of the plates or the periodicity of space for the Casimir topology) 
identifying the spatial extent where negative energy
density prevails are useful
for studying  quantum field effects in worm holes, baby universe and the 
design feasibility of `time-machines'.
This study also raises new questions into the very 
meaning of  regularization of the stress-energy tensor 
for quantum fields in curved or ordinary flat spacetimes. 
It corroborates the view held by one of us \cite{stogra} that 
at high energies it is more useful to consider spacetime as possesing
an extended structure, with the stress-energy bi-tensor
of quantum matter fields 
as the source. To compare with the more familiar point-defined
quantum field theory
operative at low energies, one needs to address the issue of
regularization of the
noise kernel. This we discuss following the work of Ref. \cite{PH2}
which uses
the  point-separation method to derive a general expression for a
regularized
noise-kernel for quantum fields in general curved spacetimes.
{}From these expressions describing the behavior of fluctuations
of quantum fields
one could investigate a host of important problems in early
universe and black hole
physics. This includes addressing the viability of a
vacuum-dominated phase which
inflationary cosmology is predicated upon, examining
how structures are seeded,
how the horizon behavior of a black hole is altered and
how the late time dynamics 
of black holes is affected with backreaction of  Hawking radiation. 
We intend to discuss these applications in a later report.

\end{abstract}
%\baselineskip=15pt
%\pacs{PACS number(s):-04.62.+v, 05.40.+j, 98.80.Cq}
%\draft
%%%%%%%%%%%%%%%%%%%%%%%%%%%%%%%%%%
%%%%%%%%%%%%%%%%%%%%%%%%%%%%%%%%%%
\section{OVERVIEW}
\label{sec1}
%%%%%%%%%%%%%%%%%%%%%%%%%%%%%%%%%%
Stochastic semiclassical gravity (STG) is a theory developed
in the Nineties using 
semiclassical gravity (SCG, quantum fields in classical
spacetimes, solved self-consistently) 
as the starting point and aiming at a theory of quantum
gravity (QG)  as the goal. 
While semiclassical gravity is based on the semiclassical
Einstein equation 
with the source given by the expectation value of the
stress-energy tensor of quantum fields, 
stochastic gravity (we will often use the shortened term 
{\it stochastic gravity} as there is no confusion in the
source of stochasticity in gravity  being due to quantum
fields here, and not of a classical origin) includes 
also its fluctuations in a new stochastic semiclassical 
or the Einstein-Langevin equation. If the centerpiece in
semiclassical gravity theory is
the vacuum expectation value of the stress-energy tensor
of a quantum field, and the central 
issues being how well the vacuum is defined and how the
divergences can be controlled
by regularization and renormalization, the centerpiece in
stochastic semiclassical gravity theory  is the 
stress-energy bi-tensor and its expectation value known as
the noise kernel. The mathematical
properties of this quantity and its physical content in
relation to the behavior of 
fluctuations of quantum fields in curved spacetimes
are the central issues of this new theory.
How they induce metric fluctuations and seed the structures
of the universe, how they affect the black hole horizons and the 
backreaction of Hawking radiance in black hole dynamics, including 
implications on trans-Planckian physics, are 
new horizons to explore.  On the theoretical issues, stochastic
gravity is the necessary foundation to investigate
the validity of semiclassical
gravity and the viability of inflationary cosmology based on the appearance
and sustenance of a vacuum energy-dominated phase. It is also
a useful beachhead supported by well-established low energy 
(sub-Planckian) physics to explore the connection with high energy
(Planckian) physics in the realm of quantum gravity.

In these lectures we want to summarize our work on the theory aspects since 
1998. (A review of ideas leading to stochastic gravity  and
further developments originating from it 
can be found in Ref. \cite{Physica,stogra}; a comprehensive formal description
is given  in Refs. \cite{MV0,MV1}).
It is  in the nature of a progress report rather than a review.
In fact we will have room
only to discuss two or three theoretical topics of basic importance.
We will try to mention all related work so the reader can at
least trace out the parallel 
and sequential developments. The references at the
end of each topic below are representative work where one can
seek out further treatments.

Stochastic gravity theory is built on three pillars:  general
relativity (GR), quantum fields (QF) and
nonequilibrium (NEq) statistical mechanics.
The first two uphold semiclassical gravity,
the last two span statistical field
theory, or NEqQF.  Strictly speaking one can
understand a great deal without 
appealing to statistical mechanics, and we will
try to do so here. But concepts 
such as quantum open systems \cite{qos} and techniques such as the
influence functional \cite{if} 
(which is related to the closed-time-path
effective action \cite{ctp,ctpcst}) were a
great help in our understanding of the physical meaning of issues involved
towards the construction of this new theory, foremost  because
quantum fluctuations and correlation have become the focus.
Quantum statistical field theory and the statistical mechanics of
quantum field theory \cite{CH88,dch,cddn,stobol} also aided us in searching
for the connection with quantum gravity through the retrieval of correlations 
and coherence. We show the scope of stochastic gravity as follows:\\

\noindent I. Ingredients\\

\noindent A. From General Relativity \cite{MTW,Wald84}
to Semiclassical Gravity.\\
B. Quantum Field Theory in Curved
Spacetimes \cite{Birrell-Davies82,Fulling89,Wald94,mostepanenko}:

   Stress-energy tensor: Regularization and renormalization. 

   Self-consistent solution: Backreaction problems \cite{scg}.

   Effective action: Closed time path, initial
 value formulation \cite{ctp,ctpcst}.

   Equation of motion: Real and causal.

\noindent C. Nonequilibrium Statistical Mechanics: 

   Open quantum systems \cite{qos}.

   Influence Functional: Stochastic equations \cite{if}.

   Noise and Decoherence: Quantum to classical
  transition \cite{envdec,conhis}.

\noindent D. Decoherence in Quantum Cosmology and
Emergence of Classical Spacetimes \cite{decQC}\\

\noindent  II. Theory\\

\noindent A. Dissipation from  Particle Creation \cite{CH87,cv94}. 

 Backreaction as Fluctuation-Dissipation Relation (FDR) \cite{HuSin}.\\
 B. Noise from Fluctuations of Quantum Fields \cite{Physica,Banff,CH94}. \\
\noindent  * C. Einstein-Langevin Equations \cite{ELE,MV0,MV1}.\\
\noindent  * D. Metric Fluctuations in Minkowski spacetime \cite{MV2}.\\

\noindent III. Issues\\
  
\noindent * A. Validity of Semiclassical Gravity \cite{HP0,PH1}.

 B. Viability of Vacuum Dominance and Inflationary Cosmology.\\
\noindent * C. Stress-Energy Bitensor and Noise Kernel:
     Regularization Reassessed \cite{PH2}.\\

\noindent IV. Applications:  Early Universe and Black Holes\\

[Test field treatment:]\\
  A. Wave Propagation in Stochastic Geometry   \cite{HuShio}.\\
  B. Black Hole Horizon Fluctuations: Spontaneous/Active
versus Induced/Passive   \cite{Ford,WuFor,Sorkin,BFP00,MasPar00}.

[Including backreaction:]\\
  C. Noise induced inflation \cite{calver99}.\\
  D. Structure Formation: \cite{strfor},
     Trace anomaly-driven inflation
     \cite{Starobinsky80,Vilenkin85,Hawking01,HuVer01}.\\
  E. Black Hole Backreaction as FDR \cite{CanSci,Mottola,HRS,CamHu}.\\

\noindent V. Related Topics \\

\noindent A. Metic Fluctuations and Trans-Planckian Problem
  \cite{BFP00,Par01,NiePar01}.\\
B. Spacetime Foam \cite{STFoam}.\\
C. Universal `Metric Conductance' Fluctuations  \cite{Shiokawa}.\\

\noindent VI. Ideas \\

\noindent A. General Relativity as Geometro-Hydrodynamics \cite{grhydro}.\\
B. Semiclassical Gravity as Mesoscopic Physics \cite{meso}.\\
C. From Stochastic to Quantum Gravity: 

   via Correlation hierarchy of interacting quantum fields
  \cite{stogra,dch,stobol}.

   Possible relation to string theory and matrix theory \cite{HuPeyr6}.\\
 
{}We list only the latest work in the respective topics above describing 
ongoing research. The reader should 
consult the references therein for earlier work and the background material. 
We do not seek a complete coverage here, but
will discuss only the selected topics in theory
and issues (Parts II and III above),
i.e., those topics marked with *s  above, 
in Sections 2-5 below. In a later report we plan to discuss recent advances
in the applications (Parts IV-V above).
We use the $(+,+,+)$ sign conventions of Refs.
\cite{MTW,Wald84}, and units in which $c=\hbar=1$.

%%%%%%%%%%%%%%%%%%%%%%%%%%%%%%%%%%%
%%%%%%%%%%%%%%%%%%%%%%%%%%%%%%%%%%%
\section{STOCHASTIC SEMICLASSICAL GRAVITY}
\label{sec2}
%%%%%%%%%%%%%%%%%%%%%%%%%%%%%%%%%%%
%%%%%%%%%%%%%%%%%%%%%%%%%%%%%%%%%%%

In this section we introduce stochastic semiclassical gravity.
We will start with  semiclasical gravity which  assumes that the
stress-energy fluctuations of the matter fields are negligible.
When these fluctuations cannot be neglected we may perturbatively
correct the semiclassical theory to take them into account.
This will lead to the semiclassical  Einstein-Langevin equations \cite{ELE}
as the dynamical equations which describe the back reaction of quantum
stress-energy fluctuations on the gravitational field.
We will introduce these equations by two independent methods.
The first, based on Ref. \cite{MV0},  is axiomatic and motivated by the search of
consistent equations to describe the back reaction of the quantum
stress-energy fluctuations on the gravitational field.
The second, based on Ref. \cite{MV1}, is by functional methods. It is motivated
by the observation \cite{Physica} that in some open quantum systems
classicalization and decoherence \cite{envdec} on the system may be
brought about by interaction with an environment. The environment
being in this case the matter fields and some degrees of freedom of
the quantum gravitational field \cite{decQC}.

%%%%%%%%%%%%%%%%%%%%%%%%%%%%%%%%
\subsection{Semiclassical gravity}
%%%%%%%%%%%%%%%%%%%%%%%%%%%%%%%%

Semiclassical gravity describes the interaccion
of the gravitational field assumed to be a classical field with
quantum matter fields. This theory cannot be derived as a limit
in a quantum theory of gravity, but can be formaly derived in 
several ways. One of them is the leading $1/N$ approximation of
quatum gravity interacting with $N$ independent and identical
free quantum fields
\cite{Hartle-Horowitz81}. When one assumes that the fields interact
with gravity only and keep the value of $NG$ finite, where $G$ is
Newton's gravitational constant, one arrives at a theory in which
formaly the gravitational field can be treated as a c-number
(i.e. quatized at tree level) but matter fields
are fully quatized.

In the semiclassical theory theory the expectation value
of the stress-energy tensor
in a given quantum state is the source of the gravitational field.
This theory may be summarized as follows.
Let $({\cal M},g_{\mu\nu})$ be a globaly hyperbolic four-dimensional
spacetime with metric $g_{\mu\nu}$ and consider a real scalar
quantum field $\phi$ of mass $m$ propagating on it
(we assume a scalar field for
simplicity but this is not essential). 

The classical action $S_m$ for this matter field is given by
the functional
\begin{equation}
S_m[g,\phi]=-{1\over2}\int d^4x\sqrt{-g}\left[g^{\mu\nu}
\nabla_\mu\phi\nabla_\nu\phi+\left(m^2+\xi R\right)\phi^2\right],
\label{2.1}
\end{equation}
where $\nabla_\mu$ is the
covariant derivative associated of the metric $g_{\mu\nu}$,
$\xi$ is a couplig parameter coupling the field to the scalar curvature
$R$, and  $g$ under the square root
stands for the determinat of $g_{\mu\nu}$.

The field may be quantized in the manifold using the standard
canonical formalism \cite{Birrell-Davies82,Fulling89,Wald94}.
The field operator in the Heisenberg reperesentation
$\hat\phi$ is an operator valued distribution solution of the
Klein-Gordon equation, the field equation of Eq. (\ref{2.1}),
\begin{equation}
(\Box-m^2 -\xi R)\hat\phi=0.
\label{2.2}
\end{equation}
We will write the field operator as $\hat\phi[g](x)$
to indicate that it is a functional of the metric.

The classical stress-energy tensor is obtained by functional derivation
of this action in the usual way
$T^{\mu\nu}(x)=(2/\sqrt{-g})\delta S_m/\delta g_{\mu\nu}$, leading to
\begin{eqnarray}
T_{\mu\nu}[g,\phi]&=&\nabla_\mu\phi\nabla_\nu\phi-{1\over2}g_{\mu\nu}
\left(\nabla^\rho\phi\nabla_\rho\phi+m^2\phi^2\right)
\nonumber \\
&&+\xi\left(g_{\mu\nu}\Box-\nabla_\mu\nabla_\nu+G_{\mu\nu}
\right)\phi^2,
\label{2.3}
\end{eqnarray}
where $\Box=\nabla_\mu\nabla^\mu$ and $G_{\mu\nu}$ is the Einstein tensor.
With the notation $T_{\mu\nu}[g,\phi]$ we explicity
indicate that the stress-energy tensor is
a functional of the metric $g_{\mu\nu}$  and the field $\phi$.

The next step is to define a stress-energy tensor operator
$\hat T_{\mu\nu}[g](x)$. Naively one would replace the classical
field $\phi$ in the above functional by the quantum operator
$\hat\phi[g]$, but this procedure involves taking the product
of two distributions at the same spacetime point. This is ill-defined
and we need a regularization procedure. There are several
regularization methods which one may use, one is the point-splitting
or point-separation
regularizaton method in which one introduces a point $y$ in a neighborhood
of the point $x$ and then uses as a regulator the vector tangent
at the point $x$ of the geodesic joining $x$ and $y$;
this method is discussed in detail in section \ref{sec4}. Another well
known method is dimesional regularization in which
one works in arbitray $n$ dimensions, where $n$
is not necessarily an integer, and then uses as the regulator
the parameter $\epsilon=n-4$; this method is used in this section and
in section \ref{sec5}. The
regularized stress-energy operator using the Weyl
ordering prescription, {\it i.e.} symmetrical ordering,
can be written as
\be
\hat{T}^{\mu\nu}[g] = {1\over 2} \{
     \nabla^{\mu}\hat{\phi}[g]\, , \,
     \nabla^{\nu}\hat{\phi}[g] \}
     + {\cal D}^{\mu\nu}[g]\, \hat{\phi}^2[g],
\label{regul s-t 2}
\ee
where ${\cal D}^{\mu\nu}[g]$ is the differential operator
\be
{\cal D}^{\mu\nu}_{x} \equiv
\left(\xi-{1\over 4}\right) g^{\mu\nu}(x) 
\Box_{x}+ \xi
\left( R^{\mu\nu}(x)- \nabla^{\mu}_{x} 
\nabla^{\nu}_{x} \right).
\label{diff operator}
\ee
Note that if dimensional regularization is used,
the field operator $\hat \phi$ propagates in a $n$-dimensional
spacetime.
Once the regularization prescription has
been introduced a renormalized and regularized stress-energy
operator may be defined as
\begin{equation}
\hat T^R_{\mu\nu}[g](x)= \hat T_{\mu\nu}[g](x)+F^C_{\mu\nu}[g](x)\hat I,
\label{2.4}
\end{equation}
where $\hat I$ is the identity operator and $F^C_{\mu\nu}[g]$ are some
symmetric tensor counterterms which depend on the regulator and are
local functionals of the metric. Here we are assuming that all
these terms depend on the regulator, see Ref. \cite{MV1}
for details. These states can be chosen
in such a way that for any pair of physically acceptable states,
i.e. Hadamard states in the sense of Ref.
\cite{Wald94}, $|\psi\rangle$ and $|\varphi\rangle$ the matrix element
$\langle\psi|T^R_{\mu\nu}|\varphi\rangle$ defined as the limit of the
previous expression when the regulator takes the physical value, is
finite and satisfies Wald's axioms \cite{Fulling89,Wald77}.
These counterterms can be extracted from the singular part of a
Schwinger-DeWitt series \cite{Fulling89,Christensen76,Bunch79}.
The choice of these counterterms is not unique but this ambiguity
can be absorbed into the renormalized coupling constants which
appear in the equations of motion for the gravitational field.

The {\it semiclassical Einstein equations} for the metric $g_{\mu\nu}$
can then be written as
\begin{equation}
{1\over 8\pi G}\left(G_{\mu\nu}[g]+\Lambda g_{\mu\nu}\right)
-2(\alpha A_{\mu\nu}+\beta B_{\mu\nu})[g]=
\langle \hat T_{\mu\nu}^R\rangle [g],
\label{2.5}
\end{equation}
where $\langle \hat T_{\mu\nu}^R\rangle [g]$ is the expectation value
of the operator (\ref{2.4}) after the regulator takes the physical value
in some physically acceptable state of the field on
$({\cal M},g_{\mu\nu})$. Note that both the stress tensor and the 
quantum state are functionals of the metric, hence the notation.
The parameters $G$, $\Lambda$, $\alpha$ and $\beta$ are the
renormalized coupling constants, respectively, the gravitational
constant, the cosmological constant and two dimensionless  coupling
constants which are zero in the classical Einstein equations.
These constants must be understoood as the result of  ``dressing''
the bare constants which appear in the classical action
before renormalization. The values of these constants must be
determined by experiment.
The left hand side of Eq. (\ref{2.5}) may be derived
from the gravitational action
\begin{equation}
S_g[g]=\int d^4 x \sqrt{-g}\left[ {1\over 16\pi G }(R-2\Lambda)
+\alpha C_{\mu\nu\rho\sigma}C^{\mu\nu\rho\sigma}
+\beta R^2\right],
\label{2.6}
\end{equation}
where $C_{\mu\nu\rho\sigma}$ is the Weyl tensor. The tensors
$A_{\mu\nu}$ and $B_{\mu\nu}$ come from the functional
derivatives with respect to the metric of the terms quadratic
in the curvature in Eq. (\ref{2.6}), they are explicitly given by
\begin{eqnarray}
A_{\mu\nu}&=&{1\over2}g_{\mu\nu}C_{\tau\epsilon\rho\sigma}
C^{\tau\epsilon\rho\sigma}-2R_{\mu\rho\sigma\tau}
R_{\nu}^{\ \rho\sigma\tau}+4R_{\mu\rho}R^\rho_{\ \nu}
-{2\over3}RR_{\mu\nu}
\nonumber\\
&& -2\Box R_{\mu\nu}+{2\over3}\nabla_\mu\nabla_\nu R+
{1\over3}g_{\mu\nu}\Box R,
\label{2.7a}\\
B_{\mu\nu}&=&{1\over2}g_{\mu\nu}R^2-2R R_{\mu\nu} 
+2\nabla_\mu\nabla_\nu R-2g_{\mu\nu}\Box R,
\label{2.7b}
\end{eqnarray}
where $R_{\mu\nu\rho\sigma}$  and
$R_{\mu\nu}$ are the Riemann and Ricci tensors, respectively.
These two tensors are, like the Einstein and metric tensors,
symmetric and
divergenceless: $\nabla^\mu A_{\mu\nu}=0=\nabla^\mu B_{\mu\nu}$.
Note that a classical stress-energy  can also
be added to the right hand side of Eq.
(\ref{2.5}), but for simplicity we omit such a term.

A solution of semiclassical gravity consists of a spacetime
(${\cal M},g_{\mu\nu}$), a quantum field operator $\hat\phi[g]$
which satisfies (\ref{2.1}), and a physically acceptable state 
$|\psi\rangle [g]$ for this field, such that Eq. (\ref{2.5}) is
satisfied when the expectation value of the renormalized
stress-energy operator is evaluated in this state.

For a free quantum field  this theory is robust in the sense that
it is consistent and fairly well understood. 
As long as the gravitational field is assumed to be described by a
classical metric, the above semiclassical Einstein
equations seems to be the only plausible dynamical equation
for this metric: the metric couples to mater fields via the
stress-energy tensor and for a given quantum state the only
physically observable c-number  stress energy-tensor that one
can construct is the above renormalized expectation value.
However, lacking a full quantum gravity theory the scope and
limits of the theory are not so well understood. It is assumed
that the semiclassical theory should break down at Planck scales,
which is when simple order of magnitude estimates suggest that
the quantum effects of gravity should not be ignored because the
energy of a quantum fluctuation in
a Planck size region, as determined by the Heisenberg uncertainty
principle, is comparable to the gravitational energy of
that fluctuation.

The theory should also break down when the fluctuations of the
stress-energy tensor are large. This has been emphasized by Ford
and colaborators \cite{Ford82,Kuo-Ford93}. It is less clear, however,
how to quantify what a large fluctuation is, and different criteria
have been proposed \cite{Kuo-Ford93,Phillips-Hu97,HP0,PH1}.
In Sec. \ref{sec3} we will discuss at length the issue of 
the validity of the semiclassical theory.
 One may illustrate the problem by
the following example inspired in Ref. \cite{Ford82}.

Let us assume a quantum state formed by an isolated sytem which
consists of a superposition with  equal amplitude of one configuration
with mass $M_1$ and another with mass $M_2$. The semiclassical
theory as described by Eq. (\ref{2.5}) predicts that the gravitational
field of the system is produced by the averaged mass $(M_1+M_2)/2$.
However, one would expect that if we send a succession of test particles
to probe the gravitational field of the above system half of the time
they would react to the field of a mass $M_1$ and half of the time to
the field of a mass $M_2$. If the two masses differ substantially
the two predictions are clearly different, note that the fluctuation in
mass of the quantum state is of order of $(M_1-M_2)^2$. Altough this
example is suggestive a word of caution should be said in order not to
take it too literaly. In fact, if the previous masses are macroscopic
the quantum system decoheres very quickly \cite{Zurek91} and instead
of being described by a pure quantum state it is described by a
density matrix which
diagonalizes in a certain pointer basis. For observables associated to
such a pointer basis the matrix density description is equivalnet to that
provided by a statistical ensemble. In any case, though, the results
will differ from the semiclassical prediction.

%%%%%%%%%%%%%%%%%%%%%%%%%%%%%%%%%%%%%%%%
\subsection{Axiomatic route to stochastic semiclassical gravity}
%%%%%%%%%%%%%%%%%%%%%%%%%%%%%%%%%%%%%%
\subsubsection{The noise kernel}
%%%%%%%%%%%%%%%%%%%%%%%%%%%%%%%%%%%%%
The purpose of stochastical semiclassical gravity is to go beyond the
semiclassical theory and account for the fluctuations of
the stress-energy operator. But first, we have to
give a physical observable
that describes these fluctuations. 
To lowest order, these fluctuations
are obviously described by the following bi-tensor,
constructed with the two-point correlation
of the stress-energy operator,
\begin{equation}
4N_{\mu\nu\rho\sigma}(x,y)={1\over2}\langle\{\hat t_{\mu\nu}(x),
\hat t_{\rho\sigma}(y)\}\rangle[g],
\label{2.8}
\end{equation}
where the curly brackets mean anticommutator, and where
\begin{equation}
\hat t_{\mu\nu}(x)
\equiv \hat T_{\mu\nu}(x)-\langle \hat T_{\mu\nu}(x)\rangle.
\label{2.9}
\end{equation}
This bi-tensor
will be called {\it noise kernel} from now on. Note that we
have defined it in terms of the unrenormalized stress-tensor operator
$\hat T_{\mu\nu}[g](x)$ on a given background metric $g_{\mu\nu}$, thus
a regulator is implicitly assumed on the r.h.s. of Eq. (\ref{2.8}).
However,
for a linear quantum field the above kernel is free of ultraviolet
divergencies because the ultraviolet behaviour of
$\langle\hat T_{\mu\nu}(x)\hat T_{\rho\sigma}(y)\rangle$ is the same
as that of
$\langle\hat T_{\mu\nu}(x)\rangle
\langle\hat T_{\rho\sigma}(y)\rangle$,
thus following Eq. (\ref{2.4}) one can replace  $\hat T_{\mu\nu}$
by the renormalized operator $\hat T_{\mu\nu}^R$ in Eq. (\ref{2.8});
an alternative proof of this result
is given in section \ref{sec4}.
The noise kernel should be thought of as a distribution function,
the limit of coincidence points has meaning only in the sense
of distributions. An analysis of the noise kernel based on
the point-separation method is given in section \ref{sec4}.

The bi-tensor $N_{\mu\nu\rho\sigma}(x,y)$ is real and
positive-semidefinite, as a consequence
of $\hat T_{\mu\nu}^R$ being self-adjoint. A simple proof
can be given as follows. Let $|\psi\rangle$ be a given quantum state
and let $\hat Q$ be a selfadjoint operator, $\hat Q^\dagger=\hat Q$,
then one can write
$\langle\psi|\hat Q\hat Q|\psi\rangle=
\langle\psi|\hat Q^\dagger Q|\psi\rangle=
| \hat Q|\psi\rangle|^2\geq 0$. Now let $\hat t(x)$ be a selfadjoint
operator representing the operator $\hat t_{\mu\nu}$ of Eq.
(\ref{2.8}) but where the coordinate $x$ now carries also the tensorial
indices of Eq. (\ref{2.8}), then if we define
$\hat Q=\int dx f(x) \hat t(x)$ for an arbitrary well behaved
``function'' $f(x)$ (which
carries also tensorial indices), the previous unequality can be
written as
$\int dx dy f(x)\langle\psi|\hat t(x)\hat t(y)|\psi\rangle f(y)\geq 0$,
which is the condition for the noise kernel to be posite semi-definite.
Note that when we will consider, later on, the inverse kernel
$N_{\mu\nu\rho\sigma}^{-1}(x,y)$ we will implicitly assume that we
are working in the subspace obtained from the eigenvectors which
have strictly positive eigenvalues when we diagonalize
the noise kernel.

%%%%%%%%%%%%%%%%%%%%%%%%%%%%%%%
\subsubsection{The Einstein-Langevin equation}
%%%%%%%%%%%%%%%%%%%%%%%%%%%%%%%

Our purpose now is to perturbatively modify the
semiclassical theory. Thus
we will assume that the background spacetime metric $g_{\mu\nu}$ is
a solution of the semiclassical Einstein Eqs. (\ref{2.5}). The
stress-energy tensor will generally have quantum fluctuations
on that background spacetime.
One would expect that these fluctuations will have some effect on the
gravitational field and that this may now be described by
$g_{\mu\nu}+h_{\mu\nu}$, where we will assume that $h_{\mu\nu}$ is
a linear perturbation to the background solution. The renormalized
stress-energy operator and the state of the quatum field may now be
denoted by $\hat T_{\mu\nu}^R[g+h]$ and $|\psi\rangle[g+h]$,
respectively, and $\langle\hat T_{\mu\nu}^R\rangle [g+h]$ will be
the corresponding expectation value.

Let us now introduce a Gaussian stochastic tensor field
$\xi_{\mu\nu}$ defined by the following correlators:
\begin{equation}
\langle\xi_{\mu\nu}(x)\rangle_s=0,\ \ \ 
\langle\xi_{\mu\nu}(x)\xi_{\rho\sigma}(y)\rangle_s=
N_{\mu\nu\rho\sigma}(x,y),
\label{2.10}
\end{equation}
where $\langle\dots\rangle_s$ means statistical average. The
symmetry and positive semi-definite property of the noise kernel
guarantees that the stochastic field tensor
$\xi_{\mu\nu}$ just introduced
is well defined. Note that this stochastic
tensor does not capture the
whole quantum nature of the fluctuations of the 
stress-energy operator since it assumes that cumulants of higher order
are zero, but it does so at the Gaussian level.

An important property of this stochastic tensor is that it is
covariantly conserved in the background spacetime
$\nabla^\mu\xi_{\mu\nu}=0$. In fact, as a consequence
of the conservation of $\hat T_{\mu\nu}^R[g]$ one can seen that
$\nabla_x^\mu N_{\mu\nu\rho\sigma}(x,y)=0$. Taking the divergence in
Eq. (\ref{2.10}) one can then show that
$\langle\nabla^\mu\xi_{\mu\nu}\rangle_s=0$ and 
$\langle\nabla_x^\mu\xi_{\mu\nu}(x)
\nabla_y^\rho\xi_{\rho\sigma}(y)\rangle_s=0$ so that
$\nabla^\mu\xi_{\mu\nu}$ is detreministic and represents
with certainty the zero vector field in $\cal{M}$.

One can also see  that
for a conformal field, i.e. a field
whose classical action is conformally invariant,
$\xi_{\mu\nu}$ is traceless, {\it i.e.} $g^{\mu\nu}\xi_{\mu\nu}=0$,
so that for a conformal matter field the stochastic source gives no
correction to the trace anomaly.
In fact, from the trace anomaly result which states that
$g^{\mu\nu}\hat T^R_{\mu\nu}[g]$ is in this case a local
c-number functional of $g_{\mu\nu}$ times the identity
operator, we have that $g^{\mu\nu}(x)N_{\mu\nu\rho\sigma}(x,y)=0$.
It then follows from Eq. (\ref{2.10}) that
$\langle g^{\mu\nu}\xi_{\mu\nu}\rangle_s=0$ and
$\langle g^{\mu\nu}(x)\xi_{\mu\nu}(x)
g^{\rho\sigma}(y)\xi_{\rho\sigma(y)}\rangle_s=0$;
an alternative proof based on the point-separation
method is given in section
\ref{sec4}.

All these properties make it quite natural to incorporate into the
Einstein equations the stress-energy fluctuations by using the
stochastic tensor $\xi_{\mu\nu}$ as the source
of the metric perturbations.
Thus we will write the following equation.
\begin{equation}
{1\over 8\pi G}\left(G_{\mu\nu}[g+h]+
\Lambda (g_{\mu\nu}+h_{\mu\nu})\right)
-2(\alpha A_{\mu\nu}+\beta B_{\mu\nu})[g+h]=
\langle \hat T_{\mu\nu}^R\rangle [g+h]+2\xi_{\mu\nu}.
\label{2.11}
\end{equation}
This equation known as the {\it semiclassical Einstein-Langevin equation},
is a dynamical equation for the metric perturbation $h_{\mu\nu}$ to 
linear order. It describes the back-reaction of the metric to the
quantum fluctuations of the stress-energy tensor of matter fields, and
gives a first order correction to semiclassical gravity as described
by the semiclassical Einstein equation (\ref{2.5}). Note that the stochastic
source $\xi_{\mu\nu}$ is not dynamical, it is independent of
$h_{\mu\nu}$ since it decribes the fluctuations of the stress tensor
on the semiclassical background $g_{\mu\nu}$.

An important property of the Einstein-Langevin equation is that it is
gauge invariant. In the sense that if we change $h_{\mu\nu}\rightarrow
h_{\mu\nu}^\prime +\nabla_\mu\eta_\nu+\nabla_\nu\eta_\mu$ where
$\eta^\mu$ is a stochastic vector field on the manifold ${\cal M}$.
All the tensor which appear in the Eq. (\ref{2.11}) transform as
$R_{\mu\nu}[g+h^\prime]=R_{\mu\nu}[g+h]+{\cal L}_\eta R_{\mu\nu}[g]$
to linear order in the perturbations, where ${\cal L}_\eta $ is the Lie
derivative with respect to $\eta^\mu$. If we substitute $h$ by $h^\prime$
in Eq. (\ref{2.11}), we get Eq. (\ref{2.11}) plus the Lie derivative of
the combination of the tensor which appear in Eq. (\ref{2.5}). This last
combination vanishes when Eq. (\ref{2.5}) is satisfied,
as we have assumed. It is thus  necessary
that the background metric $g_{\mu\nu}$ be a solution of semiclassical
gravity.

A solution of Eq. (\ref{2.11}) can be formally written as 
$h_{\mu\nu}[\xi]$. This solution is characterized by the whole
family of its correlation functions. From the statistical average
of this  equation we have that $g_{\mu\nu}+\langle h_{\mu\nu}\rangle_s$
must be a solution of the semiclassical Einstein equation linearized
around the background  $g_{\mu\nu}$. The fluctuation of the metric
around this average are described by the moments of the stochastic field
$h_{\mu\nu}^s[\xi]=h_{\mu\nu}[\xi]-\langle h_{\mu\nu}\rangle_s$.
Thus the solutions of the Einstein-Langevin equation will
provide the two point metric correlators
$\langle h_{\mu\nu}^s(x)  h_{\rho\sigma}^s(y)\rangle_s$.

The stochastic theory may be understood as an intermediate step between
the semiclassical theory and the full quantum theory. In the sense
that whereas the semiclassical theory
depends on the point-like value of the sress-energy operator,
the  stochastic
theory carries information also on the two-point correlation of the
stress-energy operator.

We should also emphasize that, even if the metric fluctuations
are classical (stochastic), their
origin is  quantum not only because they
are induced by the fluctuations of quantum matter,
but also because they are supposed to describe some remnants
of the quantum gravity fluctuations after some mechanism for
decoherence and classicalization of the metric field
\cite{gell-mann-hartle,hartle,dowker,halliwell,whelan}.
{}From the formal assumption that such a mechanism is the Gell-Mann
and Hartle mechanism of environment-induced decoherence 
of suitably coarse-grained system variables 
\cite{gell-mann-hartle,hartle},
one may, in fact, derive the stochastic semiclassical theory 
\cite{MV3}.
Nevertheless, that derivation is of course formal,
given that, due to the lack of the full quantum theory of gravity, the
classicalization mechanism for the gravitational field
is not understood.

%%%%%%%%%%%%%%%%%%%%%%%%%%%%%
\subsubsection{A toy model}
%%%%%%%%%%%%%%%%%%%%%%%%%%%%

To illustrate the relation between the quantum and
stochastic descriptions it is useful to introduce the
following toy model. 
Let us
assume that the gravitational equations are described by a linear
field $h$ coupled to a scalar source $T[\phi^2]$ independent of
$h$, with the classical equations
$\Box h=T$ in  flat spacetime. We should emphasize that this model
would describe a linear theory of gravity analogous to electromagnetism
and does not mimic the linearized theory of gravity in which $T$ is also
linear in $h$. However it captures some of the key features of linearized
gravity.

In the Heisenberg representation the quantum field $\hat h$ satisfies
\begin{equation}
\Box \hat h=\hat T.
\label{2.12}
\end{equation}
The solutions of this equation, i.e. the field operator at the point $x$,
$\hat h_x$,  may be written in terms of
the retarded propagator $G_{xy}$ as,
\begin{equation}
\hat h_x=\hat h_x^0+\int dx' G_{xx'}\hat T_{x'},
\label{2.13}
\end{equation}
where $h^0$ is the free field.
{}From this solution we may compute, for instance, the following
two point quantum
correlation function (the anticommutator)
\begin{equation}
{1\over2}\langle \{\hat h_x,\hat h_y\}\rangle = 
{1\over2} \langle \{\hat h^0_x,\hat h^0_y\}\rangle +
{1\over2} \int\int dx'dy'G_{xx'}G_{yy'}
\langle\{\hat T_{x'},\hat T_{y'}\}\rangle,
\label{2.14}
\end{equation}
where the expectation value is taken in the quantum state in which both
fields $\phi$ and $h$ have been quantized
and we have used that
for the free field $\langle \hat h^0\rangle=0$.

Let us now compare with the stochastic theory for this problem.
As discussed previously we should write a classical equation for
the field $h$ where on the r.h.s. we substitute $\hat T$ by the
expectation value $\langle\hat T\rangle$ plus a Gaussian 
stochastic field $\xi$
defined by the quantum correlation function of $T$ (the noise kernel).
Note that in our model since $T$ is
independent of $h$ we may simply renormalize
the expectation value using time ordering, then for
the vacuum state of the field $\hat\phi$, we would
simply have $\langle\hat T\rangle_0=0 $. Thus $\xi$ is defined
as in (\ref{2.10}) by $\langle\xi\rangle_s=0$ and
$\langle\xi_x\xi_y\rangle_s=(1/2)\langle\{\hat T_x,\hat T_y\}\rangle
-\langle\hat T_x\rangle\langle\hat T_y\rangle$.
The semiclassical stochastic equation is thus  
\begin{equation}
\Box  h=\langle T\rangle+\xi.
\label{2.15}
\end{equation}
The solution of this equation may be
written in terms of the retarded propagator as,
\begin{equation}
h_x=h_x^0+\int dx'G_{xx'}\left(\langle\hat T_{x'}\rangle 
+\xi_{x'}\right) ,
\label{2.16}
\end{equation}
from where the two point correlation function for the classical field $h$,
after using the definition of $\xi$ and that $\langle h^0\rangle_s=0$, is
given by
\begin{equation}
\langle  h_x h_y\rangle_s = \langle  h^0_x h^0_y\rangle_s +{1\over2}
\int\int dx'dy'G_{xx'}G_{yy'}
\langle\{\hat T_{x'},\hat T_{y'}\}\rangle.
\label{2.17}
\end{equation}

Comparing (\ref{2.14}) with (\ref{2.17}) we see that the respective second
terms on the right hand side are identical provided the expectation
values are computed in the same quantum state for the field $\hat \phi$,
note that we have assumed that $T$ does not depend on $h$. The fact that
the field $h$ is also quantized in (\ref{2.14}) does not change the
previous statement. In the real theory of gravity $T$, in fact, depends
also on $h$ and then the previous statement is only true approximately,
i.e  perturbatively in $h$. The nature of the first terms on the
right hand sides of equations (\ref{2.14}) and (\ref{2.17}) is different:
in the first case it is the two point expectation value of the free
quatum field $\hat h_0$ whereas in the second case it is the average of the
two point classical average of the homogeneous field $h_0$, which depends
on the initial conditions. Now we can still make these terms to be equal to
each other if we assume for the homogeneous field $h$ a distribution of
initial conditions such that $\langle  h^0_x h^0_y\rangle_s= 
(1/2)\langle\{\hat h^0_x,\hat h^0_y\}\rangle$.
Thus, under this assumption on initial
conditions for the field $h$ the two point correlation function of
(\ref{2.17}) equal the quantum expectation value of (\ref{2.14}) exactly.

Thus in a linear theory as in the model just described one may just use
the statistical description given by (\ref{2.15}) to compute the
quantum two point function of equation (\ref{2.13}).
This does not mean that we can recover all quantum correlation functions
with the stochastic description, see Ref. \cite{Calzetta-Roura00a}
for a general discussion about this point.
Note that, for instance,
the commutator of the classical stochastic field
$h$ is obviously zero, but the commutator of
the quantum field $\hat h$ is not zero for timelike separated points;
this is the prize we pay for the introduction of the classical field
$\xi$ to describe the quantum fluctuations.
Furthermore, the statistical description is not able to
account for the graviton-graviton
effects which go beyond the linear approximation in $\hat h$.

%%%%%%%%%%%%%%%%%%%%%%%%%%%
\subsection{Functional approach}
%%%%%%%%%%%%%%%%%%%%%%%%%%

The Einstein-Langevin equation
(\ref{2.11}) may also be derived by a method
based on functional techniques  \cite{MV1}.
Functional techniques have a long history in
semiclassical gravity. It started when effective action
methods, which are so familiar in quantum field theory, were used to
study the back-reaction of quantum fields in
cosmological models \cite{Hartle}. These methods
were of great help in the study of cosmological anisotropies
since they allowed the introduction of familiar perturbative
treatments into the subject. The
most common method, the in-out effective action method,
however, led to equations of motion
which were not real because they were taylored to compute transition
elements of quantum operators rather than expectation values. Fortunately
the appropriate technique had already been developed by Schwinger
and Keldysh \cite{ctp}
in the so called Closed Time Path (CTP) or in-in
effective action. These techniques were soon adapted
to the gravitational context
\cite{ctpcst} and  were applied to different problems in
cosmology. As a result one was now able to
deduce the semiclassical Einstein equations by the CTP functional
method: starting with an action for the interaction of gravity with
matter fields, treating the matter fields as quantum fields and the
gravitational field at tree level only.

Furthermore, in this case the CTP functional formalism  
turns out to be related 
\cite{CH94,greiner,cv96,campos-hu,morikawa,MV1}
to the influence functional formalism of Feynman and Vernon
\cite{feynman-vernon}
since the full quantum system may be understood as
consisting of a distinguished subsystem 
(the ``system'' of interest)
interacting with an environment (the remaining degrees of freedom).
The integration of the environment
variables in a CTP path integral yields the influence functional, from
which one can define an effective action for the dynamics of the
system \cite{feynman-hibbs,CH94,HuSin,hu-paz-zhang,hu-matacz94,greiner}.

In our case, we consider
the metric field $g_{\mu\nu}(x)$ as the ``system'' degrees of freedom, 
and the scalar field $\phi(x)$ 
and also some ``high-momentum'' gravitational modes
\cite{whelan} as the ``environment'' variables. 
Unfortunately, since the form of a complete
quantum theory of gravity interacting with matter is unknown,
we do not know what these ``high-momentum'' gravitational modes are.
Such a fundamental quantum theory might not even be a field theory,
in which case the metric and scalar fields would not be 
fundamental objects \cite{stogra}.
Thus, in this case, we cannot attempt to evaluate 
the influence action of Feynman and Vernon 
starting from the fundamental quantum theory and 
performing the path integrations in the environment variables.
Instead, we introduce the influence action for 
an effective quantum field theory of gravity and matter
\cite{donoghue,HuSin}, 
in which such ``high-momentum'' gravitational
modes are assumed to have been already ``integrated out.'' 
Adopting the usual procedure of 
effective field theories \cite{weinberg,donoghue},
one has to take the effective action for the metric and the scalar
field of 
the most general local form compatible with general covariance:
$$S[g,\phi] \equiv  S_g[g]+S_m[g,\phi]+ \cdots ,$$
where $S_g[g]$ and $S_m[g,\phi]$ are given by Eqs. (\ref{2.6})
and (\ref{2.1}), respectively,
and the dots  
stand for terms of order higher than two 
in the curvature and in the number of
derivatives of the scalar field.
Here, we shall neglect the higher order terms as well as 
self-interaction terms for the scalar field.
The second order terms are necessary to renormalize 
one-loop ultraviolet divergencies of the scalar field 
stress-energy tensor, as we have already seen. 
Since ${\cal M}$ is a globally hyperbolic manifold, 
we can foliate it by a family of $t\!=\! {\rm constant}$ Cauchy
hypersurfaces $\Sigma_{t}$. We denote by  
${\bf x}$ the coordinates on each of these hypersurfaces, and by
$t_{i}$ and $t_{f}$ some initial and final times, respectively.
The integration domain for the action terms must be understood
as a compact region ${\cal U}$ of the manifold ${\cal M}$, bounded by
the hypersurfaces $\Sigma_{t_i}$ and $\Sigma_{t_f}$.

Assuming the form (\ref{2.1}) for 
the effective action which couples the scalar and the metric fields,  
we can now
introduce the corresponding influence functional.
This is a functional of two copies of the metric field that we denote
by $g_{\mu\nu}^+$ and $g_{\mu\nu}^-$.
Let us assume that, in the quantum effective theory, 
the state of the full system 
(the scalar and the metric fields) in the Schr\"{o}dinger picture 
at the initial time $t\! =\! t_{i}$ can be described by 
a factorizable 
density operator, {\it i.e.}, a density operator which can be written
as the tensor product of two operators on the Hilbert spaces
of the metric and of the scalar field.
Let $\hat{\rho}^{\rm \scriptscriptstyle S}(t_{i})$ be the 
density operator describing the initial state of the 
scalar field.
If we consider the theory of a scalar field quantized in the
classical background spacetime $({\cal M},g_{\mu\nu})$ through the action 
(\ref{2.1}), a
state in the Heisenberg representation described by a density operator
$\hat{\rho}[g]$ corresponds to this state.
Let $\left\{ |\varphi(\mbox{\bf x})\rangle^{\rm \scriptscriptstyle S} 
\right\}$ 
be the basis of eigenstates of the scalar
field operator $\hat{\phi}^{\rm \scriptscriptstyle S}({\bf x})$
in the Schr\"{o}dinger representation:
$\hat{\phi}^{\rm \scriptscriptstyle S}({\bf x})
\, |\varphi\rangle ^{\rm \scriptscriptstyle S}=
\varphi(\mbox{\bf x})
\, |\varphi\rangle^{\rm \scriptscriptstyle S}$.
The matrix elements of 
$\hat{\rho}^{\rm \scriptscriptstyle S}(t_{i})$ in this basis will be
written as 
$\rho_{i} \!\left[\varphi,\tilde{\varphi}\right] \equiv 
\mbox{}^{\rm \scriptscriptstyle S}
\langle \varphi|\,\hat{\rho}^{\rm \scriptscriptstyle S}(t_{i})
\, |\tilde{\varphi}\rangle^{\rm \scriptscriptstyle S}$. 
We can now introduce the {\it influence functional} as the following
path integral over two copies of the scalar field:
\begin{equation}
{\cal F}_{\rm IF}[g^+,g^-] \equiv
\int\! {\cal D}[\phi_+]\;
{\cal D}[\phi_-] \;
\rho_i \!\left[\phi_+(t_i),\phi_-(t_i) \right] \,
\delta\!\left[\phi_+(t_f)\!-\!\phi_-(t_f)  \right]\:
e^{i\left(S_m[g^+,\phi_+]-S_m[g^-,\phi_-]\right) }.
\label{path integral}
\end{equation} 
The above double path integral can be rewritten as a closed time path
(CTP) integral, namely, as an integral over a single copy of field
paths with two different time branches, one going forward in time from
$t_i$ to $t_f$, and the other going backward in time from $t_f$ 
to $t_i$. 
{}From this influence functional, 
the {\it influence action} of Feynman and Vernon,
$S_{\rm IF}[g^+,g^-]$, is defined  by
\be
{\cal F}_{\rm IF}[g^+,g^-] \equiv
e^{i S_{\rm IF}[g^+,g^-]},
\label{influence functional}
\ee
this action has all the relevant
information on the matter fields. Then we can define
the {\it effective action} for the gavitational field,
$S_{\rm eff}[g^+,g^-]$, as
\begin{equation} 
S_{\rm eff}[g^+,g^-]\equiv S_{g}[g^+]-S_{g}[g^-]
+S_{\rm IF}[g^+,g^-].
\label{2.18}
\end{equation}
This is the action for the classical gravitational field
in the CTP formalism. However, since the
gravitational field is only treated at tree level, this 
is also the effective ``classical'' action from
where the classical equations
can be derived.
 
Expression (\ref{path integral}) is ill-defined,
it must be regularized to get a meaningful 
influence functional. We shall assume that we can use 
dimensional regularization, that is, that we can give sense
to Eq. (\ref{path integral}) by
dimensional continuation of all the quantities that appear in this
expression. 
For this we
substitute the action $S_m$ in (\ref{path integral}) 
by some generalization to $n$ spacetime
dimensions, and similarly for $S_g[g]$.
In this last case the parameters $G$, $\Lambda$,
$\alpha$ and $\beta$
of Eq. (\ref{2.6})
are the bare parameters $G_B$, $\Lambda_B$,
$\alpha_B$ and $\beta_B$, and
instead of the square of the Weyl tensor one must
use $(2/3)(R_{\mu\nu\rho\sigma}R^{\mu\nu\rho\sigma}-
R_{\mu\nu}R^{\mu\nu})$ which by
the Gauss-Bonnet theorem leads
the same equations of motion as the action (\ref{2.6})
when $n \!=\! 4$.
The form of $S_g$ in $n$ dimensions is suggested by the
Schwinger-DeWitt analysis of the ultraviolet divergencies in the
matter stress-energy tensor using dimensional regularization. 
One can then
write the effective action of Feynman and Vernon, 
$S_{\rm eff}[g^+,g^-]$, in Eq. (\ref{2.18})
in dimensional regularization. 
Since both $S_m$ and $S_g$
contain second order derivatives of the metric, one should also add
some boundary terms \cite{Wald84,HuSin}. 
The effect of these
terms is to cancel out the boundary terms which appear
when taking variations of $S_{\rm eff}[g^+,g^-]$ keeping the value
of $g^+_{\mu\nu}$ and $g^-_{\mu\nu}$ fixed on the boundary of ${\cal U}$. 
Alternatively, in order to obtain the equations of motion for
the metric in the semiclassical regime, we can work with the action terms 
without boundary
terms and neglect all boundary terms when taking variations with
respect to $g^{\pm}_{\mu\nu}$. From now on, all the functional derivatives
with respect to the metric will be understood in this sense.  

Now we can derive the semiclassical Einstein equations (\ref{2.5}).
Using the definition of the stress-energy tensor
$T^{\mu\nu}(x)=(2/\sqrt{-g})\delta S_m/\delta g_{\mu\nu}$
and the definition
of the influence functional, Eqs.
(\ref{path integral}) and (\ref{influence functional}), we see that 
\begin{equation}
\langle \hat{T}^{\mu\nu}(x) \rangle [g] =
\left. {2\over\sqrt{- g(x)}} \, 
 \frac{\delta S_{\rm IF}[g^+,g^-]}
{\delta g^+_{\mu\nu}(x)} \right|_{g^+=g^-=g},
\label{s-t expect value}
\end{equation} 
where the expectation value is taken in the $n$-dimensional spacetime
generalization of the state described by
$\hat{\rho}[g]$.
Therefore, differentiating 
$S_{\rm eff}[g^+,g^-]$ in Eq. (\ref{2.18}) 
with respect to $g^+_{\mu\nu}$, and then setting
$g^+_{\mu\nu}=g^-_{\mu\nu}=g_{\mu\nu}$, we get
the semiclassical Einstein
equation in $n$ dimensions. This equation is then
renormalized by absorbing the divergencies in
the regularized
$\langle\hat T^{\mu\nu}\rangle[g]$ into the bare
parameters, and taking the limit $n\to\infty$ we get the
physical semiclassical Einstein equations (\ref{2.5}).

%%%%%%%%%%%%%%%%%%%%%%%%%%%%%%%%%%%%%%%%%%%%

\subsection{Influence functional route to stochastic semiclassical
gravity}

%%%%%%%%%%%%%%%%%%%%%%%%%%%%%%%%%%%%%%%%%%%%%

In this subsection we derive the semiclassical Einstein-Langevin equation
(\ref{2.11}) by means of the influence functional.
We also work out the semiclassical Einstein-Langevin
equations more explicitly,
in a form more suitable for specific calculations.

In the spirit of the previous derivation of the Einstein-Langevin
equations, we now seek a dynamical
equation for a linear perturbation $h_{\mu\nu}$ to a semiclassical 
metric $g_{\mu\nu}$, solution of
Eq. (\ref{2.5}). Strictly speaking if we use dimensional
regularization we must consider the $n$-dimensional version of that
equation; see Ref. \cite{MV1} for details.
{}From the result of the previous subsection, if such equation were
simply a linearized semiclassical Einstein equation, it could be
obtained from an expansion of the effective action
$S_{\rm eff}[g+h^+,g+h^-]$. In particular, since, from 
Eq. (\ref{s-t expect value}), we have that
\begin{equation}
\langle \hat{T}_n^{\mu\nu}(x) \rangle [g+h]=
\left. {2\over\sqrt{-\det (g\!+\!h)(x)}} \, 
 \frac{\delta S_{\rm IF}
   [g\!+\!h^+,g\!+\!h^-]}{\delta h^+_{\mu\nu}(x)} 
 \right|_{h^+=h^-=h},
\label{perturb s-t expect value}
\end{equation}
the expansion of $\langle \hat{T}_{n}^{\mu\nu}\rangle [g\!+\!h]$
to linear order in $h_{\mu\nu}$ can be obtained from an expansion of the
influence action $S_{\rm IF}[g+h^+,g+h^-]$ up to second order
in $h^{\pm}_{\mu\nu}$.

To perform the expansion of the influence action, 
we have to compute the first and second order
functional derivatives of $S_{\rm IF}[g^+,g^-]$
and then set $g^+_{\mu\nu}\!=\!g^-_{\mu\nu}\!=\!g_{\mu\nu}$.
If we do so using the path integral representation
(\ref{path integral}), we can interpret these derivatives as
expectation values of operators.
The relevant second order derivatives are
\begin{eqnarray}
\left. {1\over\sqrt{- g(x)}\sqrt{- g(y)} } \,   
 \frac{\delta^2 S_{\rm IF}[g^+,g^-]}
{\delta g^+_{\mu\nu}(x)\delta g^+_{\rho\sigma}(y)}
 \right|_{g^+=g^-=g} \!\!
&=& -H_{\scriptscriptstyle \!
{\rm S}}^{\mu\nu\rho\sigma}[g](x,y)
-K^{\mu\nu\rho\sigma}[g](x,y)+
i N^{\mu\nu\rho\sigma}[g](x,y),      \nonumber \\
\left. {1\over\sqrt{- g(x)}\sqrt{- g(y)} } \, 
 \frac{\delta^2 S_{\rm IF}[g^+,g^-]}
{\delta g^+_{\mu\nu}(x)\delta g^-_{\rho\sigma}(y)} 
 \right|_{g^+=g^-=g} \!\!
&=& -H_{\scriptscriptstyle \!
{\rm A}}^{\mu\nu\rho\sigma}
[g](x,y)
-i N^{\mu\nu\rho\sigma}[g](x,y),  
\label{derivatives}  
\end{eqnarray}
where  
$$
N^{\mu\nu\rho\sigma}[g](x,y) \equiv
{1\over 8} \left\langle  \bigl\{
 \hat{t}^{\mu\nu}(x) , \,
 \hat{t}^{\rho\sigma}(y)
 \bigr\} \right\rangle [g],
\ \ \ \
H_{\scriptscriptstyle \!
{\rm S}}^{\mu\nu\rho\sigma}
[g](x,y) \equiv 
{1\over 4}\:{\rm Im} \left\langle {\rm T}^*\!\!
\left( \hat{T}^{\mu\nu}(x) \hat{T}^{\rho\sigma}(y) 
\right) \right\rangle \![g],       
$$
$$
H_{\scriptscriptstyle \!
{\rm A}}^{\mu\nu\rho\sigma}
[g](x,y) \equiv 
-{i\over 8} \left\langle 
\bigl[ \hat{T}^{\mu\nu}(x), \, \hat{T}^{\rho\sigma}(y)
\bigr] \right\rangle \![g],       
\ \ \ 
K^{\mu\nu\rho\sigma}[g](x,y) \equiv 
\left. {-1\over\sqrt{- g(x)}\sqrt{- g(y)} } \, \left\langle 
\frac{\delta^2 S_m[g,\phi]}
{\delta g_{\mu\nu}(x)\delta g_{\rho\sigma}(y)} 
\right|_{\phi=\hat{\phi}}\right\rangle \![g], 
$$  
with $\hat{t}^{\mu\nu}$ defined in Eq. (\ref{2.9}), and    
where we use a Weyl ordering prescription
for the operators in the last of these expressions. 
Here, $[ \; , \: ]$ means the commutator, and we
use the symbol ${\rm T}^*$ 
to denote that, first,
we have to time order the field operators $\hat{\phi}$ and then
apply the derivative operators which appear in each term 
of the product $T^{\mu\nu}(x) T^{\rho\sigma}(y)$,
where $T^{\mu\nu}$ is
the functional (\ref{2.3}).
This ${\rm T}^{*}$ ``time ordering'' arises because
we have path integrals containing products of derivatives of the
field, which can be expressed as derivatives of the path
integrals which do not contain such derivatives. 
Notice, from their definitions, that all the kernels
which appear 
in expressions (\ref{derivatives}) are real and also that 
$H_{\scriptscriptstyle \!{\rm A}}^{\mu\nu\rho\sigma}$ is
free of ultraviolet divergencies in the limit 
$n \to 4$. 

{}From Eqs. (\ref{derivatives})  
it is clear that the imaginary part of the 
influence action, which does not contribute to the semiclassical 
Einstein equation (\ref{2.5})
because the expectation value of 
$\hat{T}^{\mu\nu}[g]$ is real, contains information on the
fluctuations of this operator. From (\ref{s-t expect value}) and
(\ref{derivatives}), taking into account that 
$S_{\rm IF}[g,g]=0$ and that 
$S_{\rm IF}[g^-,g^+]=
-S^{ {\displaystyle \ast}}_{\rm IF}[g^+,g^-]$, we can write the
expansion for the influence action 
$S_{\rm IF}[g\!+\!h^+,g\!+\!h^-]$ around a background
metric $g_{ab}$ in terms of the previous kernels.
Taking into account that 
these kernels satisfy the symmetry relations
\begin{equation}
H_{\scriptscriptstyle \!{\rm S}}^{\mu\nu\rho\sigma}(x,y)=
H_{\scriptscriptstyle \!{\rm S}}^{\rho\sigma\mu\nu}(y,x), 
\ \  
H_{\scriptscriptstyle \!{\rm A}}^{\mu\nu\rho\sigma}(x,y)=
-H_{\scriptscriptstyle \!{\rm A}}^{\rho\sigma\mu\nu}(y,x), 
\ \ 
K^{\mu\nu\rho\sigma}(x,y) = K^{\rho\sigma\mu\nu}(y,x),
\label{symmetries}
\end{equation}
and introducing the new kernel 
\begin{equation}
H^{\mu\nu\rho\sigma}(x,y)\equiv 
H_{\scriptscriptstyle \!{\rm S}}^{\mu\nu\rho\sigma}(x,y)
+H_{\scriptscriptstyle \!{\rm A}}^{\mu\nu\rho\sigma}(x,y),
\label{H}
\end{equation}
the expansion of $S_{\rm IF}$ can be finally written as
\begin{eqnarray}
S_{\rm IF}[g\!+\!h^+,g+h^-]
&=& {1\over 2} \int\! d^4x\, \sqrt{- g(x)}\:
\langle \hat{T}^{\mu\nu}(x) \rangle [g] \,
\left[h_{\mu\nu}(x) \right] \nonumber\\
&&-{1\over 2} \int\! d^4x\, d^4y\, \sqrt{- g(x)}\sqrt{- g(y)}\,
\left[h_{\mu\nu}(x)\right]
\left(H^{\mu\nu\rho\sigma}[g](x,y)\!
+\!K^{\mu\nu\rho\sigma}[g](x,y) \right)
\left\{ h_{\rho\sigma}(y) \right\}  \nonumber  \\
&&
+{i\over 2} \int\! d^4x\, d^4y\, \sqrt{- g(x)}\sqrt{- g(y)}\,
\left[h_{\mu\nu}(x) \right]
N^{\mu\nu\rho\sigma}[g](x,y)
\left[h_{\rho\sigma}(y) \right]+0(h^3),
\label{expansion 2}
\end{eqnarray} 
where we have used the notation
\begin{equation}
\left[h_{\mu\nu}\right] \equiv h^+_{\mu\nu}\!-\!h^-_{\mu\nu},
\hspace{5 ex}
\left\{ h_{\mu\nu}\right\} \equiv h^+_{\mu\nu}\!+\!h^-_{\mu\nu}.
\label{notation}
\end{equation}

We are now in the position to carry out the formal derivation of the
semiclassical Einstein-Langevin equation. 
The procedure is well known
\cite{CH94,HuSin,cv96,gleiser}, 
it consists of deriving a new
``stochastic'' effective action using the 
the following identity:
\begin{equation}
e^{-{1\over 2} \!\int\! d^4x\, d^4y \, \sqrt{- g(x)}\sqrt{- g(y)}\,
\left[h_{\mu\nu}(x) \right]\, 
N^{\mu\nu\rho\sigma}(x,y)\, \left[h_{\rho\sigma}(y)\right] }=
\int\! {\cal D}[\xi]\: {\cal P}[\xi]\, e^{i \!\int\! d^4x \,
\sqrt{- g(x)}\,\xi^{\mu\nu}(x)\,\left[h_{\mu\nu}(x) \right] },
\label{Gaussian path integral}
\end{equation}
where ${\cal P}[\xi]$ is the probability distribution
functional of a Gaussian stochastic tensor $\xi^{\mu\nu}$ 
characterized by the correlators (\ref{2.10})
with $N^{abcd}$ given by Eq.  (\ref{2.8}), 
and where
the path integration measure is assumed to be a scalar under
diffeomorphisms of $({\cal M},g_{\mu\nu})$. The above identity follows
from the identification of the right hand side of
(\ref{Gaussian path integral}) with the characteristic functional for 
the stochastic field $\xi^{\mu\nu}$.
In fact, by differentiation of this expression 
with respect to $\left[h_{\mu\nu}\right]$, it can be checked that this is
the characteristic functional of a stochastic field characterized by the
correlators (\ref{2.10}).
When $N^{\mu\nu\rho\sigma}(x,y)$ is strictly positive definite, the 
probability distribution functional for $\xi^{\mu\nu}$ is explicitly
given by  
\begin{equation}
{\cal P}[\xi]=
 \frac{e^{-{1\over2}\!\int\! d^4x\, d^4y \, \sqrt{-g(x)}\sqrt{-g(y)}\,
 \xi^{\mu\nu}(x) \, N^{-1}_{\mu\nu\rho\sigma}(x,y)\, \xi^{\rho\sigma}(y)}}
 {\int\! {\cal D}\bigl[\bar{\xi}\bigr]\:
   e^{-{1\over2}\!\int\! d^4z\, d^4w \, \sqrt{-g(z)}\sqrt{-g(w)}\,
  \bar{\xi}^{\tau\omega}(z) \, 
N^{-1}_{\tau\omega\epsilon\kappa}(z,w)\, 
\bar{\xi}^{\epsilon\kappa}(w)}},
\label{gaussian probability}
\end{equation}
where $N^{-1}_{\mu\nu\rho\sigma}[g](x,y)$ is the inverse of
$N^{\mu\nu\rho\sigma}[g](x,y)$ defined by
\begin{equation}
\int\! d^4z \, \sqrt{- g(z)}\, N^{\mu\nu\tau\omega}(x,z)
N^{-1}_{\tau\omega\rho\sigma}(z,y)=
{1\over2} \left(\delta^\mu_\rho \delta^\nu_\sigma
+\delta^\mu_\sigma \delta^\nu_\rho \right)
{\delta^4(x\!-\!y) \over \sqrt{- g(x)}}.
\label{inverse}
\end{equation}

We may now introduce the {\it stochastic effective action} as
\begin{equation}
S^s_{\rm eff}[h^+,h^-;g;\xi] \equiv S_{g}[g+h^+]-S_{g}[g+h^-]+
S^s_{\rm IF}[h^+,h^-;g;\xi],
\label{stochastic eff action}
\end{equation}
where the ``stochastic'' influence action is defined as
\begin{equation}
S^s_{\rm IF}[h^+,h^-;g;\xi_n] \equiv 
{\rm Re}\, S_{\rm IF}[g\!+\!h^+,g\!+\!h^-]+\!
\int\! d^4x \,
\sqrt{- g(x)}\,\xi^{\mu\nu}(x)\left[h_{\mu\nu}(x) \right]+0(h^3).
\label{eff influence action}
\end{equation}
Note that the influence functional as defined from the influence
action (\ref{expansion 2}) can be written as a statistical average
over $\xi^{\mu\nu}$:
\begin{equation}
{\cal F}_{\rm IF}[g+h^+,g+h^-]= \left\langle 
e^{i S^s_{\rm IF}[h^+,h^-;g;\xi_n]}
\right\rangle_{\! s}.
\label{infl funt as average}
\end{equation}
Thus, the effect of the imaginary part of the influence action 
(\ref{expansion 2}) on the corresponding influence
functional is equivalent to the averaged effect of the stochastic
source $\xi^{\mu\nu}$ coupled linearly 
to the perturbations $h_{\mu\nu}^{\pm}$

The stochastic equations of motion for $h_{\mu\nu}$ can
now be derived as
\begin{equation}
\left. {1\over\sqrt{-\det (g\!+\!h)(x)}} \, 
\frac{\delta S^s_{\rm eff}[h^+,h^-;g;\xi_n]}{\delta h^+_{ab}(x)} 
\right|_{h^+=h^-=h}=0.
\label{eq of motion}
\end{equation}
Then, from (\ref{perturb s-t expect value}), taking into account that
only the real part of the influence action contributes to the
expectation value of the stress-energy tensor, we get,
to linear order in $h_{\mu\nu}$ the stochastic semiclassical
equations (\ref{2.11}). To be precise we get
first the regularized
$n$-dimensional equations with the bare parameters,
and where instead of the tensor $A^{\mu\nu}$ we get
$(2/3)D^{\mu\nu}$ where
\begin{eqnarray}
D^{\mu\nu} &\equiv& {1\over\sqrt{- g}}   \frac{\delta}{\delta g_{\mu\nu}}
          \int \! d^n x \,\sqrt{- g}
\left(R_{\rho\sigma\tau\omega}R^{\rho\sigma\tau\omega}-
                               R_{\rho\sigma}R^{\rho\sigma}  \right)
\nonumber\\
   & =& {1\over2}\, g^{\mu\nu} \! 
\left(  R_{\rho\sigma\tau\omega} R^{\rho\sigma\tau\omega}-
         R_{\rho\sigma}R^{\rho\sigma}+\Box  R \right) 
      -2R^{\mu\rho\sigma\tau}{R^\nu}_{\rho\sigma\tau}
\nonumber \\
&& 
      - 2 R^{\mu\rho\nu\sigma}R_{\rho\sigma}
      +4R^{\mu\rho}{R_\rho}^\nu
      -3 \Box  R^{\mu\nu}
  +\nabla^{\mu}\nabla^{\nu}  R.
\label{D}
\end{eqnarray}
Of course, when $n=4$ these tensors are related,
$A^{\mu\nu}=(2/3) D^{\mu\nu}$. After that
we renormalize and 
take the limit $n\to 4$ to obtain the Einstein-Langevin
equations in the physical spacetime. 

%%%%%%%%%%%%%%%%%%%%%%%

\subsubsection{Explicit linear form of the Eintein-Langevin equation}

%%%%%%%%%%%%%%%%%%%%%%%

We can write the Einstein-Langevin equation in a more explicit
form by working out the expansion of 
$\langle \hat{T}^{\mu\nu}\rangle [g\!+\!h]$
up to linear order in the
perturbation $h_{\mu\nu}$. 
{}From Eq. (\ref{perturb s-t expect value}), 
we see that this expansion can
be easily obtained from (\ref{expansion 2}).
The result is
\begin{equation}
\langle \hat{T}_n^{\mu\nu}(x) \rangle 
[g\!+\!h] =
\langle \hat{T}_n^{\mu\nu}(x) \rangle 
[g] +
\langle 
\hat{T}_n^{{\scriptscriptstyle (1)}\hspace{0.1ex} \mu\nu}
[g;h](x) \hspace{-0.1ex} \rangle  [g] -
2 \!\int\! \hspace{-0.2ex} d^ny \,
\sqrt{- g(y)} \hspace{0.2ex}  H_n^{\mu\nu\rho\sigma}[g](x,y) 
h_{\rho\sigma}(y) + 0(h^2),
\label{s-t expect value expansion}
\end{equation}
where the operator 
$\hat{T}_n^{{\scriptscriptstyle (1)}\hspace{0.1ex} \mu\nu}$ is defined
from the term of first order in the expansion of 
$T^{ab}[g+h]$ as
\be
T^{\mu\nu}[g\!+\!h]=T^{\mu\nu}[g]+
T^{{\scriptscriptstyle (1)}\hspace{0.1ex} \mu\nu}[g;h]
+0(h^2), 
\label{T(1)}
\ee
using, as always, a Weyl ordering prescription for the operators in the
last definition. Here we use a subscript $n$ on a given tensor
to indicate that we are
explicitly working in $n$-dimensions, as we use dimensional
regularization, and we also use the superindex
${\scriptstyle (1)}$ to generally
indicate that the tensor is the first order correction, linear
in $h_{\mu\nu}$, in a perturbative expansion around the 
background $g_{\mu\nu}$.

Using the Klein-Gordon equation (\ref{2.2}), and
expressions (\ref{2.3})  for the
stress-energy tensor and the corresponding
operator operator, we can write
\be
\hat{T}_n^{{\scriptscriptstyle (1)}\hspace{0.1ex} \mu\nu}
[g;h]=\left({1\over 2}\, g^{\mu\nu}h_{\rho\sigma}-
\delta^\mu_\rho h^\nu_\sigma-
\delta^\nu_\rho h^\mu_\sigma  \right) \hat{T}_{n}^{\rho\sigma}[g]
+{\cal F}^{\mu\nu}[g;h]\, \hat{\phi}_{n}^2[g],  
\label{T(1) operator}
\ee
where ${\cal F}^{\mu\nu}[g;h]$ is the differential operator
\bea
{\cal F}^{\mu\nu} &\equiv& \left(\xi\!-\!{1\over 4}\right)\!\! 
\left(h^{\mu\nu}\!-\!{1\over 2}\, g^{\mu\nu} h^\rho_\rho \right)\!
\Box+
{\xi \over 2} \left[ 
\nabla^{\rho}\! \nabla^{\mu}\! h^\nu_\rho+
\nabla^{\rho}\! \nabla^{\nu}\! h^\mu_\rho- 
\Box h^{\mu\nu}-
\nabla^{\mu}\! \nabla^{\nu}\!  h^\rho_\rho-
g^{\mu\nu}\! \nabla^{\rho}\!
\nabla^{\sigma} h_{\rho\sigma}
\right.   \nn \\
&&+\left. g^{\mu\nu} \Box h^\rho_\rho 
+\left( \nabla^{\mu} h^\nu_\rho+
\nabla^{\nu} h^\mu_\rho-\nabla_{\rho} 
\hspace{0.2ex} h^{\mu\nu}-
2 g^{\mu\nu}\! \nabla^{\sigma}\! h_{\rho\sigma} +
g^{\mu\nu}\! \nabla_{\rho} \! h^\sigma_\sigma
\right)\! \nabla^{\rho}
-g^{\mu\nu} h_{\rho\sigma} \nabla^{\rho}\!
\nabla^{\sigma} 
\right]. 
\label{diff operator F}
\eea
It is understood that indices are raised 
with the background inverse metric $g^{\mu\nu}$ and that all the
covariant derivatives are associated to the metric $g_{\mu\nu}$.

Substituting (\ref{s-t expect value expansion}) into 
the $n$-dimensional version of the Einstein-Langevin
Eq. (\ref{2.12}),
taking into account that
$g_{\mu\nu}$ satisfies the semiclassical Einstein equation
(\ref{2.5}), and substituting expression (\ref{T(1) operator}) 
we can write the Einstein-Langevin
equation in dimensional regularization as

\bea
&&{1\over 8 \pi G_{B}}\Biggl[
G^{{\scriptscriptstyle (1)}\hspace{0.1ex} \mu\nu}\!-\!
{1\over 2}\, g^{\mu\nu} G^{\rho\sigma}
h_{\rho\sigma}+ G^{\mu\rho} h^\nu_\rho+G^{\nu\rho} h^\mu_\rho+ 
\Lambda_{B} \left( h^{\mu\nu}\!-\!{1\over 2}\,
g^{\mu\nu} h^\rho_\rho \right) 
\Biggr](x) 
   \nn \\
&&
- \, 
{4\over 3}\, \alpha_{B} \left( D^{{\scriptscriptstyle
(1)}\mu\nu}
-{1\over 2}\, g^{\mu\nu} D^{\rho\sigma} h_{\rho\sigma}+
D^{\mu\rho} h^\nu_\rho+D^{\nu\rho} h^\mu_\rho
\right)\! (x)
-2\beta_{B}\left( B^{{\scriptscriptstyle (1)}\mu\nu}\!-\!
{1\over 2}\, g^{\mu\nu} B^{\rho\sigma}
h_{\rho\sigma}+ B^{\mu\rho} h^\nu_\rho+B^{\nu\rho} h^\mu_\rho 
\right)\! (x)   \nn \\
&&- \, \mu^{-(n-4)}\, {\cal F}^{\mu\nu}_x 
\langle \hat{\phi}_{n}^2(x) \rangle [g]
+2 \!\int\! d^ny \, \sqrt{- g(y)}\, \mu^{-(n-4)} 
H_n^{\mu\nu\rho\sigma}[g](x,y)\, h_{\rho\sigma}(y)
=2 \mu^{-(n-4)} \xi^{\mu\nu}_n(x),  
\label{Einstein-Langevin eq 3} 
\eea
where the tensors $G^{\mu\nu}$, $D^{\mu\nu}$ and
$B^{\mu\nu}$ are computed from the semiclassical metric $g_{\mu\nu}$,
and where we have omitted the functional dependence on $g_{\mu\nu}$ and
$h_{\mu\nu}$ in $G^{{\scriptscriptstyle (1)}\mu\nu}$,
$D^{{\scriptscriptstyle (1)}\mu\nu}$,
$B^{{\scriptscriptstyle (1)}\mu\nu}$ and
${\cal F}^{\mu\nu}$ to simplify the notation. The parameter
$\mu$ is a mass scale which relates the dimensions of the
physical field $\phi$ with the dimensions of the
corresponding field in
$n$-dimensions, $\phi_n=\mu^{(n-4)/2}\phi$.
Notice that, in Eq. (\ref{Einstein-Langevin eq 3}), 
all the ultraviolet divergencies in
the limit $n\to 4$, which must be removed by
renormalization of the coupling constants, are in 
$\langle \hat{\phi}_{n}^2(x) \rangle$ and the
symmetric part 
$H_{\scriptscriptstyle \!
{\rm S}_{\scriptstyle n}}^{\mu\nu\rho\sigma}(x,y)$ of the
kernel  $H_n^{\mu\nu\rho\sigma}(x,y)$, whereas the
kernels $N_n^{\mu\nu\rho\sigma}(x,y)$ and
$H_{\scriptscriptstyle \!
{\rm A}_{\scriptstyle n}}^{\mu\nu\rho\sigma}(x,y)$ are
free of ultraviolet divergencies.
If we introduce the bi-tensor
$F_{n}^{\mu\nu\rho\sigma}[g](x,y)$ defined by
\begin{equation}
F_{n}^{\mu\nu\rho\sigma}[g](x,y) \equiv
\left\langle \hat{t}_n^{\mu\nu}(x)\,
\hat{t}_n^{\rho\sigma}(y)
  \right\rangle \![g]
\label{bitensor F}
\end{equation}
where $\hat t^{\mu\nu}$ is defined by Eq. (\ref{2.9}),
then the kernels $N$ and $H_A$ can be written
as
\be
N_n^{\mu\nu\rho\sigma}[g](x,y)=
{1\over 4}\,{\rm Re} \, F_{n}^{\mu\nu\rho\sigma}[g](x,y), 
\hspace{7ex}
H_{\scriptscriptstyle \!
{\rm A}_{\scriptstyle n}}^{\mu\nu\rho\sigma}[g](x,y)=
{1\over 4}\,{\rm Im} \, F_{n}^{\mu\nu\rho\sigma}[g](x,y),   
\label{finite kernels}
\ee
where we have used that 
$2 \left\langle \hat{t}^{\mu\nu}(x)\,
\hat{t}^{\rho\sigma}(y)  \right\rangle=
\left\langle \left\{ \hat{t}^{\mu\nu}(x), \, \hat{t}^{\rho\sigma}(y)
\right\}\right\rangle +
\left\langle \left[ \hat{t}^{\mu\nu}(x), 
\, \hat{t}^{\rho\sigma}(y)\right]\right\rangle$, 
and the fact that the first term on the right hand side of this
identity is real, whereas the second one is pure imaginary.
Once we perform the renormalization procedure
in Eq.~(\ref{Einstein-Langevin eq 3}), setting
$n = 4$ will yield the physical semiclassical
Einstein-Langevin equation. 
Due to the presence of the kernel
$H_n^{\mu\nu\rho\sigma}(x,y)$,
this equation will be usually non-local in the metric perturbation. 

%%%%%%%%%%%%%%%%%%%%%%%%%%%%%%%%%%%

\subsubsection{The kernels for the vacuum state}

%%%%%%%%%%%%%%%%%%%%%%%%%%%%%%%%%%%%

When the
expectation values in the Einstein-Langevin equation 
are taken in a vacuum state $|0 \rangle$,
such as, for instance, an ``in'' vacuum,  
we can go further. Since then we can write these 
expectation values in terms of the Wightman and Feynman
functions, defined as
\be
G_n^+(x,y) \equiv \langle 0| \,
   \hat{\phi}_{n}(x)  \hat{\phi}_{n}(y) \,
   |0 \rangle [g],
\hspace{5 ex}
i G\!_{\scriptscriptstyle F_{\scriptstyle \hspace{0.1ex}  n}}
 \hspace{-0.2ex}(x,y) 
  \equiv \langle 0| \,
  {\rm T}\! \left( \hat{\phi}_{n}(x)  \hat{\phi}_{n}(y) \right)
  \hspace{-0.2ex}
  |0 \rangle [g].
\label{Wightman and Feynman functions}
\ee 
These expressions for the kernels in the Einstein-Langevin 
equation will be very useful for explicit 
calculations.
To simplify the notation, we omit the functional 
dependence on the semiclassical metric $g_{\mu\nu}$, which will be
understood in all the expressions below.

{}From Eqs. (\ref{finite kernels}), we see that the kernels
$N_n^{\mu\nu\rho\sigma}(x,y)$ and 
$H_{\scriptscriptstyle \!
{\rm A}_{\scriptstyle n}}^{\mu\nu\rho\sigma}(x,y)$ 
are the real and imaginary parts,
respectively, of the bi-tensor
$F_{n}^{\mu\nu\rho\sigma}(x,y)$.
{}From the expression (\ref{regul s-t 2})
we see that
the stress-energy operator $\hat{T}_n^{\mu\nu}$
can be written as a
sum of terms of the form $\left\{ {\cal A}_x \hat{\phi}_{n}(x), 
\,{\cal B}_x \hat{\phi}_{n}(x)\right\}$, where ${\cal A}_x$ and 
${\cal B}_x$ are some differential operators. It  then follows
that we can express the bi-tensor 
$F_{n}^{\mu\nu\rho\sigma}(x,y)$ in
terms of the Wightman function as 

\bea
F_{n}^{\mu\nu\rho\sigma}(x,y)
&=&\nabla^{\mu}_x\!
 \nabla^{\rho}_y\! G_n^+(x,y) 
 \nabla^{\nu}_x\!
 \nabla^{\sigma}_y G_n^+(x,y)
+\nabla^{\mu}_x\!
 \nabla^{\sigma}_y\! G_n^+(x,y) 
 \nabla^{\nu}_x\!
 \nabla^{\rho}_y G_n^+(x,y)
   \nn \\
&& 
+\, 2\, {\cal D}^{\mu\nu}_{x}  \bigl(
  \nabla^{\rho}_y G_n^+(x,y)
  \nabla^{\sigma}_y\! G_n^+(x,y) \bigr)
+2\, {\cal D}^{\rho\sigma}_{y} \bigl(
  \nabla^{\mu}_x G_n^+(x,y)
  \nabla^{\nu}_x\! G_n^+(x,y) \bigr)
+2\, {\cal D}^{\mu\nu}_{x} 
   {\cal D}^{\rho\sigma}_{y}  \bigl(
 G_n^{+ 2}(x,y)  \bigr),  
\label{Wightman expression 2}
\eea
where ${\cal D}^{\mu\nu}_{x}$ is the differential
operator (\ref{diff operator}).
{}From this expression and the relations
(\ref{finite kernels}), we get expressions for the kernels
$N_n$ and 
$H_{\scriptscriptstyle \!{\rm A}_{\scriptstyle n}}$ in
terms of the Wightman function $G_n^+(x,y)$. 

Similarly the kernel 
$H_{\scriptscriptstyle \!
{\rm S}_{\scriptstyle n}}^{\mu\nu\rho\sigma}(x,y)$,
can be written in terms of the Feynman
function as
\bea
H_{\scriptscriptstyle \!
{\rm S}_{\scriptstyle n}}^{\mu\nu\rho\sigma}(x,y)=
- {1 \over 4} \, {\rm Im} \Bigl[ &&
 \nabla^{\mu}_{{x}}\!
 \nabla^{\rho}_{{y}}\!
     G\!_{\scriptscriptstyle F_{\scriptstyle \hspace{0.1ex}  n}}
 \hspace{-0.2ex}(x,y)
 \nabla^{\nu}_{{x}}\!
 \nabla^{\sigma}_{{y}}
     G\!_{\scriptscriptstyle F_{\scriptstyle \hspace{0.1ex}  n}}
 \hspace{-0.2ex}(x,y)
+\nabla^{\mu}_{{x}}\!
 \nabla^{\sigma}_{{y}}\! 
     G\!_{\scriptscriptstyle F_{\scriptstyle \hspace{0.1ex}  n}}
 \hspace{-0.2ex}(x,y)
 \nabla^{\nu}_{{x}}\!
 \nabla^{\rho}_{{y}}
     G\!_{\scriptscriptstyle F_{\scriptstyle \hspace{0.1ex}  n}}
 \hspace{-0.2ex}(x,y)   \nn \\
&& 
-\,g^{\mu\nu}(x) \nabla^{\tau}_{{x}}\!
 \nabla^{\rho}_{{y}}
     G\!_{\scriptscriptstyle F_{\scriptstyle \hspace{0.1ex}  n}}
 \hspace{-0.2ex}(x,y)
 \nabla_{\!\!\tau}^{{x}}\!
 \nabla^{\sigma}_{{y}}
     G\!_{\scriptscriptstyle F_{\scriptstyle \hspace{0.1ex}  n}}
 \hspace{-0.2ex}(x,y)
-g^{\rho\sigma}(y) \nabla^{\mu}_{{x}}\!
 \nabla^{\tau}_{{y}}
     G\!_{\scriptscriptstyle F_{\scriptstyle \hspace{0.1ex}  n}}
 \hspace{-0.2ex}(x,y)
 \nabla^{\nu}_{{x}}
 \nabla_{\!\!\tau}^{{y}}
     G\!_{\scriptscriptstyle F_{\scriptstyle \hspace{0.1ex}  n}}
 \hspace{-0.2ex}(x,y)    \nn  \\
&& 
+\,{1 \over 2}\, g^{\mu\nu}(x) g^{\rho\sigma}(y) 
 \nabla^{\tau}_{{x}}\!
 \nabla^{\kappa}_{{y}}
     G\!_{\scriptscriptstyle F_{\scriptstyle \hspace{0.1ex}  n}}
 \hspace{-0.2ex}(x,y)
 \nabla_{\!\!\tau}^{{x} }\!
 \nabla_{\!\!\kappa}^{{y} }
     G\!_{\scriptscriptstyle F_{\scriptstyle \hspace{0.1ex}  n}}
 \hspace{-0.2ex}(x,y)
+{\cal K}^{\mu\nu}_{ x}  \bigl(
 2 \hspace{-0.2ex} \nabla^{\rho}_{{y}}\!
   G\!_{\scriptscriptstyle F_{\scriptstyle \hspace{0.1ex}  n}}
   \hspace{-0.2ex}(x,y)
 \nabla^{\sigma}_{{y}}\!
   G\!_{\scriptscriptstyle F_{\scriptstyle \hspace{0.1ex}  n}}
   \hspace{-0.2ex}(x,y)
     \nn   \\
&&   
  -\, g^{\rho\sigma}(y) \nabla^{\tau}_{{y}}\!
   G\!_{\scriptscriptstyle F_{\scriptstyle \hspace{0.1ex}  n}}
   \hspace{-0.2ex}(x,y)
\nabla_{\!\!\tau}^{{y} }\!
     G\!_{\scriptscriptstyle F_{\scriptstyle \hspace{0.1ex}  n}}
     \hspace{-0.2ex}(x,y) \bigr)
+{\cal K}^{\rho\sigma}_{y}  \bigl(
 2 \hspace{-0.2ex} \nabla^{\mu}_{{x}}\!
   G\!_{\scriptscriptstyle F_{\scriptstyle \hspace{0.1ex}  n}}
   \hspace{-0.2ex}(x,y)
 \nabla^{\nu}_{{x}}\!
   G\!_{\scriptscriptstyle F_{\scriptstyle \hspace{0.1ex}  n}}
   \hspace{-0.2ex}(x,y)
     \nn   \\
&&  
-\, g^{\mu\nu}(x) \nabla^{\tau}_{{x}}\!
   G\!_{\scriptscriptstyle F_{\scriptstyle \hspace{0.1ex}  n}}
   \hspace{-0.2ex}(x,y)
\nabla_{\!\!\tau}^{{x} }\!
     G\!_{\scriptscriptstyle F_{\scriptstyle \hspace{0.1ex}  n}}
     \hspace{-0.2ex}(x,y) \bigr) 
+2\, {\cal K}^{\mu\nu}_{x}
   {\cal K}^{\rho\sigma}_{y}  \bigl(
   G\!_{\scriptscriptstyle F_{\scriptstyle \hspace{0.1ex}  n}}^{\;\: 2}
   \hspace{-0.2ex}(x,y)  \bigr) \Bigr],
\label{Feynman expression 2}
\eea 
where ${\cal K}^{\mu\nu}_{x}$ is the differential
operator
\be
{\cal K}^{\mu\nu}_{x} \equiv 
\xi \left( g^{\mu\nu}(x) \Box_{x}
  -\nabla^{\mu}_{{x}}\!
   \nabla^{\nu}_{{x}}+\,
G^{\mu\nu}(x) \right)
-{1 \over 2}\, m^2 g^{\mu\nu}(x).
\label{diff operator K}
\ee
Note that, in the vacuum state
$|0 \rangle$, the term
$\langle \hat{\phi}_{n}^2 (x) \rangle$ in
equation (\ref{Einstein-Langevin eq 3}) can also be written as
$\langle \hat{\phi}_{n}^2(x) \rangle=
i G\!_{\scriptscriptstyle F_{\scriptstyle \hspace{0.1ex}  n}}
      \hspace{-0.2ex}(x,x)=G_n^+(x,x)$. 

Finally, the causality of the Einstein-Langevin equation
(\ref{Einstein-Langevin eq 3})
can be explicitly seen as follows. The non-local terms in that
equation are due to the kernel $H(x,y)$ which is defined in Eq.
(\ref{H}) as the sum of $H_S(x,y)$ and $H_A(x,y)$. Now,
when the points $x$ and $y$ are spacelike
separated, $\hat{\phi}_{n}(x)$ and $\hat{\phi}_{n}(y)$ commute and,
thus, $G_n^+(x,y) \!=\! 
i G\!_{\scriptscriptstyle F_{\scriptstyle \hspace{0.1ex}  n}}
 \hspace{-0.2ex}(x,y) \!=\! 
(1/2) \langle 0| \,
\{ \hat{\phi}_{n}(x) , \hat{\phi}_{n}(y) \} \,  |0 \rangle$, which is
real. Hence, from the above expressions, we have that 
$H_{\scriptscriptstyle \!
{\rm A}_{\scriptstyle n}}^{\mu\nu\rho\sigma}(x,y) \!=\!
H_{\scriptscriptstyle \!
{\rm S}_{\scriptstyle n}}^{\mu\nu\rho\sigma}(x,y) \!=\!
0$, and thus $H_n^{\mu\nu\rho\sigma}(x,y)=0$ 
This fact is not surprising since, from the causality 
of the expectation value of the stress-energy operator, we know that  
the non-local dependence on the metric perturbation in the
Einstein-Langevin equation, see Eq. (\ref{2.11}), must be causal.

%%%%%%%%%%%%%%%%%%%%%%%%%%%%%%%%%%%%%%%%%%%
\subsubsection{Discussion}
%%%%%%%%%%%%%%%%%%%%%%%%%%%%%%%%%%%%%%%%%%

In this section, based on Refs. \cite{MV0,MV1} 
we have shown how a consistent stochastic semiclassical
theory of gravity can be formulated. This theory is a perturbative
generalization of semiclassical gravity which describes the back
reaction of the lowest order stress-energy fluctuations of quantum
matter fields on the gravitational field through the semiclassical
Einstein-Langevin equation. We have shown that this equation 
can be formally derived with a method
based on the influence functional of Feynman and Vernon, where one
considers the metric field as the ``system'' of interest and the
matter fields as part of its ``environment''.
An explicit linear form of the Einstein-Langevin equations
has been given in terms of some kernels which depend
on the Wightman and Feynman functions when a
vacuum state is considered.

In Ref. \cite{MV1}
the fluctuation-dissipation relations and particle creation in
the stochastic gravity context was also discussed.
When the background solution of semiclassical gravity consists of a
stationary spacetime and a scalar field in a thermal equilibrium state,
it is possible to identify a  dissipation kernel in the Einstein-Langevin
equation which is related to the noise kernel by a
fluctuation-dissipation relation \cite{kubo,martin}.
The same relation was
previously derived by Mottola \cite{Mottola} using a linear response
theory approach, and it is in agreement with previous findings
on a Minkowski background \cite{campos-hu};
see, however, comments in \cite{HRS}.
This result was also  generalized to the case of a
conformal scalar field in a conformally stationary background solution
of semiclassical gravity. 

Particle creation by stochastic metric
perturbations in stationary and conformally stationary
background solutions of
semiclassical gravity was also considered.
It is possible to express the
total probability of particle creation and the number of created
particles (the expectation value of the number operator for 
``out'' particles in the ``in'' vacuum) in terms of the vacuum noise 
kernel. Remarkably the averaged value of those quantities
is enhanced, over the semiclassical result,
by the presence of stochastic metric fluctuations.
In the particular cases of a Minkowski
background and a conformal field in a spatially flat RW background, 
the energy of the created particles can 
be expressed in terms of the vacuum dissipation kernels;
see also Refs. \cite{CH94,RouVer99}.

%%%%%%%%%%%%%%%%%%%%%%%%%%%%%%%%%%%%%%%%
%%%%%%%%%%%%%%%%%%%%%%%%%%%%%%%%%%%%%%%%
\section{METRIC FLUCTUATIONS IN MINKOWSKI SPACETIME}
\label{sec5}
%%%%%%%%%%%%%%%%%%%%%%%%%%%%%%%%%%%%%%%%
%%%%%%%%%%%%%%%%%%%%%%%%%%%%%%%%%%%%%%%%

In this section we describe the first application of the full
stochastic 
semiclassical theory of gravity, where we evaluate the stochastic
gravitational fluctuations in a Minkowski background.
In order to do so, we first 
use the method developed in section \ref{sec2} to 
derive the semiclassical 
Einstein-Langevin equation around a class of 
solutions of semiclassical gravity consisting of 
Minkowski spacetime and a linear real scalar field
in its vacuum state, which may be considered the ground
state of semiclassical gravity. 
Although the Minkowski vacuum is an eigenstate of the total
four-momentum operator of a field in Minkowski spacetime, it 
is not an eigenstate of the stress-energy operator. 
Hence, even for these solutions of semiclassical gravity,
for which the expectation value of the stress-energy operator can
always be chosen to be zero, the fluctuations of this operator are
non-vanishing. This fact leads to consider the stochastic corrections
to these solutions described by the semiclassical Einstein-Langevin
equation. 

We then solve the Einstein-Langevin equation for the linearized
Einstein tensor and compute the associated two-point correlation
functions. Even though, in this case, we expect to have negligibly small
values for these correlation functions for points separated by lengths 
larger than the Planck length, there are several reasons why
it is worth carrying out this calculation.

On the one hand, these are the first
back-reaction solutions of
the full semiclassical Einstein-Langevin
equation. 
There are analogous solutions to a ``reduced'' version 
of this equation inspired in a ``mini-superspace'' model
\cite{ccv97}, and there is also a previous
attempt to obtain a solution to
the Einstein-Langevin equation in Ref.\cite{cv96},
but, there, the non-local terms in the 
Einstein-Langevin equation were neglected.

On the other hand, the results of this calculation, which confirm our
expectations that gravitational fluctuations are negligible at length
scales larger than the Planck length,
but also predict that the fluctuations are strongly
suppressed on small scales, can be considered a first test
of stochastic semiclassical gravity.
In addition, we can extract
conclusions on the possible qualitative behavior of the solutions to
the Einstein-Langevin equation. Thus, it is interesting to
note that the correlation functions at short
scales are characterized by
correlation lengths of the order of the Planck length;
furthermore, such correlation lengths enter in a non-analytic
way in the correlation functions. This kind of non-analytic behavior
is actually quite common in the solutions to Langevin-type equations
with dissipative terms
and hints at the possibility that correlation functions for other
solutions to the Einstein-Langevin equation are
also non-analytic in their characteristic correlation lengths.

%%%%%%%%%%%%%%%%%%%%%%%%%%%%%%%%%%%%%%%%%%%

\subsection{Perturbations around Minkowski spacetime}

%%%%%%%%%%%%%%%%%%%%%%%%%%%%%%%%%%%%%%%%%%%

The Minkowski metric $\eta_{\mu\nu}$ in a manifold
${\cal M}$ which is topologically ${\rm I\hspace{-0.4 ex}R}^{4}$
and the usual Minkowski vacuum, denoted as $|0 \rangle$, 
are the class of simplest solutions 
to the semiclassical Einstein equation (\ref{2.5}),
the so called trivial solutions of semiclassical gravity
\cite{flanagan}. Note that each possible value
of the parameters $(m^2, \xi)$ leads to a different solution.
In fact, 
we can always choose a renormalization scheme in which
the renormalized expectation value 
$\langle 0|\, \hat{T}_{R}^{\mu\nu}\, |0 \rangle [\eta]=0$. Thus,
Minkowski spacetime 
$({\rm I\hspace{-0.4 ex}R}^{4},\eta_{\mu\nu})$
and the vacuum state $|0 \rangle$ are a
solution to the semiclassical Einstein equation
with renormalized cosmological constant $\Lambda\!=\!0$. 
The fact that the vacuum expectation 
value of the renormalized stress-energy operator in Minkowski
spacetime should vanish was originally proposed by Wald \cite{Wald77} 
and it may be understood as a renormalization convention 
\cite{Fulling89,mostepanenko}. 
There are other possible renormalization prescriptions
in which such vacuum expectation
value is proportional to $\eta^{\mu\nu}$, and this would determine the
value of the cosmological constant $\Lambda$ in the semiclassical
equation. Of course, all these 
renormalization schemes give physically equivalent results: 
the total effective cosmological constant, {\it i.e.},  
the constant of proportionality in the sum of all the 
terms proportional to the metric in the semiclassical Einstein and
Einstein-Langevin equations, has to be zero. 

As we have already mentioned the
vacuum $|0 \rangle$ is an eigenstate of the total four-momentum
operator in Minkowski spacetime, but
not an eigenstate of $\hat{T}^{R}_{\mu\nu}[\eta]$. Hence, even in 
the Minkowski background, there are quantum
fluctuations in the stress-energy tensor and, as a result,
the noise kernel does not vanish. 
This fact leads to consider the stochastic corrections 
to this class of trivial solutions of semiclassical
gravity.
Since, in this case, the Wightman and Feynman functions 
(\ref{Wightman and Feynman functions}), their values in the two-point
coincidence limit, and the products of derivatives of two of such
functions appearing in expressions (\ref{Wightman expression 2}) and
(\ref{Feynman expression 2})
are known in dimensional regularization, 
we can compute the semiclassical Einstein-Langevin
equation using the method outlined in section \ref{sec2}.

In Minkowski spacetime, the components of the classical stress-energy
tensor (\ref{2.3}) reduce to
\be
T^{\mu\nu}[\eta,\phi]=\partial^{\mu}\phi
\partial^{\nu} \phi - {1\over 2}\, \eta^{\mu\nu} \hspace{0.2ex}
\partial^{\rho}\phi \partial_{\rho} \phi 
-{1\over 2}\, \eta^{\mu\nu}\hspace{0.2ex} m^2 \phi^2 
+\xi \left( \eta^{\mu\nu} \Box
-\partial^{\mu} \partial^{\nu} \right) \phi^2,
\label{flat class s-t} 
\ee
where $\Box \!\equiv\! \partial_{\mu}\partial^{\mu}$, and the formal
expression for the 
components of the corresponding ``operator'' 
in dimensional regularization, see Eq. (\ref{regul s-t 2}), is
\be
\hat{T}_{n}^{\mu\nu}[\eta] = {1\over 2} \{
     \partial^{\mu}\hat{\phi}_{n} , 
     \partial^{\nu}\hat{\phi}_{n} \}
     + {\cal D}^{\mu\nu} \hat{\phi}_{n}^2,
\label{flat regul s-t}
\ee
where ${\cal D}^{\mu\nu}$ is the differential operator
(\ref{diff operator}), with $g_{\mu\nu}=\eta_{\mu\nu}$,
$R_{\mu\nu}=0$, and $\nabla_\mu=\partial_\mu$.
The field
$\hat{\phi}_{n}(x)$ is the field operator in the Heisenberg
representation in 
a $n$-dimensional Minkowski spacetime, which satisfies the
Klein-Gordon equation (\ref{2.2}).
We use here a stress-energy tensor which differs from the
canonical one, which corresponds to $\xi=0$, both tensors,
however, define the same total momentum. 

The Wightman and Feynman functions 
(\ref{Wightman and Feynman functions}) when
$g_{\mu\nu}=\eta_{\mu\nu}$, are well
known:
\be
G_n^+(x,y) = i \hspace{0.2ex}\Delta_n^+(x-y),
\ \ \ \ 
G\!_{\scriptscriptstyle F_{\scriptstyle \hspace{0.1ex}  n}}
 \hspace{-0.2ex}(x,y) 
  = 
  \Delta_{\scriptscriptstyle F_{\scriptstyle \hspace{0.1ex} n}}
  \hspace{-0.2ex}(x-y),
\label{flat Wightman and Feynman functions}
\ee
with
\bea
&&\Delta_n^+(x)=-2 \pi i \int \! {d^n k \over (2\pi)^n} \,
e^{i kx}\, \delta (k^2+m^2) \,\theta (k^0),
\nn   \\
&&\Delta_{\scriptscriptstyle F_{\scriptstyle \hspace{0.1ex} n}}
  \hspace{-0.2ex}(x)=- \int \! {d^n k \over (2\pi)^n} \, 
{e^{i kx}  \over k^2+m^2-i \epsilon} , 
\hspace{5ex} \epsilon \!\rightarrow \! 0^+,
\label{flat propagators}
\eea
where 
$k^2 \equiv \eta_{\mu\nu} k^{\mu} k^{\nu}$ and
$k x \equiv \eta_{\mu\nu} k^{\mu} x^{\nu}$. 
Note that the derivatives of these functions satisfy
$\partial_{\mu}^{x}\Delta_n^+(x-y)
= \partial_{\mu}\Delta_n^+(x-y)$ and
$\partial_{\mu}^{y}\Delta_n^+(x-y)=
 - \partial_{\mu}\Delta_n^+(x-y)$,
and similarly for the Feynman propagator 
$\Delta_{\scriptscriptstyle F_{\scriptstyle \hspace{0.1ex} n}}
 \hspace{-0.2ex}(x-y)$.

To write down the semiclassical Einstein equation 
(\ref{2.5}) in $n$-dimensions for this case, we need to
compute the vacuum expectation value of the 
stress-energy operator components  
(\ref{flat regul s-t}). Since, from 
(\ref{flat Wightman and Feynman functions}), we have that
$\langle 0 |\hat{\phi}_{n}^2(x)|0 \rangle=
i\Delta_{\scriptscriptstyle F_{\scriptstyle \hspace{0.1ex} n}}
\hspace{-0.2ex}(0)
=i \Delta_n^+(0)$, which is a constant (independent
of $x$), we have simply 
\be
\langle 0 |\, \hat{T}_{n}^{\mu\nu}\, |0 \rangle [\eta]=
-i \left( \partial^{\mu} \partial^{\nu} 
\Delta_{\scriptscriptstyle F_{\scriptstyle \hspace{0.1ex} n}} 
\right) \!(0)=-i \int {d^n k \over (2\pi)^n} \, 
{k^{\mu} k^{\nu} \over k^2+m^2-i \epsilon}
= {\eta^{\mu\nu} \over 2} \left( {m^2 \over 4 \pi} \right)^{\! n/2} 
\! \Gamma \!\left(- {n \over 2}\right),
\label{vev}
\ee
where the integrals in dimensional regularization have been computed
in the standard way (see Ref. \cite{MV2}) 
and where $\Gamma (z)$ is the 
Euler's gamma function. The semiclassical Einstein equation 
(\ref{2.5}) in $n$-dimensions before renormalization
reduces now to
\be
{\Lambda_{B} \over 8 \pi G_{B}}\, \eta^{\mu\nu}
= \mu^{-(n-4)}
\langle 0 | \hat{T}_{n}^{\mu\nu}|0 \rangle [\eta] . 
\label{flat semiclassical eq} 
\ee
This equation, thus, simply sets the value
of the bare coupling constant 
$\Lambda_{B}/G_{B}$.
Note, from (\ref{vev}), that in order to have 
$\langle 0|\, \hat{T}_{R}^{\mu\nu}\, |0 \rangle [\eta]\!=\! 0$, 
the renormalized and regularized stress-energy tensor 
``operator'' for a scalar field in Minkowski spacetime,
see Eq. (\ref{2.4}),
has to be defined as 
\be
\hat{T}_{R}^{\mu\nu}[\eta] =  
\mu^{-(n-4)}\, \hat{T}_{n}^{\mu\nu}[\eta]
-{ \eta^{\mu\nu} \over 2} \, {m^4 \over (4\pi)^2}  
\left( {m^2 \over 4 \pi \mu^2}
\right)^{\!_{\scriptstyle n-4 \over 2}}
\! \Gamma \!\left(- {n \over 2}\right),
\label{flat renorm s-t operator}
\ee 
which corresponds to a renormalization of the cosmological constant
\be
{\Lambda_{B} \over G_{B}}={\Lambda \over G}
-{2 \over \pi} \, {m^4 \over n \hspace{0.2ex}(n\!-\!2)} 
\: \kappa_n 
+O(n-4),
\label{cosmological ct renorm 2}
\ee
where
\be
\kappa_n \equiv {1 \over (n\!-\!4)} 
\left({e^\gamma m^2 \over 4 \pi \mu^2} \right)
^{\!_{\scriptstyle n-4 \over 2}}=
{1 \over n\!-\!4}
+{1\over 2}\, 
\ln \!\left({e^\gamma m^2 \over 4 \pi \mu^2} \right)+O (n-4), 
\label{kappa}
\ee
being $\gamma$ the Euler's constant. In the case of a
massless scalar field, $m^2\!=\!0$, one simply has  
$\Lambda_{B} / G_{B}=\Lambda / G$. Introducing this renormalized
coupling constant into Eq.~(\ref{flat semiclassical eq}), we can 
take the limit $n \!\rightarrow \! 4$.
We find again that, 
for $({\rm I\hspace{-0.4 ex}R}^{4}, \eta_{ab},|0 \rangle )$ to 
satisfy the semiclassical Einstein equation, 
we must take $\Lambda\!=\!0$.

We can now write down the  
Einstein-Langevin equations for the components 
$h_{\mu\nu}$ of the stochastic metric perturbation 
in dimensional regularization. 
In our case, using $\langle 0 |\hat{\phi}_{n}^2(x)|0 \rangle=
i\Delta_{\scriptscriptstyle F_{\scriptstyle \hspace{0.1ex} n}}
\hspace{-0.2ex}(0)$ 
and the explicit expression of 
Eq. (\ref{Einstein-Langevin eq 3}) 
we obtain 
\bea
&&{1\over 8 \pi G_{B}}\Biggl[
G^{{\scriptscriptstyle (1)}\hspace{0.1ex} \mu\nu} + 
\Lambda_{B} \left( h^{\mu\nu}
\!-\!{1\over 2}\, \eta^{\mu\nu} h \right) 
\Biggr](x) -
{4\over 3}\, \alpha_{B} D^{{\scriptscriptstyle
(1)}\hspace{0.1ex} \mu\nu}(x)
-2\beta_{B}B^{{\scriptscriptstyle (1)}\hspace{0.1ex} \mu\nu}(x) 
\nn   \\
&&-\, \xi\, G^{{\scriptscriptstyle (1)}\hspace{0.1ex} \mu\nu}(x)
\mu^{-(n-4)}\, 
i\Delta_{\scriptscriptstyle F_{\scriptstyle \hspace{0.1ex} n}}
\hspace{-0.2ex}(0) 
+2 \!\int\! d^ny \, \mu^{-(n-4)} 
H_n^{\mu\nu\alpha\beta}(x,y)\, h_{\alpha\beta}(y)
=2 \xi^{\mu\nu}(x). 
\label{flat Einstein-Langevin eq}
\eea 
The indices in $h_{\mu\nu}$ are raised with the Minkowski 
metric and $h \equiv h_{\rho}^{\rho}$, and here
a superindex ${\scriptstyle (1)}$  denotes the components of a
tensor linearized around the flat metric.

Note that in $n$-dimensions the correlator for the field
$\xi^{\mu\nu}$ is written as
\be
\langle \xi^{\mu\nu}(x)\xi^{\alpha\beta}(y) \rangle_s
=\mu^{-2 \hspace{0.2ex} (n-4)}  N_n^{\mu\nu\alpha\beta}(x,y),
\label{correlator}
\ee

Explicit expressions for 
$D^{{\scriptscriptstyle (1)}\hspace{0.1ex} \mu\nu}$ and 
$B^{{\scriptscriptstyle (1)}\hspace{0.1ex} \mu\nu}$ are
given by
\be
D^{{\scriptscriptstyle (1)}\hspace{0.1ex} \mu\nu}(x)=
{1 \over 2}\, (3 {\cal F}^{\mu\alpha}_{x} {\cal F}^{\nu\beta}_{x}
-{\cal F}^{\mu\nu}_{x} {\cal F}^{\alpha\beta}_{x})
\, h_{\alpha\beta}(x),
\hspace{7.2ex}
B^{{\scriptscriptstyle (1)}\hspace{0.1ex} \mu\nu}(x)= 
2  {\cal F}^{\mu\nu}_{x} {\cal F}^{\alpha\beta}_{x} 
h_{\alpha\beta}(x),
\label{D, B tensors}
\ee
where ${\cal F}^{\mu\nu}_{x}$ is the differential operator
${\cal F}^{\mu\nu}_{x} \equiv \eta^{\mu\nu} \Box_x
-\partial^\mu_{x} \partial^\nu_{x}$.

%%%%%%%%%%%%%%%%%%%%%%%%%%%%%%%%%%%%%%%%%%%%
\subsection{The kernels in the Minkowski background}
%%%%%%%%%%%%%%%%%%%%%%%%%%%%%%%%%%%%%%%%%%%%
%%%%%%%%%%%%%%%%%%%%%%%%%%%%%%%%%%%%%%%%%%%
\subsubsection{The noise and dissipation kernels}
%%%%%%%%%%%%%%%%%%%%%%%%%%%%%%%%%%%%%%%%%%%%

Since the two kernels (\ref{finite kernels}) are free of ultraviolet
divergencies in the limit
$n\!\rightarrow \! 4$, we can deal directly with the
$F^{\mu\nu\alpha\beta}(x-y)\equiv
\lim_{n \rightarrow 4} \mu^{-2 \hspace{0.2ex} (n-4)} \,
F^{\mu\nu\alpha\beta}_n$ in Eq. (\ref{bitensor F}).
The kernels 
$4 N^{\mu\nu\alpha\beta}(x,y)
={\rm Re}\, F^{\mu\nu\alpha\beta}(x-y)$ and 
$4 H_{\scriptscriptstyle \!{\rm A}}^{\mu\nu\alpha\beta}(x,y) 
= {\rm Im}\, F^{\mu\nu\alpha\beta}(x-y)$ 
are actually 
the components of the ``physical''
noise and dissipation kernels that will appear in the 
Einstein-Langevin equations once the renormalization procedure has
been carried out. 
The bi-tensor $F^{\mu\nu\alpha\beta}$ can be
expressed in terms of the Wightman function in four spacetime
dimensions,
in the following way:
\bea
F^{\mu\nu\alpha\beta}(x)=-  2 \, &&\left[ 
\partial^\mu \partial^{( \alpha} \Delta^+(x) \,
\partial^{\beta )} \partial^\nu \Delta^+(x)
+{\cal D}^{\mu\nu} \!\left( \partial^\alpha \Delta^+(x) \,
\partial^\beta \Delta^+(x) \right)
\right.
\nn   \\
&&\left. \hspace{2ex}
+\, {\cal D}^{\alpha\beta} \!\left( \partial^\mu \Delta^+(x) \,
\partial^\nu \Delta^+(x) \right)
+{\cal D}^{\mu\nu} {\cal D}^{\alpha\beta} 
\!\left( \Delta^{+ \hspace{0.2ex} 2}(x) \right)
\right].  
\label{M 2}
\eea
The different terms in Eq.~(\ref{M 2}) can be easily computed using
the integrals
\be
I(p) \equiv \int\! {d^4 k \over (2\pi)^4} \:
\delta (k^2+m^2) \,\theta (-k^0)  \,
\delta [(k-p)^2+m^2]\,\theta (k^0-p^0),
\label{integrals}
\ee
and $I^{\mu_1 \dots \mu_r}(p)$ which are defined as the
previous one by inserting
the momenta $k^{\mu_1}\dots k^{\mu_r}$
with $r \!=\! 1, 2, 3 ,4$ in the integrand. All these integral can
be expressed in terms of $I(p)$; see Ref. \cite{MV2}
for the explicit expressions.
It is convenient to separate  $I(p)$ in its even
and odd parts with respect to the variables $p^{\mu}$ as
\be
I(p)=I_{\scriptscriptstyle {\rm S}}(p)
+I_{\scriptscriptstyle {\rm A}}(p),
\label{I}
\ee
where $I_{\scriptscriptstyle {\rm S}}(-p)=
I_{\scriptscriptstyle {\rm S}}(p)$ and 
$I_{\scriptscriptstyle {\rm A}}(-p)=
-I_{\scriptscriptstyle {\rm A}}(p)$. These two functions are
explicitly given by
\bea
&&I_{\scriptscriptstyle {\rm S}}(p)={1 \over 8 \, (2 \pi)^3} \;
\theta (-p^2-4m^2) \, \sqrt{1+4 \,{m^2 \over p^2} },
\nn  \\
&&I_{\scriptscriptstyle {\rm A}}(p)={-1 \over 8 \, (2 \pi)^3} \;
{\rm sign}\,p^0 \;
\theta (-p^2-4m^2) \, \sqrt{1+4 \,{m^2 \over p^2} }.
\label{S and A parts of I}
\eea
After some manipulations, we find
\bea
F^{\mu\nu\alpha\beta}(x)= &&{\pi^2 \over 45}\, 
(3 {\cal F}^{\mu (\alpha}_{x}{\cal F}^{\beta )\nu}_{x}-
{\cal F}^{\mu\nu}_{x}{\cal F}^{\alpha\beta}_{x})
\int\! {d^4 p \over (2\pi)^4} \,
e^{-i px}\hspace{0.1ex} 
\left(1+4 \,{m^2 \over p^2} \right)^2 I(p)
\nn   \\
&&+\,{8 \pi^2 \over 9 } \, 
{\cal F}^{\mu\nu}_{x}{\cal F}^{\alpha\beta}_{x}
\int\! {d^4 p \over (2\pi)^4} \,
e^{-i px}\hspace{0.1ex} 
\left(3 \hspace{0.3ex}\Delta \xi+{m^2 \over p^2} \right)^2 I(p),
\label{M 3}
\eea
where $\Delta \xi \equiv \xi - 1/6$. The real and imaginary parts of 
the last expression, which
yield the noise and dissipation kernels, are easily recognized as
the terms containing $I_{\scriptscriptstyle {\rm S}}(p)$ and
$I_{\scriptscriptstyle {\rm A}}(p)$, respectively. To write them
explicitly, it is useful to introduce the new kernels
\bea
&&N_{\rm A}(x;m^2) \equiv
{1 \over 1920 \pi} \int\! {d^4 p \over (2\pi)^4} \,
e^{i px}\,
\theta (-p^2-4m^2) \, \sqrt{1+4 \,{m^2 \over p^2} } 
\left(1+4 \,{m^2 \over p^2} \right)^2,
\nn \\
&&N_{\rm B}(x;m^2,\Delta \xi) \equiv
{1 \over 288 \pi} \int\! {d^4 p \over (2\pi)^4} \,
e^{i px}\, 
\theta (-p^2-4m^2) \, \sqrt{1+4 \,{m^2 \over p^2} } 
\left(3 \hspace{0.3ex}\Delta \xi+{m^2 \over p^2} \right)^2,
\nn \\
&&D_{\rm A}(x;m^2) \equiv
{-i \over 1920 \pi} \int\! {d^4 p \over (2\pi)^4} \,
e^{i px}\, {\rm sign}\,p^0 \;
\theta (-p^2-4m^2) \, \sqrt{1+4 \,{m^2 \over p^2} } 
\left(1+4 \,{m^2 \over p^2} \right)^2,
\nn \\
&&D_{\rm B}(x;m^2,\Delta \xi) \equiv
{-i \over 288 \pi} \int\! {d^4 p \over (2\pi)^4} \,
e^{i px}\, {\rm sign}\,p^0 \;
\theta (-p^2-4m^2) \, \sqrt{1+4 \,{m^2 \over p^2} } 
\left(3 \hspace{0.3ex}\Delta \xi+{m^2 \over p^2} \right)^2,
\label{N and D kernels}
\eea
and we finally get
\bea
&&N^{\mu\nu\alpha\beta}(x,y)=
{1 \over 6}\, 
(3 {\cal F}^{\mu (\alpha}_{x}{\cal F}^{\beta )\nu}_{x}-
{\cal F}^{\mu\nu}_{x}{\cal F}^{\alpha\beta}_{x}) \,
N_{\rm A}(x\!-\!y;m^2)
+{\cal F}^{\mu\nu}_{x}{\cal F}^{\alpha\beta}_{x} 
N_{\rm B}(x\!-\!y;m^2,\Delta \xi),
\nn \\
&&H_{\scriptscriptstyle \!{\rm A}}^{\mu\nu\alpha\beta}(x,y)=
{1 \over 6}\, 
(3 {\cal F}^{\mu (\alpha}_{x}{\cal F}^{\beta )\nu}_{x}-
{\cal F}^{\mu\nu}_{x}{\cal F}^{\alpha\beta}_{x}) \,
D_{\rm A}(x\!-\!y;m^2)
+{\cal F}^{\mu\nu}_{x}{\cal F}^{\alpha\beta}_{x} 
D_{\rm B}(x\!-\!y;m^2,\Delta \xi).
\label{noise and dissipation kernels 2}
\eea
Notice that the noise and dissipation kernels defined in 
(\ref{N and D kernels}) are actually real because, for the noise
kernels, only the $\cos px$ terms of the exponentials $e^{i px}$
contribute to the integrals, and, for the dissipation kernels, the
only contribution of such exponentials comes from the $i \sin px$
terms. 

%%%%%%%%%%%%%%%%%%%%%%%%%%%%%%%%%%%%%%
\subsubsection{The kernel $H_S^{\mu\nu\alpha\beta}(x,y)$}
%%%%%%%%%%%%%%%%%%%%%%%%%%%%%%%%%%%%%%

The evaluation of the kernel components
$H_{\scriptscriptstyle \!{\rm S}_{\scriptstyle n}}
^{\mu\nu\alpha\beta}(x,y)$ is a much more cumbersome task. 
Since these quantities contain divergencies in the limit
$n\!\rightarrow \! 4$, we shall compute them using dimensional
regularization. Using Eq.~(\ref{Feynman expression 2}), 
these components can be written in terms of the Feynman propagator 
(\ref{flat propagators}) as
\be
\mu^{-(n-4)}
H_{\scriptscriptstyle \!{\rm S}_{\scriptstyle n}}
^{\mu\nu\alpha\beta}(x,y)= 
{1\over 4}\,{\rm Im}\, K^{\mu\nu\alpha\beta}(x-y),
\label{kernel H_S}
\ee
where
\bea
&& K^{\mu\nu\alpha\beta}(x) \equiv - \mu^{-(n-4)} \biggl\{
2 \partial^\mu \partial^{( \alpha} 
\Delta_{\scriptscriptstyle F_{\scriptstyle \hspace{0.1ex} n}}
   \hspace{-0.2ex}(x) \,
\partial^{\beta )} \partial^\nu 
\Delta_{\scriptscriptstyle F_{\scriptstyle \hspace{0.1ex} n}}
   \hspace{-0.2ex}(x) 
+2 {\cal D}^{\mu\nu} \!\left( \partial^\alpha 
\Delta_{\scriptscriptstyle F_{\scriptstyle \hspace{0.1ex} n}}
   \hspace{-0.2ex}(x)  \,
\partial^\beta 
\Delta_{\scriptscriptstyle F_{\scriptstyle \hspace{0.1ex} n}}
   \hspace{-0.2ex}(x)  \right)
\nn   \\ 
&& \hspace{0.85ex}
+ \, 2 {\cal D}^{\alpha\beta}  \Bigl( \partial^\mu 
\Delta_{\scriptscriptstyle F_{\scriptstyle \hspace{0.1ex} n}}
   \hspace{-0.2ex}(x) \,
\partial^\nu 
\Delta_{\scriptscriptstyle F_{\scriptstyle \hspace{0.1ex} n}}
   \hspace{-0.2ex}(x) \Bigr)
+2 {\cal D}^{\mu\nu} {\cal D}^{\alpha\beta} 
\!\left(  
\Delta_{\scriptscriptstyle F_{\scriptstyle \hspace{0.1ex} n}}^2
   \hspace{-0.2ex}(x) \right)
+\biggl[ \eta^{\mu\nu} \partial^{( \alpha} 
\Delta_{\scriptscriptstyle F_{\scriptstyle \hspace{0.1ex} n}}
   \hspace{-0.2ex}(x)  \,
\partial^{\beta )} 
+ \eta^{\alpha\beta} \partial^{( \mu} 
 \Delta_{\scriptscriptstyle F_{\scriptstyle \hspace{0.1ex} n}}
   \hspace{-0.2ex}(x)  \,
\partial^{\nu )} 
\nn   \\ 
&& \hspace{0.85ex} \left.  
+\, \Delta_{\scriptscriptstyle F_{\scriptstyle \hspace{0.1ex} n}}
   \hspace{-0.2ex}(0) \left( \eta^{\mu\nu} 
{\cal D}^{\alpha\beta}+ \eta^{\alpha\beta} 
{\cal D}^{\mu\nu}  \right)
+{1 \over 4}\, \eta^{\mu\nu} \eta^{\alpha\beta}
\left( 
\Delta_{\scriptscriptstyle F_{\scriptstyle \hspace{0.1ex} n}}
   \hspace{-0.2ex}(x) \Box 
-m^2 
\Delta_{\scriptscriptstyle F_{\scriptstyle \hspace{0.1ex} n}}
   \hspace{-0.2ex}(0)  \right) \right] \delta^n (x)
\biggr\}.  
\label{K}
\eea
Let us define the integrals 
\be
J_n(p) \equiv 
\mu^{-(n-4)} \!\int\! {d^n k \over (2\pi)^n} \:
{1 \over (k^2+m^2-i \epsilon) \,
[(k-p)^2+m^2-i \epsilon] },
\label{integrals in n dim}
\ee
and $J_n^{\mu_1 \dots \mu_r}(p)$ obtained by inserting
the momenta $k^{\mu_1}\dots k^{\mu_r}$ into
the previous integral, together with
\be
I_{0_{\scriptstyle n}} \equiv 
\mu^{-(n-4)} \!\int\! {d^n k \over (2\pi)^n} \:
{1 \over (k^2+m^2-i \epsilon) },
\label{constant integrals in n dim}
\ee
and $I_{0_{\scriptstyle n}}^{\mu_1 \dots \mu_r}$ which are
also obtained by inserting momenta in the integrand.
Then, the different terms in 
Eq.~(\ref{K}) can be computed. These integrals
are explicitly given in Ref. \cite{MV2}. It is found that
$I_{0_{\scriptstyle n}}^{\mu}=0$ and the remaining integrals can be
written in terms of $I_{0_{\scriptstyle n}}$ and $J_n(p)$.
It is useful to introduce the projector 
$P^{\mu\nu}$ orthogonal to $p^\mu$
as
\be 
p^2 P^{\mu\nu} \!\equiv \! \eta^{\mu\nu}
p^2- p^\mu p^\nu
\label{projector}
\ee
then the action of the operator ${\cal F}^{\mu\nu}_{x}$
is simply written as ${\cal F}^{\mu\nu}_{x} \int\! d^n p \,
e^{i p x}\, f(p)
= - \!\int\! d^n p \, e^{i p x}\, f(p) \, p^2 P^{\mu\nu}$ 
where $f(p)$ is an arbitrary function of $p^\mu$.

After a rather long but straightforward calculation, we get, 
expanding around $n\!=\!4$, 
\bea
&&K^{\mu\nu\alpha\beta}(x)={i \over (4\pi)^2} \,
\Biggl\{ \kappa_n \left[ {1 \over 90} \, 
(3 {\cal F}^{\mu (\alpha}_{x}{\cal F}^{\beta )\nu}_{x}-
{\cal F}^{\mu\nu}_{x}{\cal F}^{\alpha\beta}_{x}) \, \delta^n (x)
+4 \hspace{0.3ex} \Delta \xi^2 \,
{\cal F}^{\mu\nu}_{x}{\cal F}^{\alpha\beta}_{x}
 \delta^n (x)
\right. 
\nn  \\
&& \hspace{2ex} 
+\,{2 \over 3}\, {m^2 \over (n\!-\!2)} \:
\bigr( \eta^{\mu\nu} \eta^{\alpha\beta} \Box_x
-\eta^{\mu (\alpha } \eta^{\beta )\nu} \Box_x
+\eta^{\mu (\alpha } \partial^{\beta )}_x \partial^\nu_x 
+\eta^{\nu (\alpha } \partial^{\beta )}_x \partial^\mu_x 
-\eta^{\mu\nu} \partial^\alpha_x \partial^\beta_x
-\eta^{\alpha\beta} \partial^\mu_x \partial^\nu_x \bigl)
\, \delta^n (x)
\nn  \\
&& \hspace{2ex}  
+\, {4 \hspace{0.2ex} m^4 \over n (n\!-\!2)} \:
(2 \hspace{0.2ex}\eta^{\mu (\alpha } \eta^{\beta )\nu}
\!- \eta^{\mu\nu} \eta^{\alpha\beta}) \, \delta^n (x) 
\biggr]
+{1 \over 180} \, 
(3 {\cal F}^{\mu (\alpha}_{x}{\cal F}^{\beta )\nu}_{x}-
{\cal F}^{\mu\nu}_{x}{\cal F}^{\alpha\beta}_{x})
\nn  \\
&& \hspace{2ex}
\times \!\int \! {d^n p \over (2\pi)^n} \,
e^{i p x} \left(1+4 \,{m^2 \over p^2} \right)^2 \!
\bar\phi (p^2)
+{2 \over 9}  \,
{\cal F}^{\mu\nu}_{x}{\cal F}^{\alpha\beta}_{x}
\!\int \! {d^n p \over (2\pi)^n} \, e^{i p x} 
\left(3 \hspace{0.2ex}\Delta \xi+{m^2 \over p^2} \right)^2 \!
\bar\phi (p^2)
\nn  \\
&& \hspace{2ex}
- \left[ {4 \over 675} \, 
(3 {\cal F}^{\mu (\alpha}_{x}{\cal F}^{\beta )\nu}_{x}-
{\cal F}^{\mu\nu}_{x}{\cal F}^{\alpha\beta}_{x}) 
+{1 \over 270} \, (60 \hspace{0.1ex}\xi \!-\! 11) \,
{\cal F}^{\mu\nu}_{x}{\cal F}^{\alpha\beta}_{x}
\right] \delta^n (x)
\nn  \\
&& \hspace{2ex}
-\, m^2 \left[ {2 \over 135} \, 
(3 {\cal F}^{\mu (\alpha}_{x}{\cal F}^{\beta )\nu}_{x}-
{\cal F}^{\mu\nu}_{x}{\cal F}^{\alpha\beta}_{x}) 
+{1 \over 27} \, {\cal F}^{\mu\nu}_{x}{\cal F}^{\alpha\beta}_{x}
\right] \Delta_n(x)
\Biggr\}+ O(n-4),
\label{result for K}
\eea
where $\kappa_n$ has been defined in (\ref{kappa}),
and $\bar\phi (p^2)$ and 
$\Delta_n(x)$ are given by
\bea
\bar\phi (p^2) &\equiv& \int_0^1 d\alpha \: \ln \biggl(1+{p^2 \over m^2}
\, \alpha (1\!-\!\alpha)-i \epsilon \biggr)
= -i \pi \,\theta (-p^2-4m^2) \, \sqrt{1+4 \,{m^2 \over p^2} }
+\varphi (p^2),
\label{phi}\\
\Delta_n(x)&\equiv&
\int\! {d^n p \over (2\pi)^n} \: e^{i p x}\; {1 \over p^2},
\label{Delta_n}
\eea
where $ 
\varphi (p^2) \equiv  \int_0^1 d\alpha \: \ln 
| 1+{p^2 \over m^2}
\, \alpha (1\!-\!\alpha)|
$.

The imaginary part of (\ref{result for K}) [which,
using (\ref{kernel H_S}), gives the kernel components
$\mu^{-(n-4)}
H_{\scriptscriptstyle \!{\rm S}_{\scriptstyle n}}
^{\mu\nu\alpha\beta}(x,y)$] can be easily obtained multiplying 
this expression by $-i$ and retaining only the real part,
$\varphi (p^2)$, of the function $\bar\phi (p^2)$. Making use of this
result, it is easy to compute the contribution of these 
kernel components
to the Einstein-Langevin equation.

%%%%%%%%%%%%%%%%%%%%%%%%%%%%%%%%%%%%%%%
\subsection{The Einstein-Langevin equations}
%%%%%%%%%%%%%%%%%%%%%%%%%%%%%%%%%%%%%%%

With the results of the previous subsections we can 
now write the $n$-dimensional Einstein-Langevin equation
(\ref{flat Einstein-Langevin eq}),
previous to the renormalization.
Taking into account
that, from Eqs.~(\ref{vev}) and (\ref{flat semiclassical eq}),
\be
{\Lambda_{B} \over 8 \pi G_{B}}= -{1 \over 4 \pi^2} \, 
{m^4 \over n (n\!-\!2)} \: \kappa_n + O(n-4),
\label{bare cosmological ct 2}
\ee
we may finally write:
\bea
&&{1\over 8 \pi G_{B}} \, 
G^{{\scriptscriptstyle (1)}\hspace{0.1ex} \mu\nu}(x)
-{4\over 3}\, \alpha_{B} D^{{\scriptscriptstyle
(1)}\hspace{0.1ex} \mu\nu}(x)
-2\beta_{B}B^{{\scriptscriptstyle (1)}\hspace{0.1ex} \mu\nu}(x)
+{\kappa_n \over (4\pi)^2} \, \Biggl[ 
-4 \hspace{0.2ex}\Delta \xi \, {m^2 \over (n\!-\!2)} \, 
G^{{\scriptscriptstyle (1)}\hspace{0.1ex} \mu\nu}
+{1 \over 90} \, 
D^{{\scriptscriptstyle (1)}\hspace{0.1ex} \mu\nu}
\nn  \\
&&+\, \Delta \xi^2  
B^{{\scriptscriptstyle (1)}\hspace{0.1ex} \mu\nu}
\Biggr]\hspace{-0.2ex} (x)
+{1 \over 2880 \pi^2} \, \Biggl\{ 
-{16 \over 15} \, 
D^{{\scriptscriptstyle (1)}\hspace{0.1ex} \mu\nu}(x)
+\left({1 \over 6}-\! 10\hspace{0.2ex} \Delta \xi \right) \!
B^{{\scriptscriptstyle (1)}\hspace{0.1ex} \mu\nu}(x)
\nn  \\
&& 
+ \int\! d^n y \!
\int\! {d^n p \over (2\pi)^n} \, e^{i p (x-y)} \,
\varphi (p^2) \,
\Biggl[\left(1+4 \,{m^2 \over p^2} \right)^2 \!
D^{{\scriptscriptstyle (1)}\hspace{0.1ex} \mu\nu}(y)
+10 \! 
\left(3 \hspace{0.2ex}\Delta \xi+{m^2 \over p^2} \right)^2 \!
B^{{\scriptscriptstyle (1)}\hspace{0.1ex} \mu\nu}(y)
\Biggr]
\nn  \\
&& 
-\, {m^2 \over 3} \!\int\! d^n y \, \Delta_n(x\!-\!y) \,
\Bigl( 8 D^{{\scriptscriptstyle (1)}\hspace{0.1ex} \mu\nu}
+ 5 B^{{\scriptscriptstyle (1)}\hspace{0.1ex} \mu\nu}
\Bigr)\hspace{-0.2ex} (y)
\Biggr\} \!
+2 \!\int\! d^ny \, \mu^{-(n-4)} 
H_{\scriptscriptstyle \!{\rm A}_{\scriptstyle n}}
^{\mu\nu\alpha\beta}(x,y)\, h_{\alpha\beta}(y)
+O(n\!-\!4)
\nn \\
&& =2 \xi^{\mu\nu}(x). 
\label{flat Einstein-Langevin eq 2}
\eea
Notice that the terms containing the bare cosmological constant have
canceled. These equations can now be renormalized, that is,
we can now write the bare coupling constants as renormalized coupling
constants plus some suitably chosen counterterms and take 
the limit $n\!\rightarrow \! 4$. In order to carry out such a
procedure, it is convenient to distinguish between   
massive and massless scalar fields. We shall evaluate these two
cases in different subsections.

%%%%%%%%%%%%%%%%%%%%%%%%%%%%%%%%%%%%%%%%%%%%%%%

\subsubsection{Massive field}

%%%%%%%%%%%%%%%%%%%%%%%%%%%%%%%%%%%%%%%%%%%%%%%

In the case of a scalar field with mass $m \neq 0$, we can use,
as we have done in Eq.~(\ref{cosmological ct renorm 2})
for the cosmological constant,  the renormalized coupling
constants $1/G$, $\alpha$ and $\beta$ as
\bea
&&{1 \over G_B}={1 \over G} 
+{2 \over \pi} \, \Delta \xi \, {m^2 \over (n\!-\!2)} \: \kappa_n 
+O(n-4),
\nn  \\
&&\alpha_{B}=\alpha 
+{1 \over (4 \pi)^2} \, {1 \over 120} \: \kappa_n + O(n-4),
\nn  \\
&&\beta_{B}=\beta 
+ {\Delta \xi^2 \over 32 \pi^2} \: \kappa_n + O(n-4).
\label{massive renormalization}
\eea
Note that for conformal coupling, $\Delta \xi=0$, one
has $1/G_B=1/G$ and $\beta_{B}=\beta$, that is, only
the coupling constant $\alpha$ and the cosmological constant need
renormalization.

Let us introduce the two new kernels
\bea
&&H_{\rm A}(x;m^2) \equiv
{1 \over 1920 \pi^2}\! \int\! {d^4 p \over (2\pi)^4} \,
e^{i px}\,
\Biggl\{ \! \left(1+4 \,{m^2 \over p^2} \right)^{\! 2} 
\Biggl[ - i \pi \, {\rm sign}\,p^0 \;
\theta (-p^2\!-\!4m^2) \, \sqrt{1+4 \,{m^2 \over p^2} } 
\nn  \\
&& \hspace{47.8ex}
+\, \varphi(p^2) \Biggr]
-{8 \over 3} \, {m^2 \over p^2} \Biggr\},
\nn \\
&&H_{\rm B}(x;m^2,\Delta \xi) \equiv
{1 \over 288 \pi^2}\! \int\! {d^4 p \over (2\pi)^4} \,
e^{i px}\, 
\Biggl\{ \! 
\left(3 \hspace{0.3ex}\Delta \xi+{m^2 \over p^2} \right)^{\! 2} 
\Biggl[ - i \pi \, {\rm sign}\,p^0 \;
\theta (-p^2\!-\!4m^2) \, \sqrt{1+4 \,{m^2 \over p^2} } 
\nn  \\
&& \hspace{52.5ex}
+\, \varphi(p^2) \Biggr]
-{1 \over 6} \, {m^2 \over p^2} \Biggr\},
\label{H kernels}
\eea
where $\varphi(p^2)$ is given by the restriction to $n=4$ of 
expression (\ref{phi}).
Substituting  expressions (\ref{massive renormalization}) into 
Eq.~(\ref{flat Einstein-Langevin eq 2}), we can now take the 
limit $n\!\rightarrow \! 4$, using the fact that, for $n=4$,
$D^{{\scriptscriptstyle (1)}\hspace{0.1ex} \mu\nu}(x)=
(3/ 2) \, 
A^{{\scriptscriptstyle (1)}\hspace{0.1ex} \mu\nu}(x)$,
we obtain the semiclassical 
Einstein-Langevin equations for the physical stochastic
perturbations $h_{\mu\nu}$ in the four-dimensional manifold 
${\cal M} \!\equiv \!{\rm I\hspace{-0.4 ex}R}^{4}$: 
\bea
&&{1\over 8 \pi G} \, 
G^{{\scriptscriptstyle (1)}\hspace{0.1ex} \mu\nu}(x)
\hspace{-0.2ex}-\hspace{-0.1ex} 2 \left( \alpha 
A^{{\scriptscriptstyle (1)}\hspace{0.1ex} \mu\nu}(x)
\hspace{-0.1ex}+\hspace{-0.1ex}
\beta B^{{\scriptscriptstyle (1)}\hspace{0.1ex} \mu\nu}(x)
\right) 
\hspace{-0.2ex}+\hspace{-0.1ex}
{1 \over 2880 \pi^2} \left[ -{8 \over 5} \, 
A^{{\scriptscriptstyle (1)}\hspace{0.1ex} \mu\nu}(x)
\!+\!
\left({1 \over 6}\!-\! 10\hspace{0.2ex} \Delta \xi \right) 
\! \hspace{-0.1ex}
B^{{\scriptscriptstyle (1)}\hspace{0.1ex} \mu\nu}(x) \right] 
\nn  \\
&&+ \int\! d^4y 
\left[ H_{\rm A}(x\!-\!y;m^2) 
A^{{\scriptscriptstyle (1)}\hspace{0.1ex} \mu\nu}(y)
+H_{\rm B}(x\!-\!y;m^2,\Delta \xi)
B^{{\scriptscriptstyle (1)}\hspace{0.1ex} \mu\nu}(y) \right]
=2 \xi^{\mu\nu}(x),
\label{massive Einstein-Langevin eq}
\eea
where $\xi^{\mu\nu}$ are the components of a Gaussian stochastic
tensor of vanishing mean value and 
two-point correlation function
$\langle\xi^{\mu\nu}(x)\xi^{\alpha\beta}(y) \rangle_s
=N^{\mu\nu\alpha\beta}(x,y)$, given by one of the Eqs.
(\ref{noise and dissipation kernels 2}).
Note that the two kernels defined in (\ref{H kernels})
are real and can be split into an even part 
and an odd part with respect to the variables $x^\mu$, with the 
odd terms being the dissipation kernels $D_{\rm A}(x;m^2)$ and 
$D_{\rm B}(x;m^2,\Delta \xi)$ defined in (\ref{N and D kernels}).
In spite of appearances, one can show that the Fourier transforms of
the even parts of these kernels are finite in the limit
$p^2 \! \rightarrow \! 0$ and, hence, the kernels $H_{\rm A}$ and
$H_{\rm B}$ are well defined distributions.

We should mention that in
Ref.~\cite{lomb-mazz}, the same Einstein-Langevin equations were
calculated using rather different methods. 
The way in which the result is written makes difficult a
direct comparison with our equations 
(\ref{massive Einstein-Langevin eq}). For instance, it is not obvious
that there is some analog of the
dissipation kernels.

%%%%%%%%%%%%%%%%%%%%%%%%%%%%%%%%%%%%%%%%%%%%%%%
\subsubsection{Massless field}
%%%%%%%%%%%%%%%%%%%%%%%%%%%%%%%%%%%%%%%%%%%%%%%

In this subsection, we consider the limit 
$m \! \rightarrow \! 0$ of equations 
(\ref{flat Einstein-Langevin eq 2}).
The renormalization scheme used in the previous
subsection becomes singular in the massless limit 
because the expressions 
(\ref{massive renormalization}) for $\alpha_{B}$ and $\beta_{B}$ 
diverge when $m \! \rightarrow \! 0$. Therefore, a different
renormalization scheme is needed in this case. 
First, note that we may separate $\kappa_n$ in (\ref{kappa}) as
$\kappa_n=\tilde{\kappa}_n +{1 \over 2}\ln (m^2/\mu ^2)+O(n\!-\!4)$,
where
\be
\tilde{\kappa}_n \equiv {1 \over (n\!-\!4)} 
\left({e^\gamma \over 4 \pi} \right)
^{\!_{\scriptstyle n-4 \over 2}}=
{1 \over n\!-\!4}
+{1\over 2}\, 
\ln \!\left({e^\gamma \over 4 \pi } \right)+O (n-4), 
\label{kappa tilde}
\ee
and that, from Eq.~(\ref{phi}), we have
\be
\lim_{m^2 \rightarrow 0} \left[ \varphi (p^2)+\ln (m^2/\mu ^2)
\right]=-2+\ln 
\left| \hspace{0.2ex} {p^2 \over \mu^2} \hspace{0.2ex}\right|.
\label{massless limit}
\ee
Hence, in the massless limit, equations 
(\ref{flat Einstein-Langevin eq 2}) reduce to
\bea
&&{1\over 8 \pi G_{B}} \, 
G^{{\scriptscriptstyle (1)}\hspace{0.1ex} \mu\nu}(x)
-{4\over 3}\, \alpha_{B} D^{{\scriptscriptstyle
(1)}\hspace{0.1ex} \mu\nu}(x)
-2\beta_{B}B^{{\scriptscriptstyle (1)}\hspace{0.1ex} \mu\nu}(x)
+{1 \over (4\pi)^2} \, (\tilde{\kappa}_n\!-\!1) \hspace{-0.1ex}
\left[ {1 \over 90} \, 
D^{{\scriptscriptstyle (1)}\hspace{0.1ex} \mu\nu}
+\Delta \xi^2  
B^{{\scriptscriptstyle (1)}\hspace{0.1ex} \mu\nu} \right]
\!\hspace{-0.2ex} (x)
\nn  \\
&&
+{1 \over 2880 \pi^2} \, \Biggl\{ 
-{16 \over 15} \, 
D^{{\scriptscriptstyle (1)}\hspace{0.1ex} \mu\nu}(x)
+\left({1 \over 6}-\! 10\hspace{0.2ex} \Delta \xi \right) \!
B^{{\scriptscriptstyle (1)}\hspace{0.1ex} \mu\nu}(x)
+\! \int\! d^n y \!
\int\! {d^n p \over (2\pi)^n} \, e^{i p (x-y)} \,
\ln \left| \hspace{0.2ex} {p^2 \over \mu^2} \hspace{0.2ex}\right|
\biggl[ D^{{\scriptscriptstyle (1)}\hspace{0.1ex} \mu\nu}(y)
\nn  \\
&& 
+\, 90 \hspace{0.2ex}\Delta \xi^2 
B^{{\scriptscriptstyle (1)}\hspace{0.1ex} \mu\nu}(y)
\biggr] \Biggr\} 
+\lim_{m^2 \rightarrow 0}
2 \!\int\! d^ny \, \mu^{-(n-4)} 
H_{\scriptscriptstyle \!{\rm A}_{\scriptstyle n}}
^{\mu\nu\alpha\beta}(x,y)\, h_{\alpha\beta}(y)
+O(n\!-\!4)
=2 \xi^{\mu\nu}(x).
\label{massless Einstein-Langevin eq}
\eea
These equations can be renormalized by introducing the renormalized
coupling constants $1/G$, $\alpha$ and $\beta$ as
\be
{1 \over G_B}\hspace{-0.2ex}=\hspace{-0.2ex} {1 \over G},
\hspace{4ex}
\alpha_{B} \hspace{-0.2ex} =\hspace{-0.2ex} \alpha 
\hspace{-0.1ex}+\hspace{-0.1ex}
{1 \over (4 \pi)^2} \, {1 \over 120} \, (\tilde{\kappa}_n\!-\!1)  
\hspace{-0.1ex}+\hspace{-0.1ex} O(n\!-\!4),
\hspace{4ex}
\beta_{B}\hspace{-0.2ex}=\hspace{-0.2ex} \beta 
\hspace{-0.1ex}+\hspace{-0.1ex} 
{\Delta \xi^2 \over 32 \pi^2} \, (\tilde{\kappa}_n\!-\!1)  
\hspace{-0.1ex}+\hspace{-0.1ex} O(n\!-\!4).
\label{massless renormalization}
\ee
Thus, in the massless limit, the Newtonian gravitational constant 
is not renormalized and, in the conformal coupling
case, $\Delta \xi=0$, we have again that $\beta_{B}\!=\! \beta$. 
Introducing the last expressions
into Eq.~(\ref{massless Einstein-Langevin eq}), we can take the 
limit $n \! \rightarrow \! 4$. 
Note that, by making $m \!=\!0$ in (\ref{N and D kernels}), 
the noise and dissipation kernels can be written as
\bea
&&N_{\rm A}(x;m^2\!=\!0)=N(x),
\hspace{7ex}
N_{\rm B}(x;m^2\!=\!0,\Delta \xi)
=60 \hspace{0.2ex} \Delta \xi^2 \hspace{0.2ex}  N(x),
\nn \\
&&D_{\rm A}(x;m^2\!=\!0)=D(x),
\hspace{7ex}
D_{\rm B}(x;m^2\!=\!0,\Delta \xi)
=60 \hspace{0.2ex} \Delta \xi^2 \hspace{0.2ex} D(x),
\label{massless N and D kernels}
\eea
where
\be
N(x) \equiv {1 \over 1920 \pi} \!\int \! {d^4 p \over (2\pi)^4} \,
e^{i px}\, \theta (-p^2),
\hspace{7ex}
D(x) \equiv {-i \over 1920 \pi} \!\int \! {d^4 p \over (2\pi)^4} \,
e^{i px}\, {\rm sign}\,p^0 \;
\theta (-p^2).
\label{N and D}
\ee
It is now convenient to introduce the new kernel
\bea
H(x;\mu^2) &\equiv & {1 \over 1920 \pi^2} 
\!\int \! {d^4 p \over (2\pi)^4} \, e^{i px}
\left[ 
\ln \left| \hspace{0.2ex} {p^2 \over \mu^2} \hspace{0.2ex}\right|
- i \pi \, {\rm sign}\,p^0 \; \theta (-p^2) \right]
\nn  \\
&=& {1 \over 1920 \pi^2} \lim_{\epsilon \rightarrow 0^+}
\!\int \! {d^4 p \over (2\pi)^4} \, e^{i px} \,
\ln\! \left( {-(p^0+i \epsilon)^2+p^i p_i \over \mu^2}
\right).
\label{Hnew}
\eea
Again, this kernel is real and can be written as the
sum of an even part and an odd part in the variables $x^\mu$, 
where the odd part is the dissipation kernel $D(x)$. 
The Fourier transforms (\ref{N and D}) and (\ref{Hnew}) can actually 
be computed
and, thus, in this case, we have explicit expressions for
the kernels in position space. For $N(x)$ and $D(x)$, we get
(see, for instance, Ref.~\cite{jones})
\be
N(x)= {1 \over 1920 \pi} \left[ {1 \over \pi^3} \, 
{\cal P}\hspace{-0.4ex}f  \hspace{-0.5ex}
\left( {1 \over (x^2)^2} \right) +
\delta^4(x) \right],
\hspace{10ex}
D(x)= {1 \over 1920 \pi^3} \: {\rm sign}\,x^0 \: 
\frac{d}{d(x^{2})}\, \delta(x^{2}),
\label{N and D 2}
\ee
where ${\cal P}\hspace{-0.4ex}f$ denotes a distribution generated by
the Hadamard finite part of a divergent integral, 
see Refs.~\cite{schwartz} for the definition of these distributions.
The expression for the kernel $H(x;\mu^2)$ can be found in 
Refs.~\cite{cmv95,horowitz} and it is given by
\bea
H(x;\mu^2) &=& {1 \over 960 \pi^2} 
\left\{ {\cal P}\hspace{-0.4ex} f \hspace{-0.5ex}
\left( \frac{1}{\pi}\,
\theta(x^{0})\,
\frac{d}{d(x^{2})}\, \delta(x^{2}) \right)+
\left( 1 \!-\!\gamma -\ln \! \mu \right) \delta^{4}(x)
\right\}
\nn    \\
&=& {1 \over 960 \pi^2} \,
\lim_{\lambda \rightarrow 0^{+}} \!\left\{ \frac{1}{\pi}\,
\theta(x^{0})\,
\theta( |{\bf x}| \!-\! \lambda )\,\frac{d}{d(x^{2})}\,
\delta(x^{2}) +
\left[ 1 \!-\!\gamma -\ln  (\mu \lambda) \right] \delta^{4}(x)
\right\}.
\eea
See Ref.~\cite{cmv95} for the details on how this last 
distribution acts on a test function.
 
Finally, the semiclassical Einstein-Langevin
equations for the physical stochastic perturbations $h_{\mu\nu}$ in
the massless case are 
\bea
&&{1\over 8 \pi G} \, 
G^{{\scriptscriptstyle (1)}\hspace{0.1ex} \mu\nu}(x)
\hspace{-0.2ex}-\hspace{-0.1ex} 2 \left(\alpha 
A^{{\scriptscriptstyle (1)}\hspace{0.1ex} \mu\nu}(x)
\hspace{-0.1ex}+\hspace{-0.1ex}
\beta B^{{\scriptscriptstyle (1)}\hspace{0.1ex} \mu\nu}(x)
\right) 
\hspace{-0.2ex}+\hspace{-0.1ex}
{1 \over 2880 \pi^2} \left[ -{8 \over 5} \, 
A^{{\scriptscriptstyle (1)}\hspace{0.1ex} \mu\nu}(x)
\!+\!
\left({1 \over 6}\!-\! 10\hspace{0.2ex} \Delta \xi \right) 
\! \hspace{-0.1ex}
B^{{\scriptscriptstyle (1)}\hspace{0.1ex} \mu\nu}(x) \right] 
\nn  \\
&&+\int\! d^4y \, H(x\!-\!y;\mu^2) 
\left[ 
A^{{\scriptscriptstyle (1)}\hspace{0.1ex} \mu\nu}(y)
+60 \hspace{0.2ex}  \Delta \xi^2
B^{{\scriptscriptstyle (1)}\hspace{0.1ex} \mu\nu}(y) \right]
=2 \xi^{\mu\nu}(x),
\label{massless Einstein-Langevin eq 2}
\eea
where the Gaussian stochastic source components 
$\xi^{\mu\nu}$ have zero mean value and 
\be
\langle\xi^{\mu\nu}(x)\xi^{\alpha\beta}(y) \rangle_s
=\lim_{m \rightarrow 0}\! N^{\mu\nu\alpha\beta}(x,y)=
\left[ {1 \over 6}\, 
(3 {\cal F}^{\mu (\alpha}_{x}{\cal F}^{\beta )\nu}_{x}\!-\!
{\cal F}^{\mu\nu}_{x}{\cal F}^{\alpha\beta}_{x}) 
+ 60 \hspace{0.2ex}  \Delta \xi^2
{\cal F}^{\mu\nu}_{x}{\cal F}^{\alpha\beta}_{x} 
\right]\! N(x\!-\!y).
\label{massless noise}
\ee

It is interesting to consider the conformally coupled scalar field, 
{\it i.e.}, the case $\Delta \xi\!=\!0$, of particular interest
because of its similarities with the electromagnetic field,
and also because of its interest in cosmology: massive fields become
conformally invariant when their masses are negligible
compared to the spacetime curvature.
We have already mentioned that for a conformally coupled, 
field, the
stochastic source tensor must be ``traceless'' (up to first order
in perturbation theory around semiclassical gravity), in the sense that
the stochastic variable 
$\xi^\mu_\mu \!\equiv \!\eta_{\mu\nu}\xi^{\mu\nu}$ behaves
deterministically as a vanishing scalar field. 
This can be directly checked by noticing, from 
Eq.~(\ref{massless noise}), that, when $\Delta \xi\!=\!0$, one has 
$\langle\xi^\mu_\mu(x)\xi^{\alpha\beta}(y) \rangle_s
=0$, since ${\cal F}^\mu_\mu\!=\! 3 \hspace{0.2ex}\Box $ and 
${\cal F}^{\mu \alpha}{\cal F}^\beta_\mu \!=\! 
\Box {\cal F}^{\alpha\beta}$. 
The semiclassical Einstein-Langevin equations for this 
particular case (and generalized to a spatially flat 
Robertson-Walker background)
were first obtained in Ref.~\cite{cv96},
where the coupling constant
$\beta$ was fixed to be  zero. 

Note that the expectation value of the renormalized
stress-energy tensor for a scalar field can be obtained
by identification of Eqs.~(\ref{massive Einstein-Langevin eq})
and (\ref{massless Einstein-Langevin eq 2}) with the components of
the physical Einstein-Langevin equation (\ref{2.11}).
The explicit expressions are given in Ref. \cite{MV2}
The results agree with the
general form found by Horowitz \cite{horowitz,horowitz81} 
using an axiomatic
approach and coincides with that given in Ref.~\cite{flanagan}.
The particular cases of conformal coupling, $\Delta \xi \!=\!0$, and
minimal coupling, $\Delta \xi \!=\!-1/6$, are also in agreement with
the results for this cases given in 
Refs.~\cite{horowitz,horowitz81,horowitz_wald,cv94,jordan} 
(modulo local terms proportional to 
$A^{{\scriptscriptstyle (1)}\hspace{0.1ex} \mu\nu}$ and
$B^{{\scriptscriptstyle (1)}\hspace{0.1ex} \mu\nu}$ due to different
choices of the renormalization scheme).
For the case of a massive minimally coupled scalar field,
$\Delta \xi \!=\!-1/6$, our result 
is equivalent to that of Ref.~\cite{tichy}.

%%%%%%%%%%%%%%%%%%%%%%%%%%%%%%%%%%%%%%%%%%%%%%%%%%%%%%%%%%%%%%%%%%%%%%%%%%%%
%%%%%%%%%%%%%%%%%%%%%%%%%%%%%%%%%%%%%%%%%%%%%%%%%%%%%%%%%%%%%%%%%%%%%%%%%%%%

%%%%%%%%%%%%%%%%%%%%%%%%%%%%%%%%%%%%%%%%%%%%%%%%%%%

\subsection{Correlation functions for gravitational
perturbations}

%%%%%%%%%%%%%%%%%%%%%%%%%%%%%%%%%%%%%%%%%%%%%%%%%%%

In this section, we solve the semiclassical Einstein-Langevin
equations (\ref{massive Einstein-Langevin eq}) and 
(\ref{massless Einstein-Langevin eq 2}) for the components 
$G^{{\scriptscriptstyle (1)}\hspace{0.1ex} \mu\nu}$ 
of the linearized Einstein tensor.
In the first subsection
we use these solutions to
compute the corresponding two-point correlation functions,
which give a measure of the gravitational fluctuations
predicted by the stochastic semiclassical theory of gravity in the
present case. Since the linearized Einstein tensor 
is invariant under gauge transformations 
of the metric perturbations, these two-point correlation functions are
also gauge invariant. Once we have computed the two-point correlation
functions for the linearized Einstein tensor,
we find the solutions for the
metric perturbations in the next subsection
and we show how the associated two-point correlation functions can be
computed.
This procedure to solve the Einstein-Langevin equations is similar to
the one used by Horowitz \cite{horowitz}, see also
Ref.~\cite{flanagan}, to analyze the stability of Minkowski spacetime 
in semiclassical gravity.

We first note that the tensors
$A^{{\scriptscriptstyle (1)}\hspace{0.1ex} \mu\nu}$ and 
$B^{{\scriptscriptstyle (1)}\hspace{0.1ex} \mu\nu}$ can
be written in terms of 
$G^{{\scriptscriptstyle (1)}\hspace{0.1ex} \mu\nu}$ as
\be
A^{{\scriptscriptstyle (1)}\hspace{0.1ex} \mu\nu} =
{2 \over 3} \, ({\cal F}^{\mu\nu} 
G^{{\scriptscriptstyle (1)}}\mbox{}^{\alpha}_\alpha
-{\cal F}^{\alpha}_\alpha 
G^{{\scriptscriptstyle (1)}\hspace{0.1ex} \mu\nu}),
\hspace{10ex}
B^{{\scriptscriptstyle (1)}\hspace{0.1ex} \mu\nu} =
2 \hspace{0.2ex} {\cal F}^{\mu\nu} 
G^{{\scriptscriptstyle (1)}}\mbox{}^{\alpha}_\alpha,
\label{A and B}
\ee
where we have used that 
$3 \hspace{0.2ex}\Box={\cal F}^{\alpha}_\alpha$.
Therefore, the Einstein-Langevin equations 
(\ref{massive Einstein-Langevin eq}) and 
(\ref{massless Einstein-Langevin eq 2}) can be seen as linear
integro-differential stochastic equations for the components
$G^{{\scriptscriptstyle (1)}\hspace{0.1ex} \mu\nu}$. 
These Einstein-Langevin equations can be
written in a unified form, in both cases for
$m \!\neq \!0$ and for $m\!=\!0$, as
\be
{1\over 8 \pi G} \, 
G^{{\scriptscriptstyle (1)}\hspace{0.1ex} \mu\nu}(x)
\hspace{-0.2ex}-\hspace{-0.2ex} 2 \left(\bar{\alpha} 
A^{{\scriptscriptstyle (1)}\hspace{0.1ex} \mu\nu}(x)
\hspace{-0.2ex}+\hspace{-0.2ex}
\bar{\beta} B^{{\scriptscriptstyle (1)}\hspace{0.1ex} \mu\nu}(x)
\right)\! 
+\!\!\int\! d^4y 
\left[ H_{\rm A}(x\!-\!y) 
A^{{\scriptscriptstyle (1)}\hspace{0.1ex} \mu\nu}(y)
\hspace{-0.2ex}+\hspace{-0.2ex}
H_{\rm B}(x\!-\!y)
B^{{\scriptscriptstyle (1)}\hspace{0.1ex} \mu\nu}(y) \right]
=2 \xi^{\mu\nu}(x),
\label{unified Einstein-Langevin eq}
\ee
where the new constants $\bar{\alpha}$ and $\bar{\beta}$, and 
the kernels $H_{\rm A}(x)$ and $H_{\rm B}(x)$ can be identified 
in each case by comparison of this last equation with 
Eqs.~(\ref{massive Einstein-Langevin eq}) and 
(\ref{massless Einstein-Langevin eq 2}). 
For instance, when $m \!=\! 0$, we have 
$H_{\rm A}(x)=H(x;\mu^2)$ and 
$H_{\rm B}(x)=60 \hspace{0.2ex} \Delta \xi^2 H(x;\mu^2)$.
In this case, we can  
use the arbitrariness of the mass scale $\mu$ to
eliminate one of the parameters $\bar{\alpha}$ or $\bar{\beta}$.

In order to find solutions to Eq.
(\ref{unified Einstein-Langevin eq}), it is convenient to 
Fourier transform them. Introducing Fourier transforms
with the following convention
$\tilde f(p)=\int d^4x e^{-ipx}f(x))$ for a given field $f(x)$, 
one finds, from (\ref{A and B}), 
\be
\tilde{A}^{{\scriptscriptstyle (1)}\hspace{0.1ex} \mu\nu}(p)=
2 p^2 \tilde{G}^{{\scriptscriptstyle (1)}\hspace{0.1ex} \mu\nu}(p)
-{2 \over 3} \, p^2 P^{\mu\nu} 
\tilde{G}^{{\scriptscriptstyle (1)}}\mbox{}^{\alpha}_\alpha(p),
\hspace{10ex}
\tilde{B}^{{\scriptscriptstyle (1)}\hspace{0.1ex} \mu\nu}(p)=
-2 p^2 P^{\mu\nu} 
\tilde{G}^{{\scriptscriptstyle (1)}}\mbox{}^{\alpha}_\alpha(p).
\ee
 
Using these relations, the Fourier transform of the
Einstein-Langevin 
Eq.~(\ref{unified Einstein-Langevin eq}) reads 
\be
F^{\mu\nu}_{\hspace{2ex}\alpha\beta}(p) \,
\tilde{G}^{{\scriptscriptstyle (1)}\hspace{0.1ex} \alpha\beta}(p)=
16 \pi G \, \tilde{\xi}^{\mu\nu}(p),
\label{Fourier transf of E-L eq}
\ee
where
\be
F^{\mu\nu}_{\hspace{2ex} \alpha\beta}(p) \equiv
F_1(p) \, \delta^\mu_{( \alpha} \delta^\nu_{\beta )}+
F_2(p) \, p^2 P^{\mu\nu} \eta_{\alpha\beta},
\label{F def} 
\ee
with
\be
F_1(p) \equiv 1+16 \pi G \, p^2 
\left[ \tilde{H}_{\rm A}(p)-2 \bar{\alpha}\right],
\hspace{7ex}
F_2(p) \equiv -{16 \over 3} \, \pi G 
\left[ \tilde{H}_{\rm A}(p)+3 \tilde{H}_{\rm B}(p)
-2 \bar{\alpha}-6 \bar{\beta}\right].
\label{F_1 and F_2}
\ee

In the Fourier transformed Einstein-Langevin
Eq.~(\ref{Fourier transf of E-L eq}), 
$\tilde{\xi}^{\mu\nu}(p)$, the Fourier transform of 
$\xi^{\mu\nu}(x)$, is a Gaussian stochastic source of zero average and
\be
\langle \tilde{\xi}^{\mu\nu}(p) 
\tilde{\xi}^{\alpha\beta}(p^\prime)
\rangle_s = (2 \pi)^4 \, \delta^4(p+p^\prime) \,
\tilde{N}^{\mu\nu\alpha\beta}(p),
\label{Fourier transf of corr funct}
\ee
where we have introduced the Fourier transform of the noise kernel.
The explicit expression for $\tilde{N}^{\mu\nu\alpha\beta}(p)$
is found from
(\ref{noise and dissipation kernels 2}) and (\ref{N and D kernels}) 
to be
\bea
\tilde{N}^{\mu\nu\alpha\beta}(p)= 
{1 \over 2880 \pi} \: 
\theta (-p^2\!-\!4m^2) \, \sqrt{1+4 \,{m^2 \over p^2} } \;
&& \left[ {1 \over 4} \left(1+4 \,{m^2 \over p^2} \right)^{\!2}
(p^2)^2 \,
\bigl( 3 P^{\mu (\alpha}P^{\beta )\nu}-P^{\mu\nu} P^{\alpha\beta}
\bigr) \right.
\nn  \\
&&\hspace{1.8ex} \left.
+\, 10 \!
\left(3 \hspace{0.2ex}\Delta \xi+{m^2 \over p^2} \right)^{\!2}
(p^2)^2 P^{\mu\nu} P^{\alpha\beta} \right],
\label{Fourier transf of noise 2}
\eea
which in the massless case reduces to
\be
\lim_{m \rightarrow 0}\! \tilde{N}^{\mu\nu\alpha\beta}(p)=
{1 \over 1920 \pi} \: 
\theta (-p^2)  \left[ {1 \over 6} \, 
(p^2)^2 \,
\bigl(3 P^{\mu (\alpha}P^{\beta )\nu}-P^{\mu\nu} P^{\alpha\beta}
\bigr)
+60 \hspace{0.2ex} \Delta \xi^2 (p^2)^2 P^{\mu\nu} P^{\alpha\beta}
\right].
\label{Fourier transf of massless noise}
\ee

%%%%%%%%%%%%%%%%%%%%%%%%%%%%%%%%%%%%%%%%%%%%%%%%%%%%%
\subsubsection{Correlation functions for the linearized
Einstein tensor}

%%%%%%%%%%%%%%%%%%%%%%%%%%%%%%%%%%%%%%%%%%%%%%%%%%%%%

In general, we can write 
$G^{{\scriptscriptstyle (1)}\hspace{0.1ex} \mu\nu}=
\langle G^{{\scriptscriptstyle (1)}\hspace{0.1ex} \mu\nu} \rangle_s
+G_{\rm f}^{{\scriptscriptstyle (1)}\hspace{0.1ex} \mu\nu}$,
where 
$G_{\rm f}^{{\scriptscriptstyle (1)}\hspace{0.1ex} \mu\nu}$
is a solution to Eqs.~(\ref{unified Einstein-Langevin eq})
(or, in the Fourier transformed version, 
(\ref{Fourier transf of E-L eq})) with zero average.
The averages 
$\langle G^{{\scriptscriptstyle (1)}\hspace{0.1ex} \mu\nu} \rangle_s$
must be a solution of the linearized semiclassical Einstein equations 
obtained by averaging Eqs.~(\ref{unified Einstein-Langevin eq})
[or (\ref{Fourier transf of E-L eq})]. 
Solutions to these equations (specially in the massless case, 
$m \!=\! 0$) have been studied by several authors
\cite{horowitz,horowitz81,horowitz-wald78,hartle_horowitz,simon,%
jordan,flanagan},
particularly in connection with the problem of the stability of the
ground state of semiclassical gravity. 
The two-point correlation functions for the linearized Einstein tensor
are defined by 
\be
{\cal G}^{\mu\nu\alpha\beta}(x,x^{\prime}) \equiv 
\langle G^{{\scriptscriptstyle (1)}\hspace{0.1ex} \mu\nu}(x)
G^{{\scriptscriptstyle (1)}\hspace{0.1ex} \alpha\beta}(x^{\prime}) 
\rangle_s 
-\langle G^{{\scriptscriptstyle (1)}\hspace{0.1ex} \mu\nu}(x)
\rangle_s 
\langle G^{{\scriptscriptstyle (1)}\hspace{0.1ex} \alpha\beta}
(x^{\prime})\rangle_s =
\langle G_{\rm f}^{{\scriptscriptstyle (1)}\hspace{0.1ex} \mu\nu}(x)
G_{\rm f}^{{\scriptscriptstyle (1)}\hspace{0.1ex} \alpha\beta}
(x^{\prime})\rangle_s.
\label{two-p corr funct}
\ee

Now we shall seek the family of solutions to the Einstein-Langevin
equations which can be written as a
linear functional of the stochastic source
and whose Fourier transform, 
$\tilde{G}^{{\scriptscriptstyle (1)}\hspace{0.1ex} \mu\nu}(p)$, 
depends locally on $\tilde{\xi}^{\alpha\beta}(p)$.  
Each of such solutions is a Gaussian stochastic field
and, thus, it can be completely characterized by the 
averages 
$\langle G^{{\scriptscriptstyle (1)}\hspace{0.1ex} \mu\nu} \rangle_s$
and the two-point correlation functions 
(\ref{two-p corr funct}).
For such a family of solutions, 
$\tilde{G}_{\rm f}^{{\scriptscriptstyle (1)}\hspace{0.1ex} \mu\nu}(p)$
is the most general solution to Eq.~(\ref{Fourier transf of E-L eq})
which is linear, homogeneous and local in 
$\tilde{\xi}^{\alpha\beta}(p)$. It can be written as
\be
\tilde{G}_{\rm f}^{{\scriptscriptstyle (1)}\hspace{0.1ex} \mu\nu}(p)
= 16 \pi G \, D^{\mu\nu}_{\hspace{2ex} \alpha\beta}(p) \,
\tilde{\xi}^{\alpha\beta}(p),
\label{G_f}
\ee
where
$D^{\mu\nu}_{\hspace{2ex} \alpha\beta}(p)$ 
are the components of a Lorentz invariant tensor 
field distribution in Minkowski spacetime
(by ``Lorentz
invariant'' we mean invariant under the transformations of the
orthochronous Lorentz subgroup; see  
Ref.~\cite{horowitz} for more details on the definition 
and properties of these tensor distributions), 
symmetric under the interchanges 
$\alpha \! \leftrightarrow \!\beta$ and  
$\mu \! \leftrightarrow \!\nu$, which is the most general solution
of
\be
F^{\mu\nu}_{\hspace{2ex} \rho\sigma}(p) \,
D^{\rho\sigma}_{\hspace{2ex} \alpha\beta}(p)=
\delta^\mu_{( \alpha} \delta^\nu_{\beta )}.
\label{eq for D}
\ee
In addition, we must impose the conservation condition to 
the solutions:
$p_\nu 
\tilde{G}_{\rm f}^{{\scriptscriptstyle (1)}\hspace{0.1ex} \mu\nu}(p)
= 0$, where this zero must be understood as a stochastic variable
which behaves deterministically as a zero vector field. 
We can write
$D^{\mu\nu}_{\hspace{2ex} \alpha\beta}(p)=
D^{\mu\nu}_{p \hspace{1.2ex} \alpha\beta}(p)+
D^{\mu\nu}_{h \hspace{1.2ex} \alpha\beta}(p)$, where
$D^{\mu\nu}_{p \hspace{1.2ex} \alpha\beta}(p)$ is a particular
solution to Eq.~(\ref{eq for D}) and 
$D^{\mu\nu}_{h \hspace{1.2ex} \alpha\beta}(p)$ is the most general
solution to the homogeneous equation. 
Consequently, see Eq.~(\ref{G_f}), we can write
$\tilde{G}_{\rm f}^{{\scriptscriptstyle (1)}\hspace{0.1ex} \mu\nu}(p)
=\tilde{G}_p^{{\scriptscriptstyle (1)}\hspace{0.1ex} \mu\nu}(p)+
\tilde{G}_h^{{\scriptscriptstyle (1)}\hspace{0.1ex} \mu\nu}(p)$. 
To find the particular solution, we try an ansatz of the form
\be
D^{\mu\nu}_{p \hspace{1.2ex} \alpha\beta}(p)=
d_1(p) \, \delta^\mu_{( \alpha} \delta^\nu_{\beta )}
+ d_2(p) \, p^2 P^{\mu\nu} \eta_{\alpha\beta}.
\label{ansatz for D}
\ee
Substituting this ansatz into
Eqs.~(\ref{eq for D}), it is easy to see that it
solves these equations if we take
\be
d_1(p)=\left[ {1 \over F_1(p)} \right]_r,
\hspace{7ex}
d_2(p)= - \left[ {F_2(p)\over F_1(p) F_3(p)} \right]_r,
\label{d's}
\ee
with
\be
F_3(p) \equiv F_1(p) + 3 p^2 F_2(p)= 1-48 \pi G \, p^2 
\left[ \tilde{H}_{\rm B}(p)-2 \bar{\beta}\right], 
\label{F_3}
\ee
and where the notation $[ \;\; ]_r$ means that the zeros of the
denominators are regulated with appropriate prescriptions
in such a way that $d_1(p)$ and $d_2(p)$ are well defined
Lorentz invariant scalar distributions. 
This yields a particular solution to the 
Einstein-Langevin equations:
\be
\tilde{G}_p^{{\scriptscriptstyle (1)}\hspace{0.1ex} \mu\nu}(p)
= 16 \pi G \, D^{\mu\nu}_{p \hspace{1.2ex} \alpha\beta}(p) \,
\tilde{\xi}^{\alpha\beta}(p),
\label{solution}
\ee
which, since the stochastic source is conserved, satisfies the
conservation condition.
Note that, in the case of a massless scalar field, $m\!=\!0$, the
above solution has a functional form analogous to that of the
solutions of linearized semiclassical gravity found in the 
Appendix of Ref.~\cite{flanagan}.
Notice also that, for a massless conformally coupled field,
$m\!=\!0$ and $\Delta \xi\!=\!0$, the second term in the right hand
side of Eq.~(\ref{ansatz for D}) will not contribute in the
correlation functions (\ref{two-p corr funct}), since
in this case the stochastic source is traceless.

Next, we can work out the general form for 
$D^{\mu\nu}_{h \hspace{1.2ex} \alpha\beta}(p)$, which is a linear
combination of terms consisting of a Lorentz invariant scalar 
distribution times one of the products
$\delta^\mu_{( \alpha} \delta^\nu_{\beta )}$,
$p^2 \hspace{-0.2ex} P^{\mu\nu} \eta_{\alpha\beta}$,
$\eta^{\mu\nu} \eta_{\alpha\beta}$,
$\eta^{\mu\nu} p^2 \hspace{-0.2ex} P_{\alpha\beta}$,
$\delta^{( \mu}_{( \alpha} \hspace{0.5ex} 
p^2 \hspace{-0.2ex} P^{\nu )}_{\beta )}$
and $p^2 \hspace{-0.2ex} P^{\mu\nu} \hspace{0.2ex}
p^2 \hspace{-0.2ex} P_{\alpha\beta}$.
However, taking into account that the stochastic source is conserved,
we can omit some terms in 
$D^{\mu\nu}_{h \hspace{1.2ex} \alpha\beta}(p)$
and simply write
\be
\tilde{G}_h^{{\scriptscriptstyle (1)}\hspace{0.1ex} \mu\nu}(p)
= 16 \pi G \, D^{\mu\nu}_{h \hspace{1.2ex} \alpha\beta}(p) \,
\tilde{\xi}^{\alpha\beta}(p),
\ee
where
\be
D^{\mu\nu}_{h \hspace{1.2ex} \alpha\beta}(p)=
h_1(p)\, \delta^\mu_{( \alpha} \delta^\nu_{\beta )}
+ h_2(p) \, p^2 P^{\mu\nu} \eta_{\alpha\beta}
+h_3(p) \,  \eta^{\mu\nu} \eta_{\alpha\beta},
\label{bar D}
\ee
with $h_1(p)$, $h_2(p)$ and $h_3(p)$ being
Lorentz invariant scalar 
distributions. {}From the fact that 
$D^{\mu\nu}_{h \hspace{1.2ex} \alpha\beta}(p)$ must satisfy the
homogeneous equation corresponding to Eq.~(\ref{eq for D}), we find
that $h_1(p)$ and $h_3(p)$ have support on the set of points 
$\{ p^\mu \}$ for which $F_1(p) \!=\! 0$, and that
$h_2(p)$ has support on the set of points 
$\{ p^\mu \}$ for which $F_1(p) \!=\! 0$ or $F_3(p) \!=\! 0$.
Moreover, the conservation condition for 
$\tilde{G}_h^{{\scriptscriptstyle (1)}\hspace{0.1ex} \mu\nu}(p)$
implies that the term with $h_3(p)$ is only allowed in the case 
of a massless conformally coupled field, 
$m\!=\!0$ and $\Delta \xi\!=\!0$. 
{}From (\ref{Fourier transf of corr funct}), we get
\be
\langle 
\tilde{G}_h^{{\scriptscriptstyle (1)}\hspace{0.1ex} \mu\nu}(p) \,
\tilde{\xi}^{\alpha\beta}(p^\prime)
\rangle_s = (2 \pi)^4 \, 16 \pi G \, \delta^4(p+p^\prime) \,
D^{\mu\nu}_{h \hspace{1.2ex} \rho\sigma}(p) \,
\tilde{N}^{\rho\sigma\alpha\beta}(p).
\label{c f}
\ee
Note, from expressions 
(\ref{Fourier transf of noise 2}) and 
(\ref{Fourier transf of massless noise}), that the support of 
$\tilde{N}^{\mu\nu\alpha\beta}(p)$ is on the set of points 
$\{ p^\mu \}$ for which $-p^2 \!\geq \! 0$ when $m\!=\!0$,
and for which $-p^2-4 m^2 \! > \! 0$ when $m \! \neq \! 0$. 
At such points, using expressions (\ref{F_1 and F_2}),
(\ref{F_3}), (\ref{Hnew}) and (\ref{H kernels}), 
it is easy to see that $F_1(p)$  and $F_3(p)$ are
always different from
zero, except for some particular values of $\Delta \xi$ and 
$\bar{\beta}$: (a) 
when $m \!=\!0$, $\Delta \xi\!=\!0$ and 
               $\bar{\beta} \! > \! 0$;
and (b) when $m \! \neq \! 0$, 
               $0 \! < \! \Delta \xi \! <\!  (1/12)$ and
               $\bar{\beta} \!=\! (\Delta \xi/ 32 \pi^2)
                \hspace{0.2ex}
                [ \pi/(G m^2)+1/36]$.

In the case (a), $F_3(p)\!=\!0$ for the set of points $\{ p^\mu \}$
satisfying $-p^2 \!=\! 1/(96 \pi G \bar{\beta})$; in the case
(b), $F_3(p)\!=\!0$ for $\{ p^\mu \}$ such that 
$-p^2 \!=\! m^2/(3 \Delta \xi)$.
Hence, except for the above cases (a) and (b), the intersection of the
supports of $\tilde{N}^{\mu\nu\alpha\beta}(p)$ and 
$D^{\rho\sigma}_{h \hspace{1.2ex} \lambda\gamma}(p)$ is an empty
set and, thus, the correlation function (\ref{c f}) is zero.
In the cases (a) and (b), we can have a contribution to 
(\ref{c f}) coming from the term with $h_2(p)$ in (\ref{bar D})
of the form
$D^{\mu\nu}_{h \hspace{1.2ex} \rho\sigma}(p) \,
\tilde{N}^{\rho\sigma\alpha\beta}(p)\!=\!
H_3(p; \{ C \}) \, p^2 P^{\mu\nu} \hspace{0.2ex} 
\tilde{N}^{\alpha\beta\rho}_{\hspace{3.3ex} \rho}(p)$,
where $H_3(p; \{ C \})$ is the most general Lorentz invariant
distribution satisfying 
$F_3(p) \hspace{0.2ex} H_3(p; \{ C \})\!=\! 0$, which depends on a set
of arbitrary parameters represented as $\{ C \}$. However,
from (\ref{Fourier transf of noise 2}), we see that
$\tilde{N}^{\alpha\beta\rho}_{\hspace{3.3ex} \rho}(p)$ is proportional
to $\theta (-p^2\!-\!4m^2) \hspace{0.2ex} 
(1+4 m^2/p^2)^{(1/2)} \hspace{0.2ex} 
(3 \Delta \xi+m^2/p^2 )^2$. Thus, in the case (a), 
we have $\tilde{N}^{\alpha\beta\rho}_{\hspace{3.3ex} \rho}(p) 
\!=\! 0$ and, in the case (b), the intersection of the supports of 
$\tilde{N}^{\alpha\beta\rho}_{\hspace{3.3ex} \rho}(p)$ and of
$H_3(p; \{ C \})$ is an empty set. 
Therefore, from the above analysis, we conclude that
$\tilde{G}_h^{{\scriptscriptstyle (1)}\hspace{0.1ex} \mu\nu}(p)$ gives
no contribution to the correlation functions 
(\ref{two-p corr funct}), since
$\langle 
\tilde{G}_h^{{\scriptscriptstyle (1)}\hspace{0.1ex} \mu\nu}(p) \,
\tilde{\xi}^{\alpha\beta}(p^\prime)
\rangle_s \!=\! 0$, and we have simply  
${\cal G}^{\mu\nu\alpha\beta}(x,x^{\prime}) \!=\!
\langle G_p^{{\scriptscriptstyle (1)}\hspace{0.1ex} \mu\nu}(x)
G_p^{{\scriptscriptstyle (1)}\hspace{0.1ex} \alpha\beta}
(x^{\prime})\rangle_s$, where 
$G_p^{{\scriptscriptstyle (1)}\hspace{0.1ex} \mu\nu}(x)$ is the
inverse Fourier transform of (\ref{solution}).

Therefore the correlation functions (\ref{two-p corr funct})
can then be computed from
\be
\langle 
\tilde{G}_p^{{\scriptscriptstyle (1)}\hspace{0.1ex} \mu\nu}(p) \,
\tilde{G}_p^{{\scriptscriptstyle (1)}\hspace{0.1ex} \alpha\beta}
(p^\prime) \rangle_s = 
64 \, (2 \pi)^6 \, G^2 \, \delta^4(p+p^\prime) \,
D^{\mu\nu}_{p \hspace{1.2ex} \rho\sigma}(p) \,
D^{\alpha\beta}_{p \hspace{1.2ex} \lambda\gamma}(-p) \,
\tilde{N}^{\rho\sigma\lambda\gamma}(p).
\ee
It is easy to see from the above analysis that the prescriptions
$[ \;\; ]_r$ in the factors $D_p$ are irrelevant in the last
expression and, thus, they can be suppressed.
Taking into account that 
$F_l(-p) \!=\! F^{\displaystyle \ast}_l(p)$, with $l \!=\! 1,2,3$, 
we get from Eqs.~(\ref{ansatz for D}) and (\ref{d's})
\bea
\langle 
\tilde{G}_p^{{\scriptscriptstyle (1)}\hspace{0.1ex} \mu\nu}(p) \,
\tilde{G}_p^{{\scriptscriptstyle (1)}\hspace{0.1ex} \alpha\beta}
(p^\prime) \rangle_s = &&
64 \, (2 \pi)^6 \, G^2 \: {\delta^4(p+p^\prime) \over 
\left| \hspace{0.1ex} F_1(p) \hspace{0.1ex}\right|^2 }
\left[ 
\tilde{N}^{\mu\nu\alpha\beta}(p)
- {F_2(p) \over F_3(p)} \: p^2 P^{\mu\nu} \hspace{0.2ex} 
\tilde{N}^{\alpha\beta\rho}_{\hspace{3.3ex} \rho}(p)
\right.     \nn  \\
&& \hspace{7ex}
\left. - \,
{F_2^{\displaystyle \ast}(p) \over F_3^{\displaystyle \ast}(p)} 
\: p^2 P^{\alpha\beta} \hspace{0.2ex} 
\tilde{N}^{\mu\nu\rho}_{\hspace{3.3ex} \rho}(p)
+ { \left| \hspace{0.1ex} F_2(p) \hspace{0.1ex}\right|^2
\over \left| \hspace{0.1ex} F_3(p) \hspace{0.1ex}\right|^2 } \:
p^2 P^{\mu\nu} \hspace{0.2ex} p^2 P^{\alpha\beta} \hspace{0.2ex}
\tilde{N}
^{\rho \hspace{0.9ex} \sigma}_{\hspace{1ex} \rho \hspace{1.1ex}
\sigma} (p) \right].   \nn  \\
\mbox{}
\eea 
This last expression is well defined as a bi-distribution
and can be easily evaluated using 
Eq.~(\ref{Fourier transf of noise 2}). The final explicit
result for the Fourier transformed correlation function for
the Einstein tensor is thus
\bea
\langle 
\tilde{G}_p^{{\scriptscriptstyle (1)}\hspace{0.1ex} \mu\nu}(p) \,
\tilde{G}_p^{{\scriptscriptstyle (1)}\hspace{0.1ex} \alpha\beta}
(p^\prime) \rangle_s = 
{2 \over 45} && \, (2 \pi)^5 \, G^2 \: 
{\delta^4(p+p^\prime) \over 
\left| \hspace{0.1ex} F_1(p) \hspace{0.1ex}\right|^2 } \:
\theta (-p^2\!-\!4m^2) \, \sqrt{1+4 \,{m^2 \over p^2} } 
\nn   \\
&&    \times \!
\left[{1 \over 4} \left(1+4 \,{m^2 \over p^2} \right)^{\!2} \!
(p^2)^2 \,
\bigl( 3 P^{\mu (\alpha}P^{\beta )\nu} \!-\!
P^{\mu\nu} P^{\alpha\beta} \bigr)  \right.
\nn   \\
&& \hspace{3.5ex} \left.
+\, 10 \!
\left(3 \hspace{0.2ex}\Delta \xi+{m^2 \over p^2} \right)^{\!2} \!
(p^2)^2 P^{\mu\nu} P^{\alpha\beta} 
\left| 1-3 p^2 \, {F_2(p) \over F_3(p)} \right|^2
\right].   
\label{Fourier tr corr funct}
\eea

To obtain the correlation functions in coordinate space, Eq.
(\ref{two-p corr funct}), we have
to take the inverse Fourier transform of the above result, 
the final result is:
\be
{\cal G}^{\mu\nu\alpha\beta}(x,x^{\prime})=
{\pi \over 45} \, G^2 \,{\cal F}^{\mu\nu\alpha\beta}_{x} \,
{\cal G}_{\rm A} (x-x^{\prime})+
{8 \pi \over 9} \, G^2 \,
{\cal F}^{\mu\nu}_{x} {\cal F}^{\alpha\beta}_{x} \,
{\cal G}_{\rm B} (x-x^{\prime}),
\label{corr funct}
\ee
with
\bea
&&\tilde{{\cal G}}_{\rm A}(p) \equiv 
\theta (-p^2-4m^2) \, \sqrt{1+4 \,{m^2 \over p^2} } 
\left(1+4 \,{m^2 \over p^2} \right)^{\!2} \!
{1 \over 
\left| \hspace{0.1ex} F_1(p) \hspace{0.1ex}\right|^2 } \, ,
\nn   \\
&&\tilde{{\cal G}}_{\rm B}(p) \equiv 
\theta (-p^2-4m^2) \, \sqrt{1+4 \,{m^2 \over p^2} } 
\left(3 \hspace{0.2ex}\Delta \xi+{m^2 \over p^2} \right)^{\!2} \!
{1 \over 
\left| \hspace{0.1ex} F_1(p) \hspace{0.1ex}\right|^2 } \,
\left| 1-3 p^2 \, {F_2(p) \over F_3(p)} \right|^2,
\label{distri}
\eea
and 
${\cal F}^{\mu\nu\alpha\beta}_{x} \equiv
3 {\cal F}^{\mu (\alpha}_{x}{\cal F}^{\beta )\nu}_{x}-
{\cal F}^{\mu\nu}_{x}{\cal F}^{\alpha\beta}_{x}$,
and where $F_l(p)$, $l=1,2,3$, are given in (\ref{F_1 and F_2}) and
(\ref{F_3}). Notice that, for a massless field ($m \!=\! 0$), we have
\bea
&&F_1(p)= 1+16 \pi G \hspace{0.2ex} p^2 \hspace{0.2ex} 
          \tilde{H}(p;\bar{\mu}^2), 
\nn  \\
&&F_2(p)= - {16 \over 3} \, \pi G \left[ 
(1 +180 \hspace{0.2ex}\Delta \xi^2 ) \, \tilde{H}(p;\bar{\mu}^2)
-6 \Upsilon \right],
\nn  \\
&&F_3(p)= 1- 48 \pi G \hspace{0.2ex} p^2
\left[ 60 \hspace{0.2ex}\Delta \xi^2 \hspace{0.2ex}
\tilde{H}(p;\bar{\mu}^2) -2 \Upsilon \right],
\eea
with $\bar{\mu} \equiv \mu\, \exp (1920 \pi^2 \bar{\alpha})$
and 
$\Upsilon \equiv \bar{\beta} 
 -60 \hspace{0.2ex}\Delta \xi^2 \hspace{0.2ex}\bar{\alpha}$, and where
$\tilde{H}(p;\mu^2)$ is the Fourier transform of
$H(x;\mu^2)$ given in (\ref{Hnew}).

%%%%%%%%%%%%%%%%%%%%%%%%%%%%%%%%%%%%%%%%%%%%%%%%%%%

\subsubsection{Correlation functions for the metric
perturbations}

%%%%%%%%%%%%%%%%%%%%%%%%%%%%%%%%%%%%%%%%%%%%%%%%%%%

Starting from the solutions found for the linearized Einstein tensor, 
which are characterized by the two-point correlation functions
(\ref{corr funct}) [or, in terms of Fourier transforms, 
(\ref{Fourier tr corr funct})], we can now solve the equations for the
metric perturbations. Working in the harmonic gauge, 
$\partial_{\nu} \bar{h}^{\mu\nu} \!=\! 0$ (this zero must be
understood in a statistical sense) where
$\bar{h}_{\mu\nu} \!\equiv \! h_{\mu\nu} 
\!-\! (1/2)\hspace{0.2ex} \eta_{\mu\nu} \hspace{0.2ex}h_\alpha^\alpha$,
the equations for the metric perturbations in terms of
the Einstein tensor are 
\be
\Box \bar{h}^{\mu\nu}(x) \!=\! -2 
G^{{\scriptscriptstyle (1)}\hspace{0.1ex} \mu\nu}(x),
\label{metric and G}
\ee
or, in terms of
Fourier transforms, 
$p^2 
\tilde{\bar{h}}^{\mbox{}_{\mbox{}_{\mbox{}_{\mbox{}_{\mbox{}
_{\scriptstyle \mu\nu}}}}}}\hspace{-0.5ex} (p) 
\!=\! 2 \tilde{G}^{{\scriptscriptstyle (1)}\hspace{0.1ex} \mu\nu}(p)$.
Similarly to the analysis of  the equation
for the Einstein tensor, we can write 
$\bar{h}^{\mu\nu} \!=\! \langle \bar{h}^{\mu\nu} \rangle_s
\!+\! \bar{h}^{\mu\nu}_{\rm f}$, where $\bar{h}^{\mu\nu}_{\rm f}$ is a
solution to these equations with zero average, and the two-point
correlation functions are defined by
\be
{\cal H}^{\mu\nu\alpha\beta}(x,x^{\prime}) \equiv 
\langle \bar{h}^{\mu\nu}(x) \bar{h}^{\alpha\beta}(x^{\prime})
\rangle_s - \langle \bar{h}^{\mu\nu}(x) \rangle_s 
\langle \bar{h}^{\alpha\beta}(x^{\prime}) \rangle_s =
\langle \bar{h}^{\mu\nu}_{\rm f}(x) 
\bar{h}^{\alpha\beta}_{\rm f}(x^{\prime}) \rangle_s.
\label{co fu}
\ee

We can now seek solutions of
the Fourier transform of Eq. (\ref{metric and G})
of the form 
$\tilde{\bar{h}}^{\mbox{}_{\mbox{}_{\mbox{}_{\mbox{}_{\mbox{}
_{\scriptstyle \mu\nu}}}}}}_{\rm f}\hspace{-0.5ex} (p) 
\!=\! 2 D(p) 
\tilde{G}^{{\scriptscriptstyle (1)}\hspace{0.1ex} \mu\nu}_{\rm f}(p)$,
where $D(p)$ is a Lorentz invariant scalar distribution in Minkowski
spacetime, which is the most general solution of $p^2 D(p) \!=\! 1$.
Note that, since the linearized Einstein tensor is conserved, 
solutions of this form automatically satisfy the harmonic
gauge condition. As in the previous subsection,
we can write $D(p) \!=\! [ 1/p^2 ]_r 
\!+\! D_h(p)$, where $D_h(p)$ is the most general solution to the
associated homogeneous equation and, correspondingly, we have
$\tilde{\bar{h}}^{\mbox{}_{\mbox{}_{\mbox{}_{\mbox{}_{\mbox{}
_{\scriptstyle \mu\nu}}}}}}_{\rm f}\hspace{-0.5ex} (p) 
\!=\! 
\tilde{\bar{h}}^{\mbox{}_{\mbox{}_{\mbox{}_{\mbox{}_{\mbox{}
_{\scriptstyle \mu\nu}}}}}}_p\hspace{-0.5ex} (p) +
 \tilde{\bar{h}}^{\mbox{}_{\mbox{}_{\mbox{}_{\mbox{}_{\mbox{}
_{\scriptstyle \mu\nu}}}}}}_h \hspace{-0.5ex} (p)$.
However, since $D_h(p)$ has support on the set of points for which
$p^2 \!=\! 0$, it is easy to see from 
Eq.~(\ref{Fourier tr corr funct})
(from the factor $\theta (-p^2-4 m^2)$) that
$\langle \tilde{\bar{h}}^{\mbox{}_{\mbox{}_{\mbox{}_{\mbox{}_{\mbox{}
_{\scriptstyle \mu\nu}}}}}}_h \hspace{-0.5ex} (p)
\tilde{G}^{{\scriptscriptstyle (1)}\hspace{0.1ex} \alpha\beta}
_{\rm f}(p^{\prime})
\rangle_s \!=\! 0$ and, thus, 
the two-point correlation functions (\ref{co fu})
can be computed from
$\langle 
\tilde{\bar{h}}^{\mbox{}_{\mbox{}_{\mbox{}_{\mbox{}_{\mbox{}
_{\scriptstyle \mu\nu}}}}}}_{\rm f}\hspace{-0.5ex} (p)
\tilde{\bar{h}}^{\mbox{}_{\mbox{}_{\mbox{}_{\mbox{}_{\mbox{}
_{\scriptstyle \alpha\beta}}}}}}_{\rm f}\hspace{-0.5ex}(p^{\prime})
\rangle_s \!=\! 
\langle 
\tilde{\bar{h}}^{\mbox{}_{\mbox{}_{\mbox{}_{\mbox{}_{\mbox{}
_{\scriptstyle \mu\nu}}}}}}_p\hspace{-0.5ex} (p)
\tilde{\bar{h}}^{\mbox{}_{\mbox{}_{\mbox{}_{\mbox{}_{\mbox{}
_{\scriptstyle \alpha\beta}}}}}}_p\hspace{-0.5ex}(p^{\prime})
\rangle_s$. {}From Eq.~(\ref{Fourier tr corr funct})
and due to the factor $\theta (-p^2-4 m^2)$, it is also easy to see
that the prescription $[ \;\; ]_r$ is irrelevant in this correlation
function and we obtain
\be
\langle 
\tilde{\bar{h}}^{\mbox{}_{\mbox{}_{\mbox{}_{\mbox{}_{\mbox{}
_{\scriptstyle \mu\nu}}}}}}_p\hspace{-0.5ex} (p)
\tilde{\bar{h}}^{\mbox{}_{\mbox{}_{\mbox{}_{\mbox{}_{\mbox{}
_{\scriptstyle \alpha\beta}}}}}}_p\hspace{-0.5ex}(p^{\prime})
\rangle_s = {4 \over (p^2)^2} \,
\langle 
\tilde{G}_p^{{\scriptscriptstyle (1)}\hspace{0.1ex} \mu\nu}(p) \,
\tilde{G}_p^{{\scriptscriptstyle (1)}\hspace{0.1ex} \alpha\beta}
(p^\prime) \rangle_s,   
\ee
where $\langle 
\tilde{G}_p^{{\scriptscriptstyle (1)}\hspace{0.1ex} \mu\nu}(p) \,
\tilde{G}_p^{{\scriptscriptstyle (1)}\hspace{0.1ex} \alpha\beta}
(p^\prime) \rangle_s$ is given by  Eq. (\ref{Fourier tr corr funct}).
The right hand side of this equation is a well defined
bi-distribution, at least for $m \!\neq \! 0$ (the $\theta$ function
provides the suitable cutoff).
In the massless field case, since the noise kernel 
is obtained as the limit $m \!\rightarrow \!0$ of the noise kernel
for a massive field, it seems that the natural prescription to avoid
the divergencies on the lightcone $p^2 \!=\! 0$ is a Hadamard finite
part, see Refs.~\cite{schwartz} for its definition.
Taking this prescription, we also get a well defined bi-distribution
for the massless limit of the last expression.

The final result for the two-point correlation function
for the field $\bar h^{\mu\nu}$ is:
\be
{\cal H}^{\mu\nu\alpha\beta}(x,x^{\prime})=
{4 \pi \over 45} \, G^2 \,{\cal F}^{\mu\nu\alpha\beta}_{x} \,
{\cal H}_{\rm A} (x-x^{\prime})+
{32 \pi \over 9} \, G^2 \,
{\cal F}^{\mu\nu}_{x} {\cal F}^{\alpha\beta}_{x} \,
{\cal H}_{\rm B} (x-x^{\prime}),
\label{corr funct 2}
\ee
where $\tilde{{\cal H}}_{\rm A}(p) \!\equiv \! 
[1/(p^2)^2]\, \tilde{{\cal G}}_{\rm A}(p)$ and 
$\tilde{{\cal H}}_{\rm B}(p) \!\equiv \! 
[1/(p^2)^2]\, \tilde{{\cal G}}_{\rm B}(p)$, with 
$\tilde{{\cal G}}_{\rm A}(p)$ and  
$\tilde{{\cal G}}_{\rm B}(p)$ given by (\ref{distri}).
The two-point correlation functions for the metric perturbations 
can be easily obtained using $h_{\mu\nu} \!=\! 
\bar{h}_{\mu\nu} 
\!-\! (1/2) \hspace{0.2ex}\eta_{\mu\nu} 
\hspace{0.2ex}\bar{h}^{\alpha}_{\alpha}$.

%%%%%%%%%%%%%%%%%%%%%%%%%%%%%%%%%%%%%%%%%%%%%%%

\subsubsection{Conformally coupled field}

%%%%%%%%%%%%%%%%%%%%%%%%%%%%%%%%%%%%%%%%%%%%%%%

For a conformally coupled field, {\it i.e.}, when
$m = 0$ and $\Delta \xi=0$, the previous correlation
functions are greatly simplified and
can be approximated explicitly in terms of analytic functions.
The detailed results are given in Ref. \cite{MV2},
here we outline the main features.

When $m=0$ and $\Delta \xi=0$ we have that 
${\cal G}_{\rm B} (x) \!=\!0$ and 
$
\tilde{{\cal G}}_{\rm A}(p)= \theta(-p^2) 
\left|\hspace{0.2ex} F_1(p) \hspace{0.2ex} \right|^{-2}$.
Thus the two-point correlations functions for
the Einstein tensor is
\be
{\cal G}^{\mu\nu\alpha\beta}(x,x^{\prime})=
{\pi \over 45} \, G^2 \,{\cal F}^{\mu\nu\alpha\beta}_{x} \,
\int {d^4p\over (2\pi)^4}\frac
{e^{ip(x-x^\prime)}\,\theta(-p^2)}
{| 1+16\pi Gp^2\tilde H(p;\bar\mu^2)|^2},
\label{corr funct conf}
\ee
where $\tilde H(p,\mu^2)=(1920\pi^2)^{-1}\ln
[-((p^0+i\epsilon)^2+p^ip_i)/\mu^2]$, see Eq. (\ref{Hnew}).

To estime this integral
let us consider spacelike separated points
$(x-x^{\prime})^\mu=(0,\!{\bf x}-{\bf x}^\prime)$, and define
${\bf y}={\bf x}-{\bf x}^\prime$. We may now
formaly change the momentum variable $p^\mu$
by the dimensionless vector $s^\mu$: $p^\mu=s^\mu/|{\bf y}|$, then
the previous integral denominator is
$|1+16\pi (L_P/|{\bf y}|)^2s^2\tilde H(s)|^2$,
where we have introduced the Planck length $L_P=\sqrt{G}$.
It is clear that we can consider two regimes: (a) when
$L_P \ll |{\bf y}|$, and (b) when $|{\bf y}|\sim L_P$.
In  case (a) the correlation funtion, for the $0000$
component, say,
will be of the order
$$
{\cal G}^{0000}({\bf y})\sim {L_P^4\over|{\bf y}|^8}.
$$ 
In case (b) when the denominator has zeros
a detailed calculation  carried out in Ref. \cite{MV2}
shows that:
$$
{\cal G}^{0000}({\bf y})\sim e^{-|{\bf y}|/L_P}\left(
{L_P\over |{\bf y}|^5}+\dots +{1\over L_P^2|{\bf y}|^2}\right)
$$
which indicates an exponiential decay at distances around
the Planck scale. Thus short scale fluctuations are
strongly suppressed.

For the two-point metric correlation the results
are similar. In this case we have
\be
{\cal H}^{\mu\nu\alpha\beta}(x,x^{\prime})=
{4\pi \over 45} \, G^2 \,{\cal F}^{\mu\nu\alpha\beta}_{x} \,
\int {d^4p\over (2\pi)^4}\frac
{e^{ip(x-x^\prime )}\theta(-p^2)}
{(p^2)^2| 1+16\pi Gp^2\tilde H(p;\bar\mu^2)|^2}.
\label{corr funct conf metric}
\ee
The integrand has the same behavior of the correlation function
of Eq. (\ref{corr funct conf})
thus matter fields tends to supress the short scale metric
perturbations.
In this case we find, as for the correlation of
the Einstein tensor,
that for case (a) above we have,
$$
{\cal H}^{0000}({\bf y})\sim {L_P^4\over|{\bf y}|^4},
$$ 
and for case (b) we have
$$
{\cal H}^{0000}({\bf y})\sim e^{-|{\bf y}|/L_P}\left(
{L_P\over |{\bf y}|}+\dots \right).
$$

It is interesting to write expression
(\ref{corr funct conf metric}) in an
alternative way.
If we introduce the dimensionless tensor
$P^{\mu\nu\alpha\beta}\equiv
3P^{\mu(\alpha}P^{\beta)\nu}-P^{\mu\nu}P^{\alpha\beta}$,
where $P^{\mu\nu}$ is the projector defined in Eq.
(\ref{projector}), to account for
the effect of the operator
${\cal F}^{\mu\nu\alpha\beta}_{\,x}$,
we can write
\be
{\cal H}^{\mu\nu\alpha\beta}(x,x^{\prime})=
{4\pi \over 45} \, G^2 \,
\int {d^4p\over (2\pi)^4}\,\frac
{e^{ip(x-x^\prime)}\,P^{\mu\nu\alpha\beta}\,\theta(-p^2)}
{| 1+16\pi Gp^2\tilde H(p;\bar\mu^2)|^2}.
\label{corr funct conf metric2}
\ee
This expression allows a direct comparison
with the graviton propagator for linearized quantum gravity
in the $1/N$ approximation found by Tomboulis \cite{Tomboulis77}.
One can see that the imaginary part of the graviton
propagator leads, in fact, to Eq.
(\ref{corr funct conf metric2}).

%%%%%%%%%%%%%%%%%%%%%%%%%%%%%%%%%%%%%%%%

\subsubsection{Discussion}

%%%%%%%%%%%%%%%%%%%%%%%%%%%%%%%%%%%%%%%%

The main results of this section are the
correlation functions  (\ref{corr funct})
and (\ref{corr funct 2}). In the case of a conformal field, the 
correlation functions of the linearized Einstein
tensor have been explicitly estimated.
{}From the exponential factors $e^{-|{\bf y}/L_P}$ in these results
for scales near the Planck length,
we see that the correlation functions of the linearized Einstein
tensor have the  Planck length as the correlation length.
A similar behavior is found for the
correlation functions of the metric perturbations.
Since these fluctuations are induced by the matter fluctuations
we infer that the effect of the matter fields is to suppress the
fluctuations of the metric at very small scales.
On the other hand,
at scales much larger than the Planck length
the induced metric fluctuations are small
compared with the free graviton propagator which goes like
$L_P^2/|{\bf y}|^2$, since the action for the free
graviton goes like $S_h\sim\int d^4 x\,L_P^{-2}h\Box h$
 
It is interesting to note, however, that these results for correlation
functions are non-analytic in their characteristic correlation
lengths. This kind of non-analytic behavior is actually quite typical
of the solutions of Langevin-type equations with dissipative
terms. An example in the context of a reduced version of the
semiclassical Einstein-Langevin equation is given in
Ref.~\cite{ccv97}.

For background solutions of semiclassical gravity with other scales
present apart from the Planck scales (for instance, for matter fields
in a thermal state), stress-energy fluctuations may be important at
larger scales. For such backgrounds, stochastic
semiclassical gravity might predict correlation functions with
characteristic correlation lengths larger than the Planck
scales. It seems quite plausible, nevertheless, that these correlation
functions would remain non-analytic in their characteristic
correlation lengths. 
This would imply that these correlation functions could not be
obtained from a calculation involving a perturbative expansion in the
characteristic correlation lengths. In particular, if these  
correlation lengths are proportional to the Planck constant 
$\hbar$, the 
gravitational correlation functions could not be obtained from an
expansion in $\hbar$. Hence, stochastic semiclassical
gravity might predict a behavior for gravitational
correlation functions different from that of the analogous functions
in perturbative quantum gravity \cite{donoghue}. 
This is not necessarily inconsistent with having neglected action
terms of higher order in $\hbar$ when considering semiclassical
gravity as an effective theory \cite{flanagan}.

%%%%%%%%%%%%%%%%%%%%%%%%%%%%%%%%%%%%%%%%%
%%%%%%%%%%%%%%%%%%%%%%%%%%%%%%%%%%%%%%%%%
\section{FLUCTUATIONS OF ENERGY DENSITY AND VALIDITY OF
         SEMICLASSICAL GRAVITY}
\label{sec3}
%%%%%%%%%%%%%%%%%%%%%%%%%%%%%%%%%%%%%%%%%
%%%%%%%%%%%%%%%%%%%%%%%%%%%%%%%%%%%%%%%%%

We now turn our attention to some basic issues involving vacuum energy
density fluctuations invoking only the simplest spacetimes, Minkowski and Casimir.
Recent years saw the beginning of serious studies of the fluctuations of the
energy momentum tensor (EMT) $\hat T_{\mu \nu}$
of quantum fields in spacetimes with boundaries
\cite{Birrell-Davies82,Fulling89,Wald94}
(such as Casimir effect \cite{Casimir})
\cite{Barton,Kuo-Ford93}, nontrivial
topology (such as imaginary time thermal field theory) or nonzero curvature
(such as the Einstein universe) \cite{Phillips-Hu97}. 
A natural measure of the strength of fluctuations is $\chi$ \cite{HP0},
the ratio of the variance $\Delta \rho^2$
of  fluctuations in the energy density (expectation value of the $\hat \rho^2$ 
operator minus the square
of the mean $\hat \rho$ taken with respect to some quantum state) to  its 
mean-squared (square of the 
expectation value of $\hat \rho$):
\beq
\chi \equiv  \frac{\left<\hat\rho^2\right>-\left<\hat\rho\right>^2}
               {\left<\hat\rho\right>^2} 
       \equiv \frac { \Delta \rho^2} {{\bar \rho}^2}
\eeq
Alternatively, we can use the quantity introduced by
Kuo and Ford \cite{Kuo-Ford93}  
\begin{equation}
\Delta \equiv \frac{\left<\hat\rho^2\right>-\left<\hat\rho\right>^2}
               {\left<\hat\rho^2\right>} = \frac {\chi} {\chi +1}
\end{equation}
Assuming a positive definite variance $\Delta \rho^2 \ge 0$, 
then $ 0 \le \chi \le \infty$  and $0 \le\Delta \le 1$ always, 
with $\Delta \ll 1$  falling in the classical domain. 
Kuo and Ford (KF)  displayed  a number of quantum states
(vacuum plus 2 particle
state, squeezed vacuum and Casimir vacuum) with respect to which the
expectation value of the energy momentum tensor (00 component) gives rise to
negative local energy density. For these states  $\Delta$ is of order unity. 
{}From this result they drew the implications, amongst other interesting
inferences, that semiclassical gravity (SCG) \cite{scg} 
could become invalid under these conditions  The validity of semiclassical
gravity in the face of fluctuations of quantum
fields as source is an important
issue which has caught the attention of many authors.
%\cite{valSCG}.
Amongst others 
Phillips and Hu (PH) \cite{PH1} hold a different viewpoint on this
issue from KF. 
This section is a summary of their investigations on this issue.

To begin with it may not be so surprising that states which are more quantum 
(e.g., squeezed states) in nature  than classical (e.g., coherent states) 
\cite{states} may lead to large fluctuations in energy density
comparable to the mean\footnote{This can be seen
even in the ratio of expectation values of
moments of the displacement operators in simple 
quantum harmonic oscillators}. Such a condition exists peacefully with the underlying spacetime
at least at the low energy of today's universe. 
PH calculated the variance of fluctuations to mean-squared ratio of a quantum 
field for the simplest case of Minkowski spacetime 
i.e., for ordinary quantum field theory to be $\Delta=2/5$. 
This is a simple counter-example to the claim of KF,
since $\Delta = O(1)$ holds also for Minkowski  space, where SCG is known to be
valid at large scales.  PH do not see sufficient ground
to question the validity of SCG at energy below the Planck energy when the
spacetime is depictable  by a manifold structure,  approximated locally by
the Minkowski space.  To them the fluctuations to mean being of the order unity
arises from  the quantum nature of the vacuum state and says little about the 
compatibility of the field source with the spacetime the
quantum field lives in\footnote{One should draw a distinction 
between quantum fields in curved spacetime QFCST and semiclassical gravity:
the former is a test field situation with quantum fields propagating in a fixed
background space while in the latter both the field and the spacetime are
determined self-consistently by solving the semiclassical Einstein equation.
The cases studied in Kuo and Ford \cite{Kuo-Ford93}
as well as many others 
\cite{Phillips-Hu97} are of a test-field nature, where
backreaction is not  considered. So KF's criterion pertains more to QFTCST
than to SCG, where in the former the central issue is compatibility, which is a
weaker condition than consistency in the latter.}.

PH pointed out that one should refer to a scale (of interaction or for probing
accuracy) when measuring the validity of SCG. The conventional belief is that
when reaching the Planck scale from below, QFTCST will break down because,
amongst other things happening,  graviton production at that energy will become
significant so as to render the classical background spacetime unstable, and
the mean value of quantum field taken as a source for the Einstein equation
becomes inadequate. 
To address this issue as well as the issue of the spatial extent
where negative energy density can exist,
PH view it necessary to introduce a scale in the
spacetime regions where quantum fields are defined
to monitor how the mean value
and the fluctuations of the energy momentum tensor change.

In conventional field theories 
the stress tensor  built from the product of a pair of field operators
evaluated at a single point in the spacetime manifold  is, strictly speaking, 
ill-defined. Point separation is a well-established
method which suits the present
concern very well, and  we will discuss this method in section \ref{sec4}.
For here we will use
a simpler method to introduce a scale in the quantum field theory, i.e.,
by introducing a (spatial) smearing  function $f({\bf x}) $
to define smeared field
operators $\hat \phi_t(f_{\bf x})$. Using a Gaussian smearing 
function (with variance
$\sigma^2$) PH derive expressions for the EM bi-tensor operator, its mean and
its fluctuations as  functions of $\sigma$, for a massless scalar field in
both the Minkowski and the Casimir spacetimes.  The interesting
result PH find 
is that while both the vacuum expectation value and the  fluctuations of energy
density grow as $\sigma \rightarrow 0$,  the ratio of the variance of the
fluctuations to its mean-squared remains a constant $\chi_d$
($d$ is the spatial
dimension of spacetime) which is independent of $\sigma$.
The measure $\Delta_d$
($=\chi_d/(\chi_d+1)$) depends on the dimension of space and is of
the order unity.
It varies only slightly for spacetimes with boundary or nontrivial topology. 
For example $\Delta$ for Minkowski is $2/5$, while for Casimir is $6/7$
(cf,  from \cite{Phillips-Hu97}). Add to this our prior result for the
Einstein Universe, 
 $ \Delta=111/112$, independent of curvature, and that for hot flat space
\cite{NPhD}, we see that invariably the fluctuations to mean
ratio is of the order unity.

These results allow us to address three interrelated issues in quantum field 
theory in curved spacetime in the light of fluctuations of
quantum stress energy :
1) Fluctuation to mean ratio of vacuum energy density  and
the validity of semiclassical
gravity.
2) The spatial extent where negative energy density can
exist and its implications for quantum effects
of worm holes, baby universes and time travel.
 3) Dependence of fluctuations on intrinsic (defined by
smearing or point-separation)
and the extrinsic scale (such as the Casimir or finite
temperature periodicity).
4) The circumstances when and how divergences appear and the 
meaning of regularization in point-defined field theories versus theories 
defined at separated points and/or smeard fields. This
includes also the issue of
the cross term.

We begin by defining the smeared field operators and
their products and construct from 
them the smeared energy density and its fluctuations. We then 
calculate the ratio of the fluctuations to the
mean for a flat space (Minkowski
geometry) followed by a Casimir geometry of one periodic spatial dimension.
Finally we discuss the meaning of our finding in
relation to the issues raised above.

%##############################################################################
%##############################################################################
%            BODY OF PAPER <- Mathematica output
%##############################################################################
%##############################################################################
%===========================================================================
%   FORMALISM AND DEFINITIONS FOR SMEARED FIELDS
%===========================================================================
\subsection{Smeared Field Operators}

Since the field operator in conventional point-defined quantum field theory is an
operator-valued distribution, products of field operators
at a point become problematic. This parallels the problem with
defining the square of a delta function $\delta^2(x)$. Distributions
are defined via their integral against a test function: they live
in the space dual to the test function space. By going from the
field operator $\hat\phi(x)$ to its integral against a test function,
$\hat\phi(f) = \int \hat\phi\,f$, we can now readily consider products.

When we take the test functions to be spatial Gaussians, we are
smearing the field operator over a finite spatial region.
Physically we 
see  smearing as representing the necessarily finite extent of an 
observer's probe, or the intrinsic limit of resolution in carrying out 
a measurement at a low energy (compared to Planck scale).
In contrast to the ordinary  point-defined quantum field theory,
where ultraviolet divergences occur in the energy momentum tensor, 
smeared fields give no ultraviolet divergence. This is because smearing is 
equivalent to a regularization scheme which imparts an exponential 
suppression to the high momentum modes  and restricts the contribution 
of the high  frequency modes in the mode sum.

With this in mind, we start by defining the spatially smeared field operator
\beq
\hat\phi_t(f_{\bf x}) = 
     \int \hat\phi(t,{\bf x'}) f_{\bf x}({\bf x}')  d{\bf x}'
\eeq
where $f_{\bf x}({\bf x}')$ is a suitably smooth function.
With this, the two point operator becomes
\beq
\left(\hat\phi_t(f_{\bf x})\right)^2 = \int\int 
 \hat\phi(t,{\bf x}') \hat\phi(t,{\bf x}'') 
        f_{\bf x}({\bf x}') f_{\bf x}({\bf x}'')
  d{\bf x}\,d{\bf x}'
\eeq
which is now finite.
In terms of the vacuum $\left|\left.0\right>\right.$
($\hat a_{\bf k}\left|\left.0\right>\right. = 0 $, for all ${\bf k}$)
we have the usual mode expansion
\beq
\hat \phi\left(t_1,{\bf x}_1\right) 
= \int d\mu\left({\bf k}_1\right) \left(
 \hat a_{{\bf k}_{1}}\,u_{{\bf k}_{1}}\!\left(t_1,{\bf x}_1\right) + 
  \hat a^{\dagger}_{{\bf k}_{1}}\,
  u^{*}_{{\bf k}_{1}}\!\left(t_1,{\bf x}_1\right)
\right)
\eeq
with
\beq
  u_{{\bf k}_{1}}\left(t_1,{\bf x}_1\right) = 
	N_{k_{1}} {e^{i\,\left( {\bf k}_{1}\cdot{\bf x}_{1} -
                       t_{1}\,\omega_{1} \right) }}, 
\quad
  \omega_1 = \left| {\bf k}_1 \right|,
\eeq
where the integration measure $\int d \mu\left({\bf k}_1\right)$ and the 
normalization constants $N_{{\bf k}_1}$ are given for a Minkowski
and Casimir spaces by (\ref{Mdu}) and  (\ref{Cdu}) respectively.

Consider a Gaussian smearing function
\beq
f_{{\bf x}_0}({\bf x}) 
 = \left(
{\frac{1}{4\,\pi \,{{\sigma}^2}}} \right)^\frac{d}{2}
{e^{- \left(\frac{{\bf x}_{0} - {\bf x}}{2\sigma}\right)^2}}
\eeq
with the properties
$\int f_{{\bf x}_0}\left({\bf x}'\right) d{\bf x}' = 
1$,
$\int {\bf x}'\,f_{{\bf x}_0}\left({\bf x}'\right) 
d{\bf x}' = {\bf x}_0$ and
$\int |{\bf x}'|^2\,f_{{\bf x}_0}\left({\bf x}'\right) 
d{\bf x}' =  2d\sigma^2 + |{\bf x}_0|^2$.
Using
\beqn
\int 
  u_{{\bf k}_{1}}\!\left(t,{\bf x}\right) 
    f_{{\bf x}_1}({\bf x})
d{\bf x} 
&=&
  N_{k_{1}}  e^{-it\omega_1}\;
  \prod_{i=1}^d\left( 
     {\frac{1}{2\,{\sqrt{\pi }}\,\sigma}}
    \int e^{+ik_{1i} x_i - 
                     \left(\frac{x_{1i}-x_{i}}{2\sigma}\right)^2
           } \,dx_{i} 
  \right) 
\nonumber \\
&=&
N_{k_{1}}
{e^{-i\,t\,w + i\,{\bf k}_{1}\cdot{\bf x}_{1} - {{\sigma}^2}\,{{k_{1}}^2}}}
\eeqn
we get the smeared field operator
\beq
\hat\phi_{t_1}\left(f_{{\bf x}_1}\right) = \int d\mu\left({\bf k}_1\right) 
N_{k_{1}}
{e^{-i\,{\bf k}_{1}\cdot{\bf x}_{1} - {{\sigma}^2}\,{{k_{1}}^2} - 
      i\,t_{1}\,\omega_{1}}}\,\left( {e^{2\,i\,{\bf k}_{1}\cdot{\bf x}_{1}}}\,
     \hat a_{{\bf k}_{1}} + {e^{2\,i\,t_{1}\,\omega_{1}}}\,
     \hat a^{\dagger}_{{\bf k}_{1}} \right) 
\eeq
%%%%%%%%%%%%
and their derivatives
\begin{mathletters}
\beqn
\left(\partial_{t_1}\hat\phi_{t_1}\right)\left(f_{{\bf x}_1}\right) &=& \int 
   \left(\partial_{t_1}\hat\phi\left(t_1,{\bf x}'\right)\right)
   f_{{\bf x}_1}\left({\bf x}'\right)
\, d{\bf x}'
\nonumber \\
&=& i \int d\mu({\bf k}_1)
N_{k_{1}}\,\omega_{1}
{e^{-i\,{\bf k}_{1}\cdot{\bf x}_{1} - {{\sigma}^2}\,{{k_{1}}^2} - 
     i\,t_{1}\,\omega_{1}}} \left( {e^{2\,i\,t_{1}\,\omega_{1}}}\,\hat 
a^{\dagger}_{{\bf k}_{1}} - {e^{2\,i\,{\bf k}_{1}\cdot{\bf x}_{1}}}\,\hat 
a_{{\bf k}_{1}} \right)
\\ \nonumber \\
\left(\vec\nabla_{{\bf x}_1}\hat\phi_{t_1}\right)\left(f_{{\bf x}_1}\right) &=& \int 
   \left(\vec\nabla_{{\bf x}'}\hat\phi\left(t_1,{\bf x}'\right)\right)
   f_{{\bf x}_1}\left({\bf x}'\right)
\, d{\bf x}'
\nonumber \\
&=& -i \int d\mu({\bf k}_1)
{\bf k}_{1}\,N_{k_{1}}
{e^{-i\,{\bf k}_{1}\cdot{\bf x}_{1} - {{\sigma}^2}\,{{k_{1}}^2} - 
     i\,t_{1}\,\omega_{1}}} \left( {e^{2\,i\,t_{1}\,\omega_{1}}}\,\hat 
a^{\dagger}_{{\bf k}_{1}} - {e^{2\,i\,{\bf k}_{1}\cdot{\bf x}_{1}}}\,\hat 
a_{{\bf k}_{1}} \right).
\eeqn
\end{mathletters}
{}From this we can calculate the two point function of the field which make
up the energy density and their correlation function.
Letting
$ x \equiv (t,{\bf x}) = (t_2,{\bf x}_2) - (t_1,{\bf x}_1)$, they are given by
\begin{equation}
\rho\left(t,{\bf x};\sigma\right) = \int d\mu\left({\bf k}\right) 
  {{N^2_{k}}}\,{{\omega^2}}\, {e^{-2\,{{\sigma}^2}\,{{k^2}}}}\, 
  \cos ({\bf x}\cdot{\bf k} -  t\,\omega)
\label{rho-xsigma}
\end{equation}

\begin{equation}
\Delta\rho^2\left(t,{\bf x};\sigma\right) =
{\frac{1}{2}}
\int d\mu\left({\bf k}_1,{\bf k}_2\right) 
{{N^2_{k_{1}}}}\,{{N^2_{k_{2}}}}\,
  {{\left( {\bf k}_{1}\cdot{\bf k}_{2} + \omega_{1}\,\omega_{2} \right) }^2}\,
{e^{
 - 2\,{{\sigma}^2}\,\left( {{k^2_{1}}} + {{k^2_{2}}} \right) 
 - i\,{\bf x}\cdot\left({\bf k}_{1} + {\bf k}_{2}\right)
 + i\,t\,\left( \omega_{1} + \omega_{2} \right) }}
\label{Drho-xsigma}
\end{equation}

Setting $x=0$ we obtain the smeared vacuum energy density at one point 
\beq
\rho\left(\sigma\right) =
\int d\mu\left({\bf k}_1\right) {{N^2_{k_{1}}}}\,{{\omega^2_{1}}} \,
{e^{-2\,{{\sigma}^2}\,{{k^2_{1}}}}}
\label{rho-sigma}
\eeq

%===========================================================================
%    MINKOWSKI ENERGY DENSITY
%===========================================================================
\subsubsection{Smeared-Field Energy Density and Fluctuations
in Minkowski Space}

We consider a  Minkowski space $R^1 \times R^d$ with $d$-spatial dimensions. For this 
space the mode density is
\beq
\int d\mu({\bf k})  = \int_0^\infty k^{d-1}\,dk\int_{S^{d-1}} d\Omega_{d-1}
\quad{\rm with}\quad
\int_{S^{d-1}} d\Omega_{d-1} = {\frac{2\,{{\pi 
}^{{\frac{d}{2}}}}}{\Gamma\!\left({\frac{d}{2}}\right)}}
\eeq
and the mode function normalization constant is
\beq
N_{k_{1}} = 1/{\sqrt{{2^{d+1}}\,{{\pi }^d}\,\omega_{1}}}.
\label{Mdu}
\eeq
We introduce the angle between two momenta in phase space, $\gamma$, via
\beq
{\bf k}_1 \cdot {\bf k}_2 = k_1 k_2 \cos(\gamma)
                          = \omega_1 \omega_2 \cos(\gamma).
\eeq
The averages of the cosine and cosine squared of this angle over a pair of unit
spheres are
\begin{mathletters}
\beqn
\int_{S^{d-1}} d\Omega_1 \int_{S^{d-1}} d\Omega_2 \cos(\gamma) &=& 0 \\
\int_{S^{d-1}} d\Omega_1 \int_{S^{d-1}} d\Omega_2 \cos^2(\gamma) &=& 
       {\frac{4\,{{\pi }^d}}{d\,{{\Gamma\!\left({\frac{d}{2}}\right)}^2}}}.
\eeqn
\end{mathletters}
The smeared energy density (\ref{rho-sigma}) becomes 
\beqn
\rho(\sigma) &=& 
{\frac{1}{{2^d}\,{{\pi }^{{\frac{d}{2}}}}\,\Gamma\!\left({\frac{d}{2}}\right)}}
\int_0^\infty {\frac{{{k_{1}}^d}}{{e^{2\,{{\sigma}^2}\,{{k_{1}}^2}}}}} \,dk_1
\nonumber \\
&=&
%{\frac{{{\sigma}^{-1 - d}}\,\Gamma\!\left({\frac{d+1}{2}}\right)}
%   {{2^{{\frac{3\,\left( d+1 \right) }{2}}}}\,{{\pi }^{{\frac{d}{2}}}}\,
%     \Gamma\!\left({\frac{d}{2}}\right)}}
\frac{
  \Gamma\!\left({\frac{d+1}{2}}\right)
     }{
  {2^{{\frac{3\,\left( d+1 \right) }{2}}}}\,{{\pi }^{{\frac{d}{2}}}}\,
     \sigma^{d+1} \Gamma\!\left({\frac{d}{2}}\right)
     }
\eeqn

%===========================================================================
%    MINKOWSKI FLUCTUATIONS
%===========================================================================
For the fluctuations of the smeared energy density operator, 
we evaluate (\ref{Drho-xsigma})  for this space and find
\beqn
\Delta\rho^2(\sigma) &=&
{\frac{1}{{2^{(2d+3)}}{{\pi }^{2\,d}}}}
\int_0^\infty \int_0^\infty
\int_{S^{d-1}}\int_{S^{d-1}}
{\frac{{{\left( 1 + {\cos(\gamma)} \right) }^2}\,{{k^d_{1}}}\,{{k^d_{2}}}}
   {{e^{2\,{{\sigma}^2}\,\left( {{k^2_{1}}} + {{k^2_{2}}} \right) }}}} 
\,d\Omega_1 \,d\Omega_2 \,dk_1\,dk_2
\nonumber \\
%&=&
%{\frac{\left( d+1 \right) }
%   {{2^{(2d+1)}}\,d\,{{\pi }^d}\,{{\Gamma\!\left({\frac{d}{2}}\right)}^2}}}
%\int_0^\infty \int_0^\infty
%{\frac{{{k_{1}}^d}\,{{k_{2}}^d}}
%   {{e^{2\,{{\sigma}^2}\,\left( {{k^2_{1}}} + {{k^2_{2}}} \right) }}}} \, 
%dk_1\,dk_2
%\nonumber \\
&=&
{\frac{\left( d+1 \right) \,
     {{\Gamma\!\left({\frac{d+1}{2}}\right)}^2}}{{2^{(3d+4)}}\,d\,{{\pi }^d}\,
     {{\sigma}^{2\,\left( d+1 \right) }}\,
     {{\Gamma\!\left({\frac{d}{2}}\right)}^2}}}
\eeqn
Putting these together we obtain
for the Minkowski space 
\beq
\Delta_{\rm Minkowski}(d) = 
  {\frac{1+d}{1 + 3\,d}} 
\eeq
which has the particular values:
$$
{1\over2}\  {\rm for}\  d=1,\ \ 
{2\over5}\  {\rm for}\  d=3,\ \
{3\over8}\  {\rm for}\  d=5,\ \
{1\over3}\  {\rm for}\  d=\infty.
$$

%%%%%%%%%%%%%%%%%%%%%%%%%%%%%
% @ I have disactivated the tables (I wrote results in Eq. above)
%%%%%%%%%%%%%%%%%%%%%%%%%%%%

%\[
%\begin{array}{|c||c|c|c|c}
%d & 1 & 3 & 5 & \infty \\ 
%\hline
%\Delta_{\rm Minkowski} & \frac{1}{2} &
%\frac{2}{5}  & \frac{3}{8}  & \frac{1}{3} 
%\\
%\end{array}
%\]

%%%%%%%%%%%%%%%%%%%%%%%%%%%%%%%

%===========================================================================
%   CASIMIR TOPOLOGY
%===========================================================================
\subsubsection{Smeared-Field in Casimir Topology}
%===========================================================================
%   CASIMIR ENERGY  %  d refers to the SPATIAL dim throughout this work
%===========================================================================
The Casimir topology is obtained from a  flat space (with $d$ spatial dimensions, i.e.,  
$R^1 \times R^{d}$ ) by imposing  periodicity $L$ in one of its spatial dimensions, 
say,  $z$, thus endowing it  with
a  $R^1\!\times\!R^{d-1}\!\times\!S^1$  topology. 
We decompose ${\bf k}$ into a component along the  periodic dimension and 
call the remaining components ${\bf k}_{\perp}$:
\begin{mathletters}
\label{Cdu}
\begin{eqnarray}
{\bf k} &=& \left({\bf k}_\perp,\frac{2 \pi n}{L}\right) 
        = \left({\bf k}_\perp, l n\right), l \equiv 2 \pi/L \\ 
\omega_{1} &=& {\sqrt{{{k^2_{1}}} + {l^2}\,{{n_{1}}^2}}} 
\end{eqnarray}
The normalization and momentum measure are
\begin{eqnarray}
\int d\mu({\bf k}) &=&\int_0^\infty k^{d-2}\,dk\int_{S^{d-2}} d\Omega_{d-2}
\sum_{n=-\infty}^\infty \\
N_{k_{1}} &=& {\frac{1}{{\sqrt{{2^d}\,L\,{{\pi }^{d-1}}\,\omega_{1}}}}}
\end{eqnarray}
\end{mathletters}
With this, the energy density (\ref{rho-sigma}) becomes
\beqn
\rho_L\left(\sigma\right) &=&
{\frac{l}
   {{2^d}\,{{\pi }^{{\frac{d+1}{2}}}}\,\Gamma\!\left({\frac{d-1}{2}}\right)}}
\sum_{n_1=-\infty}^\infty\int_0^\infty 
{{k_{1}^{d-2}}}\,\left({{k^2_{1}}} + {l^2}\,{n^2_1}\right)^{\frac{1}{2}}\,
{e^{-2\,{{\sigma}^2}\,\left( {{k^2_{1}}} + {l^2}\,{n^2_1} \right) }}\;dk_1
\eeqn
we can write this as the sum of the two smeared Green function derivatives
\beqn
\rho_L\left(\sigma\right) &=& 
      \left<0_L\left| \left(
        \left( \nabla_{\!\perp} \phi_t \right)\left(f_{\bf x}\right)
      \right)^2 \right|0_L\right>
+
      \left<0_L\left| \left(
        \left( \partial_z \phi_t \right)\left(f_{\bf x}\right)
      \right)^2 \right|0_L\right>
\nonumber \\
&=& \Gperp + \Gz
\eeqn
where $\left.\left.\right|0_L\right>$ is the Casimir vacuum.

%\subsection{Regularized Casimir Energy Density}

Since ${G_L(\sigma)}_{,i} = G_{L,i}^{\rm div} + G_{L,i}^{\rm fin}$
($i=x_\perp x_\perp$ or $zz$)
we see how to split the smeared energy density into a
$\sigma\rightarrow 0$ divergent term and the finite contribution:
\beq
\rho_L\left(\sigma\right) = \rho_L^{\rm div} + \rho_L^{\rm fin}
\eeq
where
\beqn
\rho_L^{\rm div} &=& \Gperpdiv +\Gzdiv
\nonumber \\ &=&
{\frac{\,\Gamma\!\left({\frac{d+1}{2}}\right)}
   {{2^{{\frac{3\,\left( d+1 \right) }{2}}}}\,{{\pi }^{{\frac{d}{2}}}}\,
     {{\sigma}^{d+1}}\Gamma\!\left({\frac{d}{2}}\right)}}
\nonumber \\
&=& \rho\left(\sigma\right)
\eeqn
and
\beqn
\rho_L^{\rm fin} &=& \Gperpfin + \Gzfin
\nonumber \\
&=&  
-\frac{
   d\,
\Gamma\!\left(-{\frac{d}{2}}\right)\Gamma\!\left({\frac{d}{2}}\right)\,
      }{
   (4\pi)^{(d+3)/2}\,{l^{d+1}}\, 
}
\sum_{p=1}^\infty
{{\left( -1 \right) }^p}\,
 {(2\,l)^{2p}}\,
  p\, {{\left( 2p-1 \right) }^2}\,
  {{\sigma}^{2\,\left(p-1\right) }} 
%\nonumber \\&&\hspace{10mm}\times 
{\frac{B_{2p+d-1}}{2p+d-1}} 
{\frac{\left( 2p-3 \right) !!}{\left( 2\,p \right) !}} 
{\frac{
     \Gamma\!\left(p-{\frac{1}{2}}\right)}{\Gamma\!\left(p+{\frac{d}{2}}
     \right)}}
\eeqn
With this we define the regularized energy density
\beqn
\rho_{L,{\rm reg}} &\equiv& \lim_{\sigma\rightarrow 0}\left(
 \rho_L\left(\sigma\right) - \rho\left(\sigma\right)
\right)\nonumber \\
&=&  {\frac{d\,\,{{\pi }^{{\frac{d}{2}}}}\,B_{d+1}\,
     \Gamma\!\left(-{\frac{d}{2}}\right)\,\Gamma\!\left({\frac{d}{2}}\right)}
     {2\,\left( d+1 \right) {L^{d+1}}\,\Gamma\!\left({\frac{d}{2}}+1\right)}}
\eeqn
and get the usual results:
$$ 
- {\frac{{{\pi }}}{6\,{L^2}}}\ {\rm for}\ d=1,\ \ 
- {\frac{{{\pi }^2}}{90\,{L^4}}}\ {\rm for}\ d=3,\ \ 
- {\frac{2\,{{\pi }^3}}{945\,{L^6}}} \ {\rm for}\ d=5.
$$

%%%%%%%%%%%%%%%%%%%%%%%%%%%%%%%
% @ desactivated table (results above)
%%%%%%%%%%%%%%%%%%%%%%%%%%%%%%%%%
%\[
%\begin{array}{|c||c|c|c|}
%\hline
%     d             &          1 & 3 & 5 \\ 
%\hline 
%\rho_{L,{\rm reg}} &  - {\frac{{{\pi }}}{6\,{L^2}}} &
% - {\frac{{{\pi }^2}}{90\,{L^4}}} &
%          - {\frac{2\,{{\pi }^3}}{945\,{L^6}}} 
%\\ 
%\hline
%\end{array}
%\]

%===========================================================================
%   CASIMIR FLUCTUATIONS
%===========================================================================
%\subsection{Casimir energy density fluctuations}

For the $d$-dimensional Casimir geometry, the fluctuations are
\beqn
\Delta\rho^2_L\left(\sigma\right) &=&
{\frac{{l^2}}{{2^{2d+3}}\,{{\pi }^{2\,d}}}}
\sum_{n_1=-\infty}^\infty \sum_{n_2=-\infty}^\infty
\int_0^\infty \hspace{-3mm}k_1^{d-2}\,dk_1 \int_0^\infty \hspace{-3mm}k_2^{d-2}\,dk_2
\int_{S^{d-2}}\hspace{-5mm} d\Omega_{1} \int_{S^{d-2}}\hspace{-5mm} d\Omega_{2}
\nonumber \\ && \hspace{10mm}\times
\frac{
{e^{-2\,{{\sigma}^2}\left(\omega_1^2 + \omega_2^2\right)}} 
}{\omega_{1}\,\omega_{2}}
\left( {\cos(\gamma)}\,k_{1}\,k_{2} + {l^2}\,n_{1}\,n_{2} + 
         \omega_{1}\,\omega_{2} \right)^2
\eeqn
This expression can again be written in terms of products of
the Green functions derivatives used above:
\beq                 
\Delta\rho^2_L\left(\sigma\right) = 
  {\frac{d\,\left(\Gperp\right)^2
      }{2\,\left( d-1 \right) }} 
+  
    \Gz
       \left(\Gperp  + \Gz \right)
\eeq
and split into three general terms
\beq
\Delta\rho^2_L\left(\sigma\right) = \Delta\rho^{2,{\rm div}}_L
   + \Delta\rho^{2,{\rm cross}}_L
   + \Delta\rho^{2,{\rm fin}}_L
\eeq
The full expression for each of these terms are given in the Appendix of \cite{PH1}
The first term contains only the divergent parts of the Green functions
while the last term contains only the finite parts. This is similar
to the split we used for the smeared energy density above. What is new
here is the middle term $\Delta\rho^{2,{\rm cross}}_L$. This comes about
from the products of the divergent part of one Green function and the
finite part of the other. That this term arises for computations of the
energy density fluctuations is a generic feature. We will discuss in greater detail
the meaning of this term later.

The results of Appendix A in \cite{PH1}  give
\begin{mathletters}
\beq
\Delta\rho^{2,{\rm div}}_L = \chi_d\left( \rho_L^{\rm div} \right)^2
=  \chi_d\left( \rho\left(\sigma\right)\right)^2,
\eeq
\beq
\Delta\rho^{2,{\rm cross}}_L =
2 \chi_d \, \rho_L^{\rm div} \, \rho_L^{\rm fin},
\eeq
\beq
\Delta\rho^{2,{\rm fin}}_L =
{\frac{d\,\left( d+1 \right) }{2}}
 \left( \rho_{L,{\rm reg}} \right)^2.
\eeq
\end{mathletters}

{}From this we see the divergent and cross terms can be related to
the smeared energy density via
\beq
\Delta\rho^{2,{\rm div}}_L + \Delta\rho^{2,{\rm cross}}_L = 
 \chi_d\left\{
    \left(\rho_L^{\rm div}\right)^2 + 2 \rho_L^{\rm div} \rho_L^{\rm fin}
\right\}
\eeq
where $\chi_d$ is the function that relates the fluctuations of the
energy density to the mean energy density when the boundaries are not
present, i.e., Minkowski space. This leads us to interpret these terms 
as due to the vacuum fluctuations that are always present. With this in mind, 
we define the regularized fluctuations of the energy density
\beqn
\Delta\rho^2_{L,{\rm reg}} &=&
 \lim_{\sigma\rightarrow 0}\left(
      \Delta\rho^2_L\left(f\right)
    - \chi_d\left\{
      \left(\rho_L^{\rm div}\right)^2 + 2 \rho_L^{\rm div} \rho_L^{\rm fin}
      \right\}
 \right)
\nonumber \\
&=& \chi_{d,L} \left( \rho_{L,{\rm reg}} \right)^2
\eeqn
where 
\beq
\chi_{d,L} \equiv {\frac{d\,\left( d+1 \right) }{2}}.
\eeq
We also define a regularized version of the dimensionless measure
$\Delta$:
\beq
\Delta_{L,{\rm Reg}} \equiv
  \frac{\Delta\rho^2_{L,{\rm Reg}}}
       {\Delta\rho^2_{L,{\rm Reg}} + \left(\rho_{L,{\rm Reg}}\right)^2}
=
{\frac{d\,\left( d+1 \right) }{2 + d + {d^2}}} 
\eeq
and note the values:
$$
{1\over2}\ {\rm for}\ d=1,\ \ 
{6\over7}\ {\rm for}\ d=3,\ \
{15\over16}\ {\rm for}\ d=5,\ \
1\ {\rm for}\ d=\infty.
$$

%%%%%%%%%%%%%%%%%%%%%%%%%%%%%%%%%%%
% @ desactivated table (results above)
%%%%%%%%%%%%%%%%%%%%%%%%%%%%%%%%%%%%
%\[
%\begin{array}{|c||c|c|c|c|}
%\hline
%d & 1 & 3 & 5 & \infty \\ 
%\hline 
%\Delta_{L,{\rm Reg}} & \frac{1}{2} & \frac{6}{7} & \frac{15}{16} & 1
%\\ \hline
%\end{array}
%\]
%%%%%%%%%%%%%%%%%%%%%%%%%%%%%%%%%%%%%%%
Following the procedures described in Appendix B of \cite{PH1},
Phillips and Hu have made two plots,
Fig. 1 of $\Delta (\sigma, L)$ and $\Delta_{L, Reg}$ versus
$\sigma /L$, (which 
we call  $\sigma'$ here for short); 
and Fig. 2 of $\rho_{L, Reg}$ and $\sqrt {\Delta \rho^2_{L, Reg}}$
versus $\sigma'$.
The range of $\sigma'$ is limited to $\le 0.4$ because going 
any further would make the meaning of a local energy density ill-defined, as  
the smearing of the field extends to the Casimir boundary in  space.
(The infrared limit also carry important physical meaning
in reference to the structure of spacetime.)

Let us ponder on the meaning they convey.  In Fig. 1, we
first note that both curves 
are of the order unity. But the behavior of $\Delta$
(recall that the energy density 
fluctuations thus  defined include the cross term along with 
the finite part and the  state independent divergent part)  
is relatively insensitive to the smearing width, whereas $\Delta_{L,Reg}$,
which measures only the finite part of the energy density fluctuations to
the mean, has more structure.   In particular, it saturates
its upper bound of 1
around $\sigma' = 0.24$. Note that if one adheres to the KF
criterion \cite{Kuo-Ford93}
one would say that semiclassical gravity fails, but all that
is happening here
is that $\rho_{L,{\rm Reg}}=0$ while $\Delta \rho_{L,{\rm Reg}}^2 $
shows no special 
feature.  The real difference between 
these two functions is the cross term, which is responsible for their 
markedly different structure and behavior.

In Fig. 2, the main feature to notice is that the regularized  energy density 
crosses from negative to positive values at around $\sigma' = 0.24$.
The negative
Casimir energy density  calculated in a point-wise field theory
which corresponds
to small ranges of $\sigma'$ is expected, and is usually taken to
signify the quantum nature of the Casimir state. As  $\sigma'$ increases
we are averaging the field operator over a larger region, and thus sampling
the field theory from the ultraviolet all the way to the
infrared region. At large
$\sigma'$ finite size effect begins to set in. The difference and relation 
of these two effects are explained in  \cite{HuO'C}: Casimir effect
arises from summing up the quantum fluctuations of ALL modes
(as altered by the boundary), with no insignificant short wavelength 
contributions, whereas finite size effect has dominant contributions from 
the LONGEST wavelength modes, and thus reflect the large scale 
behavior.  As the smearing moves from a small scale to the far boundary of space,
the behavior of the system is expected to shift from a Casimir-dominated to 
a finite size-dominated effect.  This could be the underlying reason in 
the  crossover behavior of $\rho_{L, Reg}$.

\subsection{Fluctuation to Mean Ratio and Spatial
Extent of Negative Energy Density}

Now that we have the results we can return to the issues raised earlier. We discuss the first 
two here, i.e.,
 1) Fluctuations of the energy density and validity of semiclassical gravity,
 2) The spatial extent where negative energy density can exist. 
 We will discuss the regularization of energy density fluctuations and the issue of the cross
term in the next subsection.

\subsubsection {Fluctuation to Mean ratio and Validity of SCG}

{}From these results 
we see that i) the fluctuations of the energy density as well as its mean both 
increase with decreasing distance (or probing scale), while ii)
the ratio of the variance of the fluctuations in EMT to its mean-squared 
is of the order unity. 
We view the first but not the second feature as linked
to the question of the validity of SCG.
The second feature represents something quite different,  
pertaining more to the quantum nature of the
vacuum state than to the validity of SCG.

If we adopt the criterion of Kuo and Ford
\cite{Kuo-Ford93} that the variance of
the fluctuation relative to the mean-squared (vev
taken with respect to the ordinary
Minkowskian vacuum)  being of the order unity be an indicator of the failure of
SCG, then all spacetimes studied above would indiscriminately fall into that
category, and SCG fails wholesale, regardless of the scale these  physical
quantities are probed. This contradicts with the common expectation  that SCG
is valid at scales below  Planck energy. We believe the
criterion for the validity or
failure of a theory should depend on the range or the energy probed.
The findings 
of Phillips and Hu \cite{PH1} related here seem to confirm this:
Both the mean (the vev of EMT with
respect to the Minkowski vacuum) AND  the fluctuations of EMT increase as the
scale deceases.  As one probes into an increasingly finer
scale or higher energy the
expectation value of EMT  grows in value and the induced  metric fluctuations
become important, leading to the failure of SCG.  A generic scale for this to
`happen is the Planck length.  At such energy densities and above, particle
creation from the quantum field vacuum would become copious and their
backreaction on the background spacetime would become
important \cite{scg}. Fluctuations
in the quantum field EMT entails these quantum processes. The
induced metric fluctuations \cite{stogra} render the smooth
manifold structure of spacetime inadequate, spacetime foams \cite{Wheeler}
including topological transitions \cite{Hawking} begin to appear and SCG no
longer can provide an adequate description of these dominant processes. This
picture first conjured by Wheeler is consistent with the common notion adopted
in SCG, and we believe it is a valid one.

\subsubsection{Extent of Negative Energy Density}

It is well known that negative energy density exists in Casimir geometry,
moving mirrors, black holes and worm holes. Proposals have also been conjured
to use the negative energy density for the design of time machines \cite{Thorne}. Our results
(Figures 1, 2) provide an explicit scale dependence of the regularized vacuum energy density 
$\rho_{L,reg}$ and its fluctuations $\Delta_{L,reg}$ , specifically $\sigma/L$, the ratio
of the smearing length (field scale) to that of the Casimir length (geometry scale).  For 
example, Fig. 2 shows that only for $\sigma /L < 0.24$ is $\rho_{L,reg}<0$.
Recall $\sigma$ gives the spatial extent the field is probed or smeared. Ordinary
pointwise quantum field theory which probes the field only at a point does not carry
information about the spatial extent where negative energy density sustains.
These results have direct implications on wormhole physics (and time machines,  if one 
gets really serious about these fictions \cite{Thorne}). 
If L is the scale characterizing the size (`throat') of the wormhole where one
thinks negative energy density might prevail, and designers of `time machines'
wish to exploit for `time-travel' , the results of Phillips and Hu can provide a limit
on the size of the probe (spaceship in the case of time-travel) in ratio to L where such conditions
may exist. It could also provide a quantum field-theoretical bound on the probability
of spontaneous creation of baby universes from quantum field energy fluctuations. 

\subsection{Dependence of fluctuations on intrinsic
and extrinsic scales}

One may ask why the fluctuations of the energy density 
to its mean for the many cases calculated by PH and KF
should be the fractional numbers as they are.
Is  there  any simple reason behind the following features observed
in Phillips and Hu's calculations?

a)  $\Delta= O (1)$

b) The specific numeric values of $\Delta$ for the different cases. 

c)  For the Minkowski vacuum  the ratio of the variance to the mean-squared, 
calculated from the
coincidence limit, is identical to the value of the Casimir case at the same
limit for spatial point separation  while identical to the value of a hot flat
space result with a temporal point-separation. 

Point a) has also been shown by earlier calculations
\cite{Kuo-Ford93,Phillips-Hu97}, and
our understanding is that this is true only for states of quantum nature,
including the vacuum and certain squeezed states, but probably not true for
states of a more classical nature like the  coherent state. We also emphasized
that this result should not be  used as a criterion for the validity  of
semiclassical gravity.

For point b), we can trace back the calculation of the fluctuations (second
moment) of the energy momentum tensor in ratio to its mean (first moment) to
the  integral of the term containing the inner product of  two momenta ${\bf
k}_1\cdot{\bf k}_2$ summed over all participating modes. The modes
contributing to this are different for different geometries, e.g., Minkowski
versus Casimir boundary --for the  Einstein universe this enters as 3j symbols
-- and could account for  the difference in the numerical
values of $\Delta$ for
the different cases.

For point c),  to begin with,  it is well-known that the regularization by 
taking the coincidence limit of a spatial versus a temporal point separation
will give different results.  The case
of temporal split involves integration of three spatial dimensions  while the
case of spatial split involve integration of two remaining spatial  and one
temporal dimension, which would give different results. The calculation of
fluctuations involves the second moment of the field and is in this regard
similar to what enters into the calculation of moments of  inertia
for rotating objects. We suspect that the difference between the
temporal and the spatial results is similar, to the extent this analogy holds, 
to the difference in the moment of inertia of the same object but taken with 
respect to different axes of rotation. 

It may be surprising that in a Minkowski calculation the result of Casimir 
geometry or thermal field should appear, as both cases involve a scale -- 
the former in the spatial dimension and the latter in the (imaginary)
 temporal dimension. (Both cases have the same topology 
$R^3 \times S^1$, with the $S^1$ in the (imaginary) time for the former
and in the space for the latter.) But if we note
that the results for Casimir geometry or thermal field are obtained at the
coincidence (ultraviolet) limit, when the scale (infrared) of the  problem does
not intercede in any major way, then the main components of  the calculations
for these two cases would be similar to the two cases of taking the coincidence
limit in the spatial and temporal directions in Minkowski space. 

All of these
cases are effectively devoid of scale as far as the pointwise field theory is
concerned. As soon as we depart from this limit the effect of the presence of a
scale shows up. The point-separated or field-smeared results for the Casimir
calculation shows clearly that the boundary scale enters in a major
way and the result for the fluctuations and the ratio are different from those
obtained at the coincident limit.  For other cases where  a scale enters
intrinsically in the problem such as   that of a massive or non-conformally
coupled field it would show a similar effect in these regards as the present
cases (of Casimir and thermal field) where a periodicity condition exists (in
the spatial and temporal directions respectively). We expect a similar strong 
disparity between point-coincident and point-separated cases. The field theory
changes its nature in a fundamental and physical way when this limit is taken.
This brings us to an even more fundamental issue made clear
in this investigation,
i.e., the appearance of divergences and the  meaning of regularization in the
light of  a point-separated versus a point-defined quantum field theory.

\subsection{Regularization in the Fluctuations of EMT and
the Issue of the Cross Term}

An equally weighty issue brought to light in the study of Phillips and Hu
is the meaning of regularization in the face of EMT fluctuations. Since the
point-separated or smeared field expressions of the EMT and its fluctuations
become available one can study how
they change as a function of separation or smearing scale in addition to
how  divergences arise at the coincidence limit. Whether certain cross terms
containing divergences have physical meaning is a question raised by the recent
studies of Wu and Ford \cite{WuFor}.  We can use these calculations to examine
these issues and  ask the broader question of what exactly
regularization means and
entails, where divergences arise and why they need to be,
and not just how they 
ought to be treated.

Recall the smeared energy density fluctuations
for the Casimir topology has the form
\beq
\Delta\rho_L^2(\sigma) = 
   \Delta\rho_L^{\rm div} + \Delta\rho_L^{\rm cross}  + \Delta\rho_L^{\rm fin}
\eeq
with
\begin{mathletters}
\beqn
\Delta\rho_L^{\rm div}   &=& \chi_d \left( \rho_L^{\rm div} \right)^2
                        =  \chi_d (\rho(\sigma))^2
\\
\Delta\rho_L^{\rm cross} &=& 2 \chi_d \rho_L^{\rm div} \rho_L^{\rm fin}
\\
\Delta\rho_L^{\rm fin} &=& 2 \chi_{d,L} \left( \rho_L^{\rm fin} \right)^2
+{\rm terms\; that\; vanish\; as}\;\sigma\rightarrow 0
\eeqn
\end{mathletters}
where $\chi_d$ is the ratio between the fluctuations for Minkowski space
and the square of the corresponding energy density:
$\Delta\rho^2 = \chi_d(\rho(\sigma))^2$.
Our results show that
$\Delta\rho_L^2(\sigma)$ diverges as the width $\sigma$ of the smearing 
function
shrinks to zero with contributions from the truly divergent and the 
cross terms.
We also note that the divergent term 
$\Delta\rho^{\rm div}$ is state independent, in the sense that 
it is independent
of $L$, while the cross term $\Delta\rho^{\rm cross}$ is state dependent, 
as is the
finite term  $\Delta\rho^{\rm fin}$.

If we want to ask about the strength of fluctuations of the energy density, the
relevant quantity to study is the energy density correlation function
$H(x,y) = \left<\hat\rho(x)\hat\rho(y)\right>
-\left<\hat\rho(x)\right>\left<\hat\rho(y)\right>$. It is finite at
 $x\ne y$ for a linear quantum theory (this happens since the
divergences for $\left<\hat\rho(x)\hat\rho(y)\right>$ are exactly the
same as the product $\left<\hat\rho(x)\right>\left<\hat\rho(y)\right>$),
but diverges as $y\rightarrow x$,  corresponding to the
coincident or unsmeared limit $\sigma\rightarrow 0$.

To define a procedure for rendering  
$\Delta\rho_L^2(\sigma)$ finite, one can see that there exists
choices -- which means ambiguities in the regularization scheme.
Three possibilities present themselves:
The first is to just drop the state independent 
$\Delta\rho^{\rm div}$. This is easily seen to fail since we are left with the
divergences from the cross term. The second is to neglect all terms that
diverge as $\sigma\rightarrow 0$.
This is too rash a move  since $\Delta\rho^{\rm cross}$
has, along with its divergent
parts, ones that are finite in the $\sigma\rightarrow 0$
limit. This comes about
since it is of the form $\rho_L^{\rm div} \rho_L^{\rm fin}$ and the negative
powers of $\sigma$ present in $\rho_L^{\rm div}$ will cancel out against the
positive powers in $\rho_L^{\rm fin}$.
Besides, they yield results in disagreement with earlier
results using well-tested
methods such as normal ordering in flat space \cite{Kuo-Ford93} 
and zeta-function regularization in curved space \cite{Phillips-Hu97}.

The third choice is the one adopted by PH \cite{PH1}.
 For the energy density, we can think of
regularization as computing the contribution
``above and beyond'' the Minkowski
vacuum contribution, same for  regularizing the fluctuations.
So we need to first determine for Minkowski space vacuum
how the fluctuations of the energy density  are related
to the vacuum energy density  $\Delta\rho^2 = \chi\;(\rho)^2$.
This we obtained  for finite smearing.
For Casimir topology the sum of the divergent and cross terms take the
form
\beq
\Delta\rho_L^{\rm div} + \Delta\rho_L^{\rm cross} =
\chi\left\{
	\left(\rho_L^{\rm div}\right)^2+2\rho_L^{\rm div}\rho_L^{\rm fin}
		\right\}
= \chi\left\{
	\left(\rho_L\right)^2-\left(\rho_L^{\rm fin}\right)^2
		\right\}
\eeq
where $\chi$ is the ratio derived for Minkowski vacuum. We take this to represent
the (state dependent) vacuum contribution. What we find interesting is that 
to regularize the smeared energy density fluctuations, a state
dependent subtraction must be used. With this, just the 
$\sigma\rightarrow 0$ limit of the finite part 
$\Delta\rho_L^{\rm fin}$ is identified as the regularized fluctuations
$\Delta\rho^2_{L,{\rm Reg}}$. The ratio $\chi_L$ thus obtained gives  exactly 
the same result as derived  by Kuo and Ford for $d=3$
via normal ordering \cite{Kuo-Ford93}
and by PH for arbitrary $d$ via the $\zeta$-function
\cite{Phillips-Hu97}.

That this procedure is the one to follow can be seen by considering the
problem from the point separation method, see section \ref{sec4}.
For this method, the energy
density expectation value is defined as the $x'\rightarrow x$ limit of
\beq
\rho(x,x') = {\cal D}_{x,x'} G(x,x')
\eeq
for the suitable Green function $G(x,x')$ and
${\cal D}_{x,x'}$ is a second order differential operator.
In the limit ${x'\rightarrow x}$, $G(x,x')$ is divergent. 
The Green function is regularized
by subtracting from it a Hadamard form $G^L(x,x')$:
$G_{\rm Reg}(x,x') = G(x,x') - G^L(x,x')$ \cite{Wald75}. With this, the regularized energy
density can be obtained
\beq
\rho_{\rm Reg}(x) = 
    \lim_{x'\rightarrow x}\left({\cal D}_{x,x'} G_{\rm Reg}(x,x')\right)
\eeq
Or, upon re-arranging terms,  define the divergent and finite pieces as
\beq
G^{\rm div}(x,x') =  G^L(x,x'), \quad
G^{\rm fin}(x,x') = G_{\rm Reg}(x,x') = G(x,x') - G^L(x,x')
\eeq
and
\beq
\rho(x,x') = \rho^{\rm div}(x,x') + \rho^{\rm fin}(x,x')
\eeq
\[
\rho^{\rm div}(x,x')  = {\cal D}_{x,x'} G^{\rm div}(x,x')
\quad{\rm and}\quad
\rho^{\rm fin}(x,x')  = {\cal D}_{x,x'} G^{\rm fin}(x,x')
\]
so that $\rho_{\rm Reg}(x) = \lim_{x'\rightarrow x}\rho^{\rm fin}(x,x')$,
which corresponds to the $\sigma\rightarrow 0$ limit in PH's computation
of the Casimir energy density.

Now turning to the fluctuations, we have the point-separated
expression for the
correlation function
\beq
H(x,y) = \lim_{x'\rightarrow x} \lim_{y'\rightarrow y}
  {\cal D}_{x,x'} {\cal D}_{y,y'} G(x,x',y,y')
\eeq
where $G(x,x',y,y')$ is the suitable four point function. For linear theories
we can use Wick's Theorem to express this in terms
of products of Green functions
$G(x,x',y,y') = G(x,y) G(x',y') +
{\rm permutations\; of}(x,x',y,y')$. Excluded from
the permutations is $G(x,x')G(y,y')$. (Details are in
\cite{NPhD,PH2}, which includes 
correct identifications of needed permutations and Green functions.)
The general form is
\beq
H(x,y) = \lim_{x'\rightarrow x} \lim_{y'\rightarrow y}
  {\cal D}_{x,x'} {\cal D}_{y,y'} G(x,y) G(x',y')
\;+\;{\rm permutations}
\eeq
The ${(x',y')\rightarrow (x,y)}$ limits are only
retained to keep track of which
derivatives act on which Green functions, but we can
see there are no divergences
for $y\ne x$. However, to get the point-wise fluctuations
of the energy density,
the divergences from $\lim_{y\rightarrow x}G(x,y)$ will present a problem.
Splitting the Green function into its finite and divergent pieces, we can
recognize terms leading to those we found for $\Delta\rho^2_L(\sigma)$:
\beq
H(x,y) = H^{\rm div}(x,y) + H^{\rm cross}(x,y) + H^{\rm fin}(x,y)
\eeq
where
\begin{mathletters}
\beqn
H^{\rm div}(x,y)  &=&  
    \lim_{x'\rightarrow x} \lim_{y'\rightarrow y}
    {\cal D}_{x,x'} {\cal D}_{y,y'} G^{\rm div}(x,y) G^{\rm div}(x',y')
\\
H^{\rm cross}(x,y)  &=&  
    2\lim_{x'\rightarrow x} \lim_{y'\rightarrow y}
    {\cal D}_{x,x'} {\cal D}_{y,y'} G^{\rm div}(x,y) G^{\rm fin}(x',y')
\\
H^{\rm fin}(x,y)  &=&  
    \lim_{x'\rightarrow x} \lim_{y'\rightarrow y}
    {\cal D}_{x,x'} {\cal D}_{y,y'} G^{\rm fin}(x,y) G^{\rm fin}(x',y'),
\eeqn
\end{mathletters}
plus permutations.
Thus we see the origin of  both the divergent and cross terms.
When the un-regularized Green function is used,
we must get a cross term, along with the expected divergent term.
If  the fluctuations of the  energy density is 
regularized via point separation, i.e. $G(x,x')$ is replaced
by $G_{\rm Reg}(x,x')=G^{\rm fin}(x,y)$, then we should do the
same replacement for the fluctuations. When this is done, it is
only the finite part above that will be left and we can define
the point-wise fluctuations as
\beq
\Delta\rho^2_{\rm Reg}=\lim_{y\rightarrow x}H^{\rm fin}(x,y)
\eeq
The parallel with the smeared-field derivation presented
here can be seen when the
analysis of $\Gz$ and $\Gperp$ (given in the Appendix
of \cite{PH1}) is considered. There it is
shown they are derivatives of Green functions and can be separated into 
state-independent divergent part and state-dependent finite contribution:
 $\Gi = \Gidiv + \Gifin$, same as 
the split hereby shown for the Green function.

When analyzing the energy density fluctuations, discarding
the divergent piece is the same as subtracting from the Green function 
its divergent part. If this is done, we also no longer have the cross term, 
just as viewing the problem
from the point separation method outlined above.
We feel this makes it
problematic to analyze the cross term without also including the 
divergent term. At the same time,  regularization of the fluctuations
involving the subtraction of state dependent terms as realized in this
calculation raises new issues on regularization which merits further
investigations. In a recent work Wu and Ford \cite{WuFor2}  showed 
a connection between the cross term and radiation pressure.

%%%%%%%%%%%%%%%%%%%%%%%%%%%%%%%%%%%%%%%%%
%%%%%%%%%%%%%%%%%%%%%%%%%%%%%%%%%%%%%%%%%
\section{NOISE KERNEL, STRESS-ENERGY BI-TENSOR AND POINT SEPARATION}
\label{sec4}
%%%%%%%%%%%%%%%%%%%%%%%%%%%%%%%%%%%%%%%%%
%%%%%%%%%%%%%%%%%%%%%%%%%%%%%%%%%%%%%%%%%

As pointed out by one of us before \cite{stogra}
the stress energy bi-tensor could be the starting point for a new quantum field
theory constructed on spacetimes with extended structures. But for comparison
with ordinary phenomena at low energy we need to find out if it behaves 
normally in the limit of  ordinary (point-defined) quantum field theory. 
The method of  point-separation is best suited for this purpose.
The task is to seek  a regularized  noise-kernel for quantum fields 
in general  curved spacetimes upon taking the coincidence limit.
This was carried out by Phillips and Hu \cite{PH2}. The following is a 
summary of their work.
  
PH began with a discussion of  the procedures 
for dealing with the quantum stress tensor bi-operator at two
separated points and the noise kernel and
end with a general expression for the noise kernel in terms of the quantum 
field's Green function and its covariant derivatives up
to the fourth order . (The stress
tensor involves up to two covariant derivatives.) This result holds for $x\ne
y$ without recourse to renormalization of the Green function, showing that
$N_{abc'd'}(x,y)$ is always finite for $x\ne y$ (and off the light cone for
massless theories). In particular for a massless conformally
coupled free scalar
field on a four dimensional manifold they computed the trace
of the noise kernel
 at both points and found this double trace vanishes identitically. This 
implies that there is no stochastic
correction to the trace anomaly  for massless conformal fields,
in agreement with 
results arrived at in Refs. \cite{CH94,cv96,MV0}.

Now to obtain the point-defined quantities, one needs to deal
with the divergences
in the coincidence limit.
For this PH adopted the ``modified'' point  separation
scheme \cite{Wald75,ALN77,Wald78}
to get a regularized Green function.   In this procedure,
the naive Green function is rendered finite by assuming the divergences
present for $y \rightarrow x$ are state independent and can be removed by
subtraction of a Hadamard form. They showed that the noise kernel  in the  
$y \rightarrow x$ limit is meaningful  for an arbitrary curved spacetime
by explicitly deriving a  general expression for the
noise kernel and its coincident form.

After following these expositions of PH,  we will end our
discussion with a  reflection
on the issues related to regularization we started
addressing in the previous section, 
reiterating the important role the point-separated quantities can play in 
a new approach to a quantum theory of gravity. There is a fundamental
shift of viewpoint in the nature and meaning of
the point separation scheme: in the 70's
it was used as a technique (many practitioners may still
view it as a trick, even a clumpsy one) for the purpose of identifying the
ultraviolet divergences. Now in the new approach 
(as advocated by one of us \cite{stogra,HuPeyr6}) 
we want  to use the point-separated expressions 
to construct a quantum field theory for extended spacetimes. 

%**************************************************************************
%***                    POINT SEPERATION REVIEW                        ****
%**************************************************************************

\subsection{Point Separation}

The point separation scheme introduced
in the 60's by DeWitt  \cite{DeWitt65}  was brought
to more popular use in the 70's  
in the context of quantum field theory in curved
spacetimes \cite{DeWitt75,Christensen76} 
as a means for obtaining a finite quantum stress
tensor\footnote{We know  there are several regularization 
methods developed for the removal of ultraviolet divergences in the stress 
energy tensor of quantum fields  in curved spacetime
\cite{Birrell-Davies82,Fulling89,Wald94}.
Their mutual relations are known, and discrepancies explained. 
This formal structure of regularization schemes for 
quantum fields in curved 
spacetime should remain intact as we apply them to the regularization
of the noise kernel in general curved spacetimes.
Specific considerations will 
of course enter for each method. But for the
methods employed so far, such as  
zeta-function, point separation, dimensional, smeared-field
\cite{Phillips-Hu97,PH2} applied to simple cases 
(Casimir, Einstein, thermal fields)
there is no new  inconsistency or discrepancy.}.
Since the stress-energy
tensor is built from the product of a pair of field operators evaluated at a
single point, it is not well-defined. In this scheme, one introduces an
artificial separation of the single point $x$ to a pair of closely separated
points $x$ and $x'$. The problematic terms involving field products such as 
$\hat\phi(x)^2$ becomes $\hat\phi(x)\hat\phi(x')$, whose expectation value is
well defined. If one is interested in the low energy behavior captured by
the point-defined quantum field theory -- as the effort in the
70's was directed --
one takes the coincidence limit. Once the divergences present are identified, 
they may be  removed (regularization) or moved (by
renormalizing the coupling constants), 
to produce  a well-defined, finite stress tensor at a single point.  

Thus the first order of business is  the construction of the stress tensor and 
then derive the symmetric stress-energy tensor two point function, the
noise kernel, in terms of the Wightman Green function.
In this section we will use
the traditional notation for index tensors in
the point-separation context.

\subsubsection{n-tensors and end-point expansions}

An object like the Green function $G(x,y)$ is an example of a
{\em bi-scalar}: it transforms as scalar at both points $x$ and $y$.
We can also define a {\em bi-tensor}\, 
$T_{a_1\cdots a_n\,b'_1\cdots b'_m}(x,y)$:
upon a coordinate transformation, this transforms as a rank $n$ tensor at 
$x$ and a rank $m$ tensor at $y$.
We will extend this up to a 
{\em quad-tensor}\, 
$T_{a_1\cdots a_{n_1}\,b'_1\cdots b'_{n_2}\,
    c''_1\cdots c''_{n_3}\,d'''_1\cdots d'''_{n_4}}$ 
which has support at
four points $x,y,x',y'$, transforming as rank $n_1,n_2,n_3,n_4$ tensors
at each of the four points. This also sets the notation we will use:
unprimed indices referring to the tangent space constructed
above $x$, single primed indices to
$y$, double primed to $x'$ and triple primed to $y'$.
For each point, there is the covariant derivative $\nabla_a$ at that point.
Covariant derivatives at different points commute and the covariant
derivative at, say, point $x'$, does not act on a bi-tensor defined at,
say,  $x$ and $y$:
\begin{equation}
T_{ab';c;d'} = T_{ab';d';c} \quad {\rm and } \quad
T_{ab';c''} = 0.
\end{equation}
To simplify notation, henceforth we will eliminate the semicolons
after the first one for multiple covariant derivatives at multiple points.

Having objects defined at different points, the {\rm coincident limit} is
defined as evaluation ``on the diagonal'',
in the sense of the spacetime support of the function or tensor,
and the usual shorthand
$\left[ G(x,y) \right] \equiv G(x,x)$ is used. This extends to $n$-tensors
as
\begin{equation}
\left[ T_{a_1\cdots a_{n_1}\,b'_1\cdots b'_{n_2}\,
    c''_1\cdots c''_{n_3}\,d'''_1\cdots d'''_{n_4}} \right] = 
T_{a_1\cdots a_{n_1}\,b_1\cdots b_{n_2}\,
    c_1\cdots c_{n_3}\,d_1\cdots d_{n_4}},
\end{equation}
{\it i.e.}, this becomes a rank $(n_1+n_2+n_3+n_4)$ tensor at $x$.
The multi-variable chain rule relates covariant derivatives acting at
different points, when we are interested in the coincident limit:
\begin{equation}
\left[ T_{a_1\cdots a_m \,b'_1\cdots b'_n} \right]\!{}_{;c} = 
\left[ T_{a_1\cdots a_m \,b'_1\cdots b'_n;c} \right] +
\left[ T_{a_1\cdots a_m \,b'_1\cdots b'_n;c'} \right].
\label{ref-Synge's}
\end{equation}
This result is referred to as {\em Synge's theorem} in this context;
we  follow Fulling's \cite{Fulling89} discussion.

The bi-tensor of {\em parallel transport}\, $g_a{}^{b'}$ is defined such
that when it acts on a vector $v_{b'}$ at $y$, it parallel transports
the vector along the geodesics connecting $x$ and $y$. This allows us to
add vectors and tensors defined at different points. We cannot directly add a
vector $v_a$ at $x$ and vector $w_{a'}$ at $y$. But by using $g_a{}^{b'}$,
we can construct the sum
$v^a + g_a{}^{b'} w_{b'}$.
We will also need the obvious property $\left[ g_a{}^{b'} \right] = g_a{}^b$.

The main bi-scalar we need  is the {\em world function}
$\sigma(x,y)$. This is defined as a half of the square of the geodesic
distance between the points $x$ and $y$.
It satisfies the equation
\begin{equation}
\sigma = \frac{1}{2} \sigma^{;p} \sigma_{;p}
\label{define-sigma}
\end{equation}
Often in the literature, a covariant derivative is implied when the
world function appears with indices: $\sigma^a \equiv \sigma^{;a}$,
{\it i.e.}taking the covariant derivative at $x$, while $\sigma^{a'}$ means
the covariant derivative at $y$.
This is done since the vector $-\sigma^a$ is the tangent vector to the
geodesic with length equal the distance between $x$ and $y$. 
As $\sigma^a$ records information about distance and direction for the two
points  this makes it ideal for constructing a series expansion of
a bi-scalar.  The {\em end point} expansion of a bi-scalar $S(x,y)$ is
of the form
\begin{equation}
S(x,y) = A^{(0)} + \sigma^p A^{(1)}_p
+ \sigma^p \sigma^q A^{(2)}_{pq}
+ \sigma^p \sigma^q  \sigma^r A^{(3)}_{pqr}
+ \sigma^p \sigma^q  \sigma^r \sigma^s A^{(4)}_{pqrs} + \cdots
\label{general-endpt-series}
\end{equation}
where, following our convention, the expansion tensors 
$A^{(n)}_{a_1\cdots a_n}$ with  unprimed indices have support at $x$
and hence the name end point expansion. Only the
symmetric part of these tensors  contribute to the expansion.
For the purposes of multiplying series expansions it is convenient to
separate the distance dependence from the direction dependence. This
is done by introducing the unit vector $p^a = \sigma^a/\sqrt{2\sigma}$.
Then the series expansion can be written
\begin{equation}
S(x,y) = A^{(0)} + \sigma^{\frac{1}{2}} A^{(1)}
+ \sigma  A^{(2)}
+ \sigma^{\frac{3}{2}} A^{(3)}
+ \sigma^2 A^{(4)} + \cdots
\end{equation}
The expansion scalars are related to the expansion tensors via
$A^{(n)} = 2^{n/2} A^{(n)}_{p_1\cdots p_n}
p^{p_1}\cdots p^{p_n}$.

The last object we need  is the {\em VanVleck-Morette}
determinant $D(x,y)$, defined as
$D(x,y) \equiv -\det\left( -\sigma_{;ab'} \right)$.
The related bi-scalar
\begin{equation}
{\Delta\!^{1/2}} = \left( \frac{D(x,y)}{\sqrt{g(x) g(y)}}\right)^\frac{1}{2}
\end{equation}
satisfies the equation
\begin{equation}
{\Delta\!^{1/2}}\left(4-\sigma_{;p}{}^p\right) -
2{\Delta\!^{1/2}}_{\,\,;p}\sigma^{;p} = 0
\label{define-VanD}
\end{equation}
with the boundary condition $\left[{\Delta\!^{1/2}}\right] = 1$.

Further details on these objects and discussions of the definitions and
properties are contained in \cite{Christensen76} and \cite{NPsc}. 
There it is shown how the defining equations for $\sigma$
and ${\Delta\!^{1/2}}$
are used to determine the coincident limit expression for
the various covariant
derivatives of the world function
($\left[ \sigma_{;a}\right]$, $\left[ \sigma_{;ab}\right]$, {\it etc.})
and how the defining differential equation for ${\Delta\!^{1/2}}$ can be used
to determine the series expansion of ${\Delta\!^{1/2}}$.
We show how the expansion tensors $A^{(n)}_{a_1\cdots a_n}$
are determined in terms of the coincident limits of covariant
derivatives of the bi-scalar $S(x,y)$.  (Ref.  \cite{NPsc} details how
point separation can be implemented on the computer to provide
easy access to a wider range of applications involving higher
derivatives of the curvature tensors. )

\subsection{Stress Energy Bi-Tensor Operator and Noise Kernel}

Even though we believe that the point-separated results are more basic in the
sense that it reflects a deeper structure of the quantum theory of spacetime,
we will nevertheless start with  quantities defined at one point because they
are more familiar in conventional quantum field theory. We will use point
separation to introduce the biquantities. The key issue here is thus the
distinction between point-defined ({\it pt}) and point-separated ({\it bi}) 
quantities.

For  a free classical scalar field $\phi$ with the action $S_m[g,\phi]$
defined in Eq. (\ref{2.1}), the classical stress-energy tensor is
\begin{eqnarray}
T_{ab}
&=&
\left( 1 - 2\,\xi  \right) \,{\phi {}_;{}_{a}}\,{\phi {}_;{}_{b}}
 + \left(2\,\xi -{1\over 2} \right) \,{\phi {}_;{}_{p}}
\,{\phi {}^;{}^{p}}\,{g{}_{a}{}_{b}}
+ 2\xi\,\phi \, \,\left({\phi {}_;{}_{p}{}^{p} -{\phi {}_;{}_{a}{}_{b}}}
\,{g{}_{a}{}_{b}} \right)  \cr
&&+ {{\phi }^2}\,\xi \,
\left({R{}_{a}{}_{b}} - {1\over 2}{ R\,{g{}_{a}{}_{b}}  }
 \right) 
 - \frac{1}{2}{{m^2}\,{{\phi }^2}\,{g{}_{a}{}_{b}}},
\label{ref-define-classical-emt}
\end{eqnarray}
which is equivalent to the tensor of Eq. (\ref{2.3}) but written
in a slightly different form for convenience. 
When we make the transition to quantum field theory,
we promote the field $\phi(x)$ to a field operator $\hat\phi(x)$.
The fundamental problem of defining a quantum operator for the 
stress tensor is immediately visible: the field operator 
appears quadratically. Since $\hat\phi(x)$ is an operator-valued
distribution, products at a single point are not well-defined.
But if the product is point separated ($\hat\phi^2(x) \rightarrow
\hat\phi(x)\hat\phi(x')$), they are finite and well-defined.

Let us first seek a point-separated extension of these classical quantities and
then consider the quantum field operators. Point separation is symmetrically
extended to products of covariant derivatives of the field according to
\begin{eqnarray}
\left({\phi {}_;{}_{a}}\right)\left({\phi {}_;{}_{b}}\right) &\rightarrow&
\frac{1}{2}\left(
g_a{}^{p'}\nabla_{p'}\nabla_{b}+g_b{}^{p'}\nabla_a\nabla_{p'}
\right)\phi(x)\phi(x'), \\
\phi \,\left({\phi {}_;{}_{a}{}_{b}}\right) &\rightarrow&
\frac{1}{2}\left(
\nabla_a\nabla_b+g_a{}^{p'}g_b{}^{q'}\nabla_{p'}\nabla_{q'}
\right)\phi(x)\phi(x').
\end{eqnarray}
The bi-vector of parallel displacement
$g_a{}^{a'}(x,x')$ is included so that we may 
have objects that are rank 2 tensors
at $x$ and scalars at $x'$.

To carry out  point separation on (\ref{ref-define-classical-emt}), we first
define the differential operator
\begin{eqnarray}
{\cal T}_{ab} &=&
  \frac{1}{2}\left(1-2\xi\right)
   \left(g_a{}^{a'}\nabla_{a'}\nabla_{b}+g_b{}^{b'}\nabla_a\nabla_{b'}\right)
+ \left(2\xi-\frac{1}{2}\right)
     g_{ab}g^{cd'}\nabla_c\nabla_{d'} \cr
&&
- \xi
     \left(\nabla_a\nabla_b+g_a{}^{a'}g_b{}^{b'}\nabla_{a'}\nabla_{b'}\right)
+ \xi g_{ab}
     \left(\nabla_c\nabla^c+\nabla_{c'}\nabla^{c'}\right) \cr
&&
+\xi\left(R_{ab} - \frac{1}{2}g_{ab}R\right)
    -\frac{1}{2}m^2 g_{ab}
\label{PSNoise-emt-diffop}
\end{eqnarray}
from which we obtain the classical stress tensor as
\begin{equation}
T_{ab}(x) = \lim_{x' \rightarrow x} {\cal T}_{ab}\phi(x)\phi(x').
\end{equation}
That the classical tensor field no longer appears as a product of scalar 
fields at a single point allows a smooth transition
to the quantum tensor field. From the viewpoint of the stress tensor, 
the separation of points is an artificial construct so when promoting the 
classical field to a
quantum one, neither point should be favored. The product of field
configurations is taken to be the symmetrized operator product,
denoted by curly brackets:
\begin{equation}
\phi(x)\phi(y) \rightarrow \frac{1}{2}
 \left\{{\hat\phi(x)},{\hat\phi(y)}\right\}
= \frac{1}{2}\left( {\hat\phi(x)} {\hat\phi(y)} +
                    {\hat\phi(y)} {\hat\phi(x)}
\right)
\end{equation}
With this, the point separated stress energy tensor operator is defined as
\begin{equation}
\hat T_{ab}(x,x') \equiv \frac{1}{2}
{\cal T}_{ab}\left\{\hat\phi(x),\hat\phi(x')\right\}.
\label{PSNoise-emt-define}
\end{equation}
While the classical stress tensor was defined at the coincidence limit
$x'\rightarrow x$, we cannot attach any physical meaning to the quantum stress 
tensor at one point
until the issue of regularization is dealt with, which will happen in the next 
section.
For now,  we will maintain point separation so as to have a mathematically 
meaningful operator. 

The expectation value of the point-separated stress tensor can now be
taken. This amounts to replacing the field operators by their expectation
value, which is given by the Hadamard (or Schwinger) function
\begin{equation}
{G^{(1)}}(x,x') =
     \langle\left\{{\hat\phi(x)},{\hat\phi(x')}\right\}\rangle.
\end{equation}
and the point-separated stress tensor is defined as
\begin{equation}
\langle \hat T_{ab}(x,x') \rangle = 
\frac{1}{2} {\cal T}_{ab}{G^{(1)}}(x,x')
\label{ref-emt-PSdefine}
\end{equation}
where, since ${\cal T}_{ab}$ is a differential operator, it can be taken
``outside'' the expectation value. The expectation value of the point-separated
quantum stress tensor for a free, massless ($m=0$)  conformally
coupled ($\xi=1/6$) scalar field on a four dimension spacetime with 
scalar curvature $R$ is
\begin{eqnarray}
\langle \hat T_{ab}(x,x') \rangle &=&
  \frac{1}{6}\left( {g{}^{p'}{}_{b}}\,{{G^{(1)}}{}_;{}_{p'}{}_{a}}
 + {g{}^{p'}{}_{a}}\,{{G^{(1)}}{}_;{}_{p'}{}_{b}} \right)
 -\frac{1}{12} {g{}^{p'}{}_{q}}\,{{G^{(1)}}{}_;{}_{p'}{}^{q}}
\,{g{}_{a}{}_{b}} \cr 
&& -\frac{1}{12}\left( {g{}^{p'}{}_{a}}\,{g{}^{q'}{}_{b}}
\,{{G^{(1)}}{}_;{}_{p'}{}_{q'}} + {{G^{(1)}}{}_;{}_{a}{}_{b}} \right)
 +\frac{1}{12}\left( \left( {{G^{(1)}}{}_;{}_{p'}{}^{p'}}
 + {{G^{(1)}}{}_;{}_{p}{}^{p}} \right) \,{g{}_{a}{}_{b}} \right) \cr 
&& +\frac{1}{12} {G^{(1)}}\,
\left({R{}_{a}{}_{b}} -{1\over 2} R\,{g{}_{a}{}_{b}} \right)
\end{eqnarray}

%**************************************************************************
%***                    NOISE KERNEL                                   ****
%**************************************************************************
%%%%%%%%%%%%%%%%%%%%%
\subsubsection{Finiteness of Noise Kernel}
%%%%%%%%%%%%%%%%%%%%%%

We now turn our attention to the noise kernel introduced in Eq.
(\ref{2.8}), which  is the symmetrized product of the (mean
subtracted) stress tensor operator:
\begin{eqnarray}
8 N_{ab,c'd'}(x,y) &=&
\langle \left\{ 
		\hat T_{ab}(x)-\langle \hat T_{ab}(x)\rangle,
	\hat T_{c'd'}(y)-\langle \hat T_{c'd'}(y) \rangle
\right\} \rangle \cr
&=&
\langle \left\{ \hat T_{ab}(x),\hat T_{c'd'}(y) \right\} \rangle
-2 \langle \hat T_{ab}(x)\rangle\langle \hat T_{c'd'}(y) \rangle
\end{eqnarray}
Since $\hat T_{ab}(x)$ defined at one point can be ill-behaved as it is 
generally divergent,
one can question the soundness of these quantities.
But as will be shown later, 
the noise kernel
is finite for $y\neq x$. All field operator
products present in the first expectation value that could be
divergent are canceled by similar products in the second term.
We will replace each of the stress tensor operators in
the above expression for 
the noise kernel
by their point separated versions,
effectively separating the two points $(x,y)$ into the four points
$(x,x',y,y')$. This will allow us to express the noise kernel in terms
of a pair of differential operators acting on a combination of
four and two point functions. Wick's theorem will allow the four
point functions to be re-expressed in terms of two point functions. From this
we see that all possible divergences for $y\neq x$ will cancel.
When the coincidence limit is taken divergences do occur. The above procedure
will allow us to isolate the divergences and obtain a finite result.

Taking the point-separated quantities as more basic,
one should replace each of 
the stress tensor operators in the above with the
corresponding point separated 
version (\ref{PSNoise-emt-define}),
with ${\cal T}_{ab}$ acting at $x$ and $x'$ and ${\cal T}_{c'd'}$ acting
at $y$ and $y'$. In this framework the noise kernel is defined as
\begin{equation}
8 N_{ab,c'd'}(x,y) =
   \lim_{x'\rightarrow x}\lim_{y'\rightarrow y}
   {\cal T}_{ab} {\cal T}_{c'd'}\, G(x,x',y,y')
\end{equation}
where the four point function is
\begin{eqnarray}
G(x,x',y,y') &=& \frac{1}{4}\left[
\langle\left\{\left\{{\hat\phi(x)},{\hat\phi(x')}\right\},\left\{{\hat\phi(y)}
,{\hat\phi(y')}\right\}\right\}\rangle
\right. \cr\cr&&\hspace{1cm}\left.
  -2\,\langle\left\{{\hat\phi(x)},{\hat\phi(x')}\right\}\rangle
  \langle\left\{{\hat\phi(y)},{\hat\phi(y')}\right\}\rangle \right].
\label{PSNoise-G4a}
\end{eqnarray}
We assume the pairs $(x,x')$ and $(y,y')$ are each 
within their respective Riemann normal coordinate
neighborhoods so as to avoid problems
that possible geodesic caustics might be present. When we later
turn our attention to computing the limit $y\rightarrow x$, after
issues of regularization are addressed, we will want to assume all four
points are within the same Riemann normal coordinate neighborhood.

Wick's theorem, for the case of free fields which we
are considering, gives the simple product four point function in terms
of a sum of products of Wightman functions (we use the shorthand notation
$G_{xy}\equiv G_{+}(x,y) = \langle{\hat\phi(x)}\,{\hat\phi(y)}\rangle$):
\begin{equation}
\langle{\hat\phi(x)}\,{\hat\phi(y)}\,{\hat\phi(x')}\,{\hat\phi(y')}\rangle =
{G_{xy'}}\,{G_{yx'}} + {G_{xx'}}\,{G_{yy'}} + {G_{xy}}\,{G_{x'y'}}
\end{equation}
Expanding out the anti-commutators in (\ref{PSNoise-G4a}) and applying
Wick's theorem, the four point function becomes
\begin{equation}
G(x,x',y,y')  =
{G_{xy'}}\,{G_{x'y}} + {G_{xy}}\,{G_{x'y'}} + {G_{yx'}}\,{G_{y'x}} + {G_{yx}}
\,{G_{y'x'}}
\end{equation}
We can now easily see that the noise kernel defined via this
function is indeed well defined for the limit $(x',y')\rightarrow (x,y)$:
\begin{equation}
G(x,x,y,y) = 2\,\left( {{{G^2_{xy}}}} + {{{G^2_{yx}}}} \right) .
\end{equation}
{}From this we can see that the noise kernel is
also well defined for $y \neq x$;
any divergence present in the first expectation value of
(\ref{PSNoise-G4a}) have been cancelled by those present in the
pair of Green functions in the second term, in agreement
with the results of section \ref{sec2}.

%%%%%%%%%%%%%%%%%
\subsubsection{Explicit Form of the Noise Kernel}
%%%%%%%%%%%%%%%%%%

We will let the points  separated for a while so we can keep track of which
covariant derivative acts on which arguments of which Wightman function. 
As an example
(the complete calculation is quite long), consider the
result of the first set of covariant derivative operators in the
differential operator (\ref{PSNoise-emt-diffop}), from both
${\cal T}_{ab}$ and ${\cal T}_{c'd'}$, acting on $G(x,x',y,y')$:
\begin{eqnarray}
&&\frac{1}{4}\left(1-2\xi\right)^2
   \left(g_a{}^{p''}\nabla_{p''}\nabla_{b}+
         g_b{}^{p''}\nabla_{p''}\nabla_{a}\right)\cr
&&\hspace{17mm}\times
   \left(g_{c'}{}^{q'''}\nabla_{q'''}\nabla_{d'}
        +g_{d'}{}^{q'''}\nabla_{q'''}\nabla_{c'}\right)
    G(x,x',y,y')
\end{eqnarray}
(Our notation is that $\nabla_a$ acts at $x$, $\nabla_{c'}$ at $y$,
$\nabla_{b''}$ at $x'$ and $\nabla_{d'''}$ at $y'$).
%{\it Null}
Expanding out the differential operator above, we can determine which
derivatives act on which Wightman function:
\begin{eqnarray}
{{{{\left( 1 - 2\,\xi  \right) }^2}}\over 4} &\times & \left[
    {g{}_{c'}{}^{p'''}}\,{g{}^{q''}{}_{a}}
 \left( {{G_{xy'}}{}_;{}_{b}{}_{p'''}}\,{{G_{x'y}}{}_;{}_{q''}{}_{d'}}
 + {{G_{xy}}{}_;{}_{b}{}_{d'}}\,{{G_{x'y'}}{}_;{}_{q''}{}_{p'''}}
 \right.\right. \cr
&&  \hspace{20mm} + \left. {{G_{yx'}}{}_;{}_{q''}{}_{d'}}
\,{{G_{y'x}}{}_;{}_{b}{}_{p'''}} + {{G_{yx}}{}_;{}_{b}{}_{d'}}
\,{{G_{y'x'}}{}_;{}_{q''}{}_{p'''}} \right) \cr
&&+ {g{}_{d'}{}^{p'''}}\,{g{}^{q''}{}_{a}}
 \left( {{G_{xy'}}{}_;{}_{b}{}_{p'''}}\,{{G_{x'y}}{}_;{}_{q''}{}_{c'}}
 + {{G_{xy}}{}_;{}_{b}{}_{c'}}\,{{G_{x'y'}}{}_;{}_{q''}{}_{p'''}} \right. \cr
&&  \hspace{20mm} + \left. {{G_{yx'}}{}_;{}_{q''}{}_{c'}}
\,{{G_{y'x}}{}_;{}_{b}{}_{p'''}} + {{G_{yx}}{}_;{}_{b}{}_{c'}}
\,{{G_{y'x'}}{}_;{}_{q''}{}_{p'''}} \right) \cr
&&+ {g{}_{c'}{}^{p'''}}\,{g{}^{q''}{}_{b}}
 \left( {{G_{xy'}}{}_;{}_{a}{}_{p'''}}\,{{G_{x'y}}{}_;{}_{q''}{}_{d'}}
 + {{G_{xy}}{}_;{}_{a}{}_{d'}}\,{{G_{x'y'}}{}_;{}_{q''}{}_{p'''}} \right. \cr
&&  \hspace{20mm} + \left. {{G_{yx'}}{}_;{}_{q''}{}_{d'}}
\,{{G_{y'x}}{}_;{}_{a}{}_{p'''}} + {{G_{yx}}{}_;{}_{a}{}_{d'}}
\,{{G_{y'x'}}{}_;{}_{q''}{}_{p'''}} \right) \cr
&&+ {g{}_{d'}{}^{p'''}}\,{g{}^{q''}{}_{b}}
 \left( {{G_{xy'}}{}_;{}_{a}{}_{p'''}}\,{{G_{x'y}}{}_;{}_{q''}{}_{c'}}
 + {{G_{xy}}{}_;{}_{a}{}_{c'}}\,{{G_{x'y'}}{}_;{}_{q''}{}_{p'''}} \right. \cr
&&  \hspace{20mm} + \left.\left. {{G_{yx'}}{}_;{}_{q''}{}_{c'}}
\,{{G_{y'x}}{}_;{}_{a}{}_{p'''}} + {{G_{yx}}{}_;{}_{a}{}_{c'}}
\,{{G_{y'x'}}{}_;{}_{q''}{}_{p'''}} \right) \right]
\end{eqnarray}
If we now  let $x'\rightarrow x$ and $y' \rightarrow y$
the contribution to the noise kernel is (including the factor of
$\frac{1}{8}$ present in the definition of the noise kernel):
\begin{eqnarray}
&&\frac{1}{8}\left\{ {{\left( 1 - 2\,\xi  \right) }^2}
\,\left( {{G_{xy}}{}_;{}_{a}{}_{d'}}\,{{G_{xy}}{}_;{}_{b}{}_{c'}}
 + {{G_{xy}}{}_;{}_{a}{}_{c'}}\,{{G_{xy}}{}_;{}_{b}{}_{d'}}
 \right)  \right. \cr
&&\hspace{20mm} \left. + {{\left( 1 - 2\,\xi  \right) }^2}
\,\left( {{G_{yx}}{}_;{}_{a}{}_{d'}}\,{{G_{yx}}{}_;{}_{b}{}_{c'}}
 + {{G_{yx}}{}_;{}_{a}{}_{c'}}\,{{G_{yx}}{}_;{}_{b}{}_{d'}} \right)  \right\}
\end{eqnarray}
That this term can be written as the sum of a part involving $G_{xy}$ and
one involving $G_{yx}$ is a general property of the
entire noise kernel. It thus
takes the form
\begin{equation}
N_{abc'd'}(x,y) = N_{abc'd'}\left[ G_{+}(x,y)\right]
                + N_{abc'd'}\left[ G_{+}(y,x)\right].
\end{equation}
We will present the form of the functional $N_{abc'd'}\left[ G \right]$
shortly. First we note, for $x$ and $y$ time-like separated, the above
split of the noise kernel allows us to express it
in terms of the Feynman (time ordered) Green function $G_F(x,y)$ and
the Dyson (anti-time ordered) Green function $G_D(x,y)$:
\begin{equation}
N_{abc'd'}(x,y) = N_{abc'd'}\left[ G_F(x,y)\right]
                + N_{abc'd'}\left[ G_D(x,y)\right]
\label{noiker}
\end{equation}
\footnote{This can be connected  with the zeta function approach to this 
problem \cite{Phillips-Hu97} as follows:
Recall when the quantum stress tensor
fluctuations determined in the Euclidean section is
analytically continued back
to Lorentzian signature ($\tau \rightarrow i t$), the time ordered product
results. On the other hand, if the continuation is
$\tau \rightarrow -i t$, the
anti-time ordered product results. With this in mind,
the noise kernel is seen
to be related to the quantum stress tensor fluctuations derived via the
effective action as
\begin{equation}
16 N_{abc'd'} =
   \left.\Delta T^2_{abc'd'}\right|_{t=-i\tau,t'=-i\tau'}
 + \left.\Delta T^2_{abc'd'}\right|_{t= i\tau,t'= i\tau'}.
\end{equation}
 }
The complete form of the functional $N_{abc'd'}\left[ G \right]$ is
%*************************************************************************
%*************************************************************************
%
%    NOISE KERNEL FUNCTIONAL FORM, GENERAL EXPRESSION
%
%*************************************************************************
%*************************************************************************
%
\begin{mathletters}
\label{ref-noise-kernel}
\begin{equation}
 N_{abc'd'}\left[ G \right]  = 
    \tilde N_{abc'd'}\left[ G \right]
  + g_{ab}   \tilde N_{c'd'}\left[ G \right]
 + g_{c'd'} \tilde N'_{ab}\left[ G \right]
 + g_{ab}g_{c'd'} \tilde N\left[ G \right]
\label{general-noise-kernel}
\end{equation}
with
\begin{eqnarray}
%=========================================================================
%          Display N_{abcd} terms,     GENERAL EXPRESSION
%=========================================================================
8 \tilde N_{abc'd'} \left[ G \right]&=& 
%
%     pf = 1 of 9
{{\left(1-2\,\xi\right)}^2}\,\left( G{}\!\,_{;}{}_{c'}{}_{b}\,
     G{}\!\,_{;}{}_{d'}{}_{a} + 
    G{}\!\,_{;}{}_{c'}{}_{a}\,G{}\!\,_{;}{}_{d'}{}_{b} \right) 
%
%     pf = 5 of 9
+ 4\,{{\xi}^2}\,\left( G{}\!\,_{;}{}_{c'}{}_{d'}\,G{}\!\,_{;}{}_{a}{}_{b} + 
    G\,G{}\!\,_{;}{}_{a}{}_{b}{}_{c'}{}_{d'} \right)  \cr
%
%     pf = 2,4 of 9
&& -2\,\xi\,\left(1-2\,\xi\right)\,
  \left( G{}\!\,_{;}{}_{b}\,G{}\!\,_{;}{}_{c'}{}_{a}{}_{d'} + 
    G{}\!\,_{;}{}_{a}\,G{}\!\,_{;}{}_{c'}{}_{b}{}_{d'} + 
    G{}\!\,_{;}{}_{d'}\,G{}\!\,_{;}{}_{a}{}_{b}{}_{c'} + 
    G{}\!\,_{;}{}_{c'}\,G{}\!\,_{;}{}_{a}{}_{b}{}_{d'} \right)  \cr
%
%     pf = 3,7 of 9
&& + 2\,\xi\,\left(1-2\,\xi\right)\,\left( 
G{}\!\,_{;}{}_{a}\,G{}\!\,_{;}{}_{b}\,
     {R{}_{c'}{}_{d'}} + G{}\!\,_{;}{}_{c'}\,G{}\!\,_{;}{}_{d'}\,
     {R{}_{a}{}_{b}} \right)  \cr
%
%     pf = 6,8 of 9
&&  -4\,{{\xi}^2}\,\left( G{}\!\,_{;}{}_{a}{}_{b}\,{R{}_{c'}{}_{d'}} + 
    G{}\!\,_{;}{}_{c'}{}_{d'}\,{R{}_{a}{}_{b}} \right)  G
%
%     pf = 9 of 9
 +  2\,{{\xi}^2}\,{R{}_{c'}{}_{d'}}\,{R{}_{a}{}_{b}} {G^2}
\end{eqnarray}
\begin{eqnarray}
%=========================================================================
%          Display N_{ab} terms       GENERAL EXPRESSION
%=========================================================================
8 \tilde N'_{ab} \left[ G \right]&=& 
%
%    pf = 1,2 of 9
2\,\left(1-2\,\xi\right) \left(
   \left(2\,\xi-{\frac{1}{2}}\right)\,G{}\!\,_{;}{}_{p'}{}_{b}\,
  G{}\!\,_{;}{}^{p'}{}_{a}
 + \xi\,\left( G{}\!\,_{;}{}_{b}\,G{}\!\,_{;}{}_{p'}{}_{a}{}^{p'} + 
    G{}\!\,_{;}{}_{a}\,G{}\!\,_{;}{}_{p'}{}_{b}{}^{p'} \right)  
\right)\cr &&
%
%    pf = 4,5
-4\,\xi \left(
    \left(2\,\xi-{\frac{1}{2}}\right)\,G{}\!\,_{;}{}^{p'}\,
  G{}\!\,_{;}{}_{a}{}_{b}{}_{p'}
  + \xi\,\left( G{}\!\,_{;}{}_{p'}{}^{p'}\,G{}\!\,_{;}{}_{a}{}_{b} + 
    G\,G{}\!\,_{;}{}_{a}{}_{b}{}_{p'}{}^{p'} \right) 
\right) \cr
%
%    pf = 3,6
&& - \left({m^2}+\xi R'\right)\,\left( 
\left(1-2\,\xi\right)\,G{}\!\,_{;}{}_{a}\,
     G{}\!\,_{;}{}_{b} - 2\,G\,\xi\,G{}\!\,_{;}{}_{a}{}_{b} \right)  \cr
%
%    pf = 7,8
&& + 2\,\xi\,\left( \left(2\,\xi-{\frac{1}{2}}\right)\,G{}\!\,_{;}{}_{p'}\,
     G{}\!\,_{;}{}^{p'} + 2\,G\,\xi\,G{}\!\,_{;}{}_{p'}{}^{p'} \right) \,
  {R{}_{a}{}_{b}} \cr
%
%    pf = 9
&& - \left({m^2}+\xi R'\right)\,\xi\,{R{}_{a}{}_{b}} {G^2}
\end{eqnarray}
\begin{eqnarray}
%=========================================================================
%          Display N terms           GENERAL EXPRESSION
%=========================================================================
8 \tilde N \left[ G \right]&=& 
2\,{{\left(2\,\xi-{\frac{1}{2}}\right)}^2}\,G{}\!\,_{;}{}_{p'}{}_{q}\,
  G{}\!\,_{;}{}^{p'}{}^{q}
+ 4\,{{\xi}^2}\,\left( G{}\!\,_{;}{}_{p'}{}^{p'}\,G{}\!\,_{;}{}_{q}{}^{q} + 
    G\,G{}\!\,_{;}{}_{p}{}^{p}{}_{q'}{}^{q'} \right)  \cr
&& + 4\,\xi\,\left(2\,\xi-{\frac{1}{2}}\right)\,
  \left( G{}\!\,_{;}{}_{p}\,G{}\!\,_{;}{}_{q'}{}^{p}{}^{q'} + 
    G{}\!\,_{;}{}^{p'}\,G{}\!\,_{;}{}_{q}{}^{q}{}_{p'} \right)  \cr
&& - \left(2\,\xi-{\frac{1}{2}}\right)\,
  \left( \left({m^2}+\xi R\right)\,G{}\!\,_{;}{}_{p'}\,G{}\!\,_{;}{}^{p'} + 
    \left({m^2}+\xi R'\right)\,G{}\!\,_{;}{}_{p}\,G{}\!\,^{;}{}^{p} \right)  \cr
&& - 2\,\xi\,\left( \left({m^2}+\xi R\right)\,G{}\!\,_{;}{}_{p'}{}^{p'} + 
    \left({m^2}+\xi R'\right)\,G{}\!\,_{;}{}_{p}{}^{p} \right)  G \cr
&& {\frac{1}{2}} \left({m^2}+\xi R\right)\,\left({m^2}+\xi R'\right) {G^2}
\end{eqnarray}
\end{mathletters}

%**************************************************************************
%***                    TRACE OF NOISE KERNEL                          ****
%**************************************************************************
\subsubsection{Trace of the Noise Kernel}

One of the most interesting and surprising results to come  out of the
investigations undertaken in the 1970's of the quantum stress
tensor was the discovery of the trace anomaly\cite{Duff74}. When the trace of
the stress tensor $T=g^{ab}T_{ab}$ is evaluated for a field
configuration that satisties the field equation (\ref{2.2})
the trace is seen to vanish for  massless conformally coupled fields.
When this analysis is carried over to the renormalized expectation
value of the quantum stress tensor, the trace no longer vanishes.
Wald \cite{Wald78}
showed this was due to the failure of the renormalized Hadamard function
$G_{\rm ren}(x,x')$ to be symmetric in $x$ and $x'$, implying it
does not necessarily
satisfy the field equation (\ref{2.2}) in the
variable $x'$. The definition of  $G_{\rm ren}(x,x')$ in the context of 
point separation will come next.)

With this in  mind, we can now determine the noise associated with the
trace. Taking the trace at both points $x$ and $y$ of the noise kernel
functional (\ref{noiker}):
\begin{eqnarray}
N\left[ G \right] &=& g^{ab}\,g^{c'd'}\, N_{abc'd'}\left[ G \right] \cr
&=&
- 3\,G\,\xi
    \left\{
        \left({m^2} + {1\over 2} \xi R \right) \,{G{}_;{}_{p'}{}^{p'}} 
      + \left({m^2} + {1\over 2} \xi R'\right) \,{G{}_;{}_{p}{}^{p}} 
    \right\} \cr
&&
   + {{9\,{{\xi }^2}}\over 2} 
       \left\{
           {G{}_;{}_{p'}{}^{p'}}\,{G{}_;{}_{p}{}^{p}}
          + G\,{G{}_;{}_{p}{}^{p}{}_{p'}{}^{p'}} 
       \right\}
    +\left({m^2} + {1\over 2} \xi R \right)  \,
     \left({m^2} + {1\over 2} \xi R'\right)   G^2
\cr
&&+ 3 \left( {1\over 6} - \xi  \right) 
      \left\{
        +3 {{\left( {1\over 6} - \xi  \right) }} 
              {G{}_;{}_{p'}{}_{p}}\,{G{}_;{}^{p'}{}^{p}}  
        -3\xi
           \left(
                {G{}_;{}_{p}}\,{G{}_;{}_{p'}{}^{p}{}^{p'}}
             +  {G{}_;{}_{p'}}\,{G{}_;{}_{p}{}^{p}{}^{p'}} 
           \right)
      \right.\cr
&&\left.\hspace{25mm}
          \left({m^2} + {1\over 2} \xi R \right) 
\,{G{}_;{}_{p'}}\,{G{}_;{}^{p'}}
        + \left({m^2} + {1\over 2} \xi R'\right) \,{G{}_;{}_{p}}\,{G{}^;{}^{p}}
\right\}
\end{eqnarray}
For the massless conformal case, this reduces to
\begin{equation}
N\left[ G \right] = \frac{1}{144}\left\{
R R' G^2 - 6G\left(R \Box' + R' \Box\right) G
  + 18\left( \left(\Box G\right)\left(\Box' G\right)+ \Box' \Box  G\right)
\right\}
\end{equation}
which holds for any function $G(x,y)$. For  $G$ being the Green function, 
it satisfies the field equation (\ref{2.2}):
\begin{equation}
\Box G = (m^2 + \xi R) G
\end{equation}
We will only assume the Green function satisfies the field equation
in its first variable. Using the fact $\Box' R=0$ (because the covariant 
derivatives
act at a different point than at which $R$ is supported), it follows that
\begin{equation}
\Box' \Box  G = (m^2 + \xi R)\Box' G.
\end{equation}
With these results, the noise kernel trace becomes
\begin{eqnarray}
N\left[ G \right] &=& \frac{1}{2} 
\left(
      {m^2}\,\left( 1 - 3\,\xi  \right)
    + 3\,R\,\left( {1\over 6} - \xi  \right) \,\xi  
\right) \cr
&&\hspace{30mm}\times
  \left\{
       {G^2}\,\left( 2\,{m^2} + {R'} \,\xi  \right)
            + \left( 1 - 6\,\xi  \right) \,{G{}_;{}_{p'}}\,{G{}_;{}^{p'}}
            - 6\,G\,\xi \,{G{}_;{}_{p'}{}^{p'}} 
  \right\} \cr
&&+ \frac{1}{2} \left( {1\over 6} - \xi  \right) 
\left\{
     3\,\left( 2\,{m^2} + {R'}\,\xi  \right) \,{G{}_;{}_{p}}\,{G{}^;{}^{p}}
  - 18\,\xi \,{G{}_;{}_{p}}\,{G{}_;{}_{p'}{}^{p}{}^{p'}} 
\right.\cr&&\hspace{30mm}\left.
  + 18\,\left( {1\over 6} - \xi  \right) \,
         {G{}_;{}_{p'}{}_{p}}\,{G{}_;{}^{p'}{}^{p}}  
\right\}
\end{eqnarray}
which vanishes for the massless conformal case.  We have thus shown,
based solely on the definition of the point separated noise kernel, there
is no noise associated with the trace anomaly. 
This result obtained in Ref. \cite{PH2} is completely general since
it is assumed that the Green function 
is only satisfying the field equations in its first variable;
an alternative proof of this result was given in Ref.
\cite{MV0}. This condition 
holds not just for the classical field case, but also for the regularized
quantum case, where one does not expect the Green function to satisfy the
field equation in both variables. One can see this result from the simple 
observation used in section \ref{sec2}:
since the trace anomaly is known to be locally determined
and quantum state independent, whereas 
the noise present in the quantum field 
is non-local, it is hard to find a noise associated with it. 
This general result is in agreement with previous
findings \cite{CH94,HuSin,cv96},
derived from the Feynman-Vernon influence
functional formalism \cite{feynman-vernon,feynman-hibbs}
for some particular cases.

%%%%%%%%%%%%%%%%%%%%%%%%%%%%%%%%%%%

\subsection{Regularization of the Noise Kernel}

We pointed out earlier that field quantities defined at two
separated points  possess important information which could be the
starting point for probes into possible extended structures of spacetime.
Moving in the other (homeward) direction, it is of interest to see how  
fluctuations of energy momentum (loosely, noise) would enter in the 
ordinary (point-wise) quantum field theory in helping
us to address a new set of issues  such as 
a) whether the fluctuations to mean 
ratio can act as a criterion for the validity of semiclassical gravity.
b) Whether the fluctuations in the vacuum energy density which drives  
inflationary cosmology violates the positive energy condition, 
c) How can we derive  structure formation from 
quantum fluctuations, or 
d) General relativity as a low energy effective theory in the 
geometro- hydrodynamic limit \cite{grhydro,stogra}. 

For these inquires we need to construct regularization procedures
to remove the ultraviolet divergences in the coincidence limit. 
The goal is to  obtain a finite expression for the noise kernel in this limit.

We can see from the point separated form of the stress tensor
(\ref{ref-emt-PSdefine}) what we need to regularize is the
Green function $G^{(1)}(x,x')$. Once the Green function has been
regularized such that it is smooth and has a well defined
$x'\rightarrow x$ limit, the stress tensor will be well defined.
In Minkowski space, this issue is easily resolved by a ''normal
ordering'' prescription, which hinges on the existence of a unique
vacuum.
Unfortunately, for a general curved spacetime, there is no unique
vacuum, and hence, no unique mode expansion on which to build a 
normal ordering prescription. But we can still ask if there is a
way to determine a contribution we can subtract to yield a unique
quantum stress tensor.
Here we follow the prescription of Wald \cite{Wald75},
and Adler {\it et. al.} \cite{ALN77} (with corrections \cite{Wald78})
summarized in \cite{Wald94}. We give a short synopsis below
as it will be referred to in subsequent discussions.

The idea builds on the
fact that for %two states $\omega_1$ and $\omega_2$ and letting
$G(x,x')_\omega = \langle \omega |\hat\phi(x)\hat\phi(x')|\omega \rangle$,
the function
\begin{equation}
F(x,x') = G(x,x')_{\omega_1} - G(x,x')_{\omega_2}
\end{equation}
is a smooth function of $x$ and $x'$,
where $\omega_1$ and $\omega_2$ denote two different states.
 This means the difference
between the stress tensor for two states is well defined for the
point separation scheme, {\it i.e.},
\begin{equation}
{\cal F}_{ab} = \frac{1}{2}\lim_{x'\rightarrow x} {\cal T}_{ab}
  \left(F(x,x') + F(x',x) \right)
\label{ref-reg-difftab}
\end{equation}
is well defined. So  a bi-distribution $G^L(x,x')$ might be useful  for the 
vacuum subtraction. 
At first, it would seem unlikely we could find such a unique 
bi-distribution. Wald found that with the introduction of four axioms for
the regularized stress tensor
\begin{equation}
\langle \hat T_{ab}(x) \rangle_{\rm ren} = \lim_{x'\rightarrow x}
\frac{1}{2} {\cal T}_{ab}\left(
{G^{(1)}}(x,x') - {G^{L}}(x,x') \right)
\label{ref-eq-STren-define}
\end{equation}
 $G^L(x,x')$ is uniquely determined,
up to a local conserved curvature term. The Wald axioms are 
\cite{Wald78,Wald94}:
\begin{enumerate}
  \item The difference between the stress tensor for two states should
        agree with (\ref{ref-reg-difftab});
  \item The stress tensor should be local with respect to the state of
        the field;
  \item For all states, the stress tensor is conserved:
                    $\nabla^a \langle T_{ab}\rangle = 0$;
  \item In Minkowski space, the result 
               $\langle 0| T_{ab}|0\rangle = 0$ is recovered.
\end{enumerate}

We are still left with the problem of determining the form of such a
bi-distribution.  Hadamard \cite{Hadamard} showed that the Green function for 
a large class of states takes the form (in four spacetime dimensions)
\begin{equation}
G^L(x,x') = \frac{1}{8\pi^2}\left(\frac{2U(x,x')}{\sigma}
  +V(x,x')\log\sigma + W(x,x') \right)
\label{ref-HadamardAnsatz}
\end{equation}
with $U(x,x')$, $V(x,x')$ and $W(x,x')$ being smooth
functions\footnote{When working in the
Lorentz sector of a field theory,
{\it i.e.}, when the metric signature is $(-,+,+,+)$, as opposed to the
Euclidean sector with the signature $(+,+,+,+)$,
we must modify the above function to
account for null geodesics, since $\sigma(x,x')=0$ for null separated
$x$ and $x'$. This problem can be overcome
by using $\sigma \rightarrow \sigma +
2i\epsilon (t-t') + \epsilon^2$. Here, we will work only with
geometries that possess Euclidean sectors and carry out our analysis
with Riemannian geometries and only at the end continue back to the
Lorentzian geometry. Nonetheless, this presents no difficulty. At any
point in the analysis the above replacement
for $\sigma$ can be performed.}.
We refer to Eqn (\ref{ref-HadamardAnsatz})
as the ``Hadamard ansatz''.

Since the functions $V(x,x')$ and $W(x,x')$ are smooth functions, they
can be expanded as
\begin{mathletters}
\label{ref-vwseries}
\begin{eqnarray}
V(x,x') = \sum_{n=0}^\infty v_n(x,x') \sigma^n, \\
W(x,x') = \sum_{n=0}^\infty w_n(x,x') \sigma^n,
\end{eqnarray}
\end{mathletters}
with the $v_n$'s and $w_n$'s themselves smooth functions. These functions
and $U(x,x')$ are determined by substituting $G^L(x,x')$ in the wave equation
$K G^L(x,x') = 0$ and equating to zero the coefficients of the explicitly
appearing powers of $\sigma^n$ and $\sigma^n\log\sigma$. Doing so, we 
get the infinite set of equations
\begin{mathletters}
\begin{eqnarray}
                              U(x,x') &=& {\Delta\!^{1/2}}; \\
                  2 H_0 v_0 + K {\Delta\!^{1/2}} &=& 0; \label{ref-v0eqn}\\
               2n H_n v_n + K v_{n-1} &=& 0, \quad n \ge 1;\label{ref-vneqn}\\
 2H_{2n} v_n + 2n H_n w_n + K w_{n-1} &=& 0, \quad n \ge 1\label{ref-wneqn}
\label{ref-vweqns}
\end{eqnarray}
with
\begin{equation}
H_n = \sigma^{;p} \nabla_p + \left(n-1 
    +\frac{1}{2}\left(\Box\sigma\right)\right).
\end{equation}
\end{mathletters}
{}From Eqs (\ref{ref-vweqns}), the functions $v_n$ are completely determined.
In fact, they are symmetric functions, and hence $V(x,x')$ is a symmetric
function of $x$ and $x'$. On the other hand, the field equations only determine
$w_n$, $n \ge 1$, leaving $w_0(x,x')$ completely  arbitrary. This reflects the
state dependence of the Hadamard form above. Moreover, even if $w_0(x,x')$ is
chosen to be symmetric, this does not guarantee that $W(x,x')$ will be. By
using axiom (4)  $w_0(x,x') \equiv 0$. With this choice, the Minkowski
spacetime
limit is
\begin{equation}
G^L = \frac{1}{(2\pi)^2} \frac{1}{\sigma},
\end{equation}
where now $2 \sigma =  (t-t')^2 - ({\bf x}-{\bf x}')^2$ and this
corresponds to the correct vacuum contribution that needs to be subtracted.

With this choice though, we are left with a $G^L(x,x')$ which is not symmetric
and hence does not satisfy the field equation at $x'$, for fixed $x$. Wald
\cite{Wald78} showed this in turn implies axiom (3) is not satisfied. He
resolved this problem by adding to the regularized stress tensor a term which 
cancels that which breaks the conservation of the old stress tensor:
\begin{equation}
\langle T^{\rm new}_{ab}\rangle = \langle T^{\rm old}_{ab}\rangle +
\frac{1}{2(4\pi)^2} g_{ab} \left[ v_1 \right],
\end{equation}
where $\left[ v_1 \right] = v_1(x,x)$ is the coincident limit of the
$n=1$ solution of Eq (\ref{ref-vweqns}). This yields a stress tensor
which satisfies all four axioms and produces the well known trace anomaly
$\langle T_{a}{}^a\rangle = \left[ v_1 \right]/8\pi^2$. We can view this
redefinition as taking place at the level of the stress tensor operator
via
\begin{equation}
\hat T_{ab} \rightarrow \hat T_{ab} 
+ \frac{\hat 1}{2(4\pi)^2} g_{ab} \left[ v_1 \right]
\end{equation}
Since this amounts to a constant shift of the stress tensor operator, it
will have no effect on the noise kernel or fluctuations, as they are
 the variance about the mean. This is further supported by
the fact that there is no noise associated with the trace. Since
this result was derived by only assuming that the Green function satisfies the
field equation in {\em one} of its variables, it is independent of the
issue of the lack of symmetry in the Hadamard ansatz 
(\ref{ref-HadamardAnsatz}).

Using the above formalism we now derive  the coincident limit expression for 
the noise kernel
(\ref{ref-noise-kernel}).
To get a meaningful result, we must work with the regularization of the 
Wightman function, obtained by following the same procedure outlined
above for the Hadamard function:
\begin{equation}
G_{\rm ren}(x,y) \equiv G_{\rm ren,+}(x,y) =
	G_+(x,y) - G^L(x,y)
\end{equation}
In doing this, we assume the singular structure of the Wightman function
is the same as that for the Hadamard function. In all applications, this
is indeed the case. Moreover, when we compute the coincident limit
of $N_{abc'd'}$, we will be working in the Euclidean section where there is no
issue of operator ordering.   For now we only consider spacetimes
with no time dependence present in the final coincident limit result, 
so there is also no issue of Wick rotation back to a Minkowski signature.
If this was the case, then care must be taken as to
whether we are considering
$\left[ N_{abc'd'} \left[ G_{\rm ren,+}(x,y) \right] \right]$ or
$\left[ N_{abc'd'} \left[ G_{\rm ren,+}(y,x) \right] \right]$.

We now have all the information we need to compute the coincident limit
of the noise kernel (\ref{ref-noise-kernel}). Since the point
separated noise kernel $N_{abc'd'}(x,y)$ involves covariant derivatives at 
the two points at which it has support, when we take the coincident limit
we can use Synge's theorem (\ref{ref-Synge's}) to move the derivatives
acting at $y$ to ones acting at $x$. Due to the long length of the noise
kernel expression, we will only  give an example by examining a single
term. 

Consider a typical term from the noise kernel functional
(\ref{ref-noise-kernel}):
\begin{equation}
G_{{\rm ren}}{}_{;}{}_{c'}{}_{b}\,G_{{\rm ren}}{}_{;}{}_{d'}{}_{a} + G_{{\rm 
ren}}{}_{;}{}_{c'}{}_{a}\,G_{{\rm ren}}{}_{;}{}_{d'}{}_{b}
\end{equation}
Recall the noise kernel itself is related to the noise kernel functional
via
\begin{equation}
N_{abc'd'} = N_{abc'd'}\left[G_{ren}(x,y)\right]
                + N_{abc'd'}\left[G_{ren}(y,x)\right].
\end{equation}
This is implemented on our typical term by adding to it the same term,
but now with the roles of $x$ and $y$ reversed, so we have to consider
\begin{equation}
G_{{\rm ren}}{}_{;}{}_{c'}{}_{b}\,G_{{\rm ren}}{}_{;}{}_{d'}{}_{a} + G_{{\rm 
ren}}{}_{;}{}_{a'}{}_{d}\,G_{{\rm ren}}{}_{;}{}_{b'}{}_{c} + G_{{\rm 
ren}}{}_{;}{}_{c'}{}_{a}\,G_{{\rm ren}}{}_{;}{}_{d'}{}_{b} + G_{{\rm 
ren}}{}_{;}{}_{a'}{}_{c}\,G_{{\rm ren}}{}_{;}{}_{b'}{}_{d}
\end{equation}
It is to this form that we can take the coincident limit:
\begin{equation}
{\left[G_{{\rm ren}}{}_{;}{}_{c'}{}_{b}\right]}\,
  {\left[G_{{\rm ren}}{}_{;}{}_{d'}{}_{a}\right]} + {\left[G_{{\rm 
ren}}{}_{;}{}_{a'}{}_{d}\right]}\,
  {\left[G_{{\rm ren}}{}_{;}{}_{b'}{}_{c}\right]} + {\left[G_{{\rm 
ren}}{}_{;}{}_{c'}{}_{a}\right]}\,
  {\left[G_{{\rm ren}}{}_{;}{}_{d'}{}_{b}\right]} + {\left[G_{{\rm 
ren}}{}_{;}{}_{a'}{}_{c}\right]}\,
  {\left[G_{{\rm ren}}{}_{;}{}_{b'}{}_{d}\right]}
\end{equation}
We can now apply Synge's theorem:
\begin{eqnarray}
&&\hspace{3mm}
\left( {\left[G_{{\rm ren}}{}_{;}{}_{a}\right]}{}_{;}{}_{d} - 
    {\left[G_{{\rm ren}}{}_{;}{}_{a}{}_{d}\right]} \right) \,
  \left( {\left[G_{{\rm ren}}{}_{;}{}_{b}\right]}{}_{;}{}_{c} - 
    {\left[G_{{\rm ren}}{}_{;}{}_{b}{}_{c}\right]} \right)  \cr
&& + \left( {\left[G_{{\rm ren}}{}_{;}{}_{d}\right]}{}_{;}{}_{a} - 
    {\left[G_{{\rm ren}}{}_{;}{}_{a}{}_{d}\right]} \right) \,
  \left( {\left[G_{{\rm ren}}{}_{;}{}_{c}\right]}{}_{;}{}_{b} - 
    {\left[G_{{\rm ren}}{}_{;}{}_{b}{}_{c}\right]} \right)  \cr
&& + \left( {\left[G_{{\rm ren}}{}_{;}{}_{a}\right]}{}_{;}{}_{c} - 
    {\left[G_{{\rm ren}}{}_{;}{}_{a}{}_{c}\right]} \right) \,
  \left( {\left[G_{{\rm ren}}{}_{;}{}_{b}\right]}{}_{;}{}_{d} - 
    {\left[G_{{\rm ren}}{}_{;}{}_{b}{}_{d}\right]} \right)  \cr
&& + \left( {\left[G_{{\rm ren}}{}_{;}{}_{c}\right]}{}_{;}{}_{a} - 
    {\left[G_{{\rm ren}}{}_{;}{}_{a}{}_{c}\right]} \right) \,
  \left( {\left[G_{{\rm ren}}{}_{;}{}_{d}\right]}{}_{;}{}_{b} - 
    {\left[G_{{\rm ren}}{}_{;}{}_{b}{}_{d}\right]} \right).
\end{eqnarray}
This is the desired form for once we have an end point expansion
of $G_{{\rm ren}}$, it will be straightforward to compute the above
expression. The details of such an evaluation in the context of symbolic 
computations can be found in \cite{NPsc}.

The final result for the coincident limit of the noise kernel is
broken down into a rank four and rank two tensor and a scalar
according to
%#########################################################################
%#########################################################################
%
%    NOISE KERNEL FUNCTIONAL FORM, COINCIDENT LIMIT
%
%#########################################################################
%#########################################################################
%
\begin{mathletters}
\label{coincident-noise}
\begin{equation}
 \left[ N_{abc'd'} \right]  = 
    N'_{abcd}
  + g_{ab} N''_{cd}
 + g_{cd} N''_{ab}
 + g_{ab}g_{cd} N'''.
\end{equation}
\end{mathletters}
The complete expression
is given in Ref. \cite{PH2}.

\subsection{Summary Statements}

\subsubsection{Further Developments}

In this section we showed how to obtain a general expression for
the noise kernel, or the vacuum expectation value of the
stress energy bi-tensor for a quantum scalar field in 
a general curved space time  using the point separation 
method. 
The general form is expressed as products of covariant derivatives of 
the quantum field's Green function. It  is finite when
the noise kernel is evaluated for distinct pairs of points
(and non-null points for a massless field). We also
have shown the trace of the noise kernel  vanishes
for massless conformal fields, confirming there is no noise
associated with the trace anomaly. This holds regardless
of issues of regularization of the noise kernel.

The noise kernel as a two point function of the
stress energy tensor diverges  as the
pair of points are brought together, representing 
the ``standard'' ultraviolet divergence present in the
(point-defined) quantum field theory. By using the modified point
separation regularization method we render the
field's Green function finite in the coincident limit.
This in turn permits the derivation of the formal expression 
for the regularized coincident limit of the noise kernel.

The general results obtained here are now applied by Phillips and Hu
to compute the regularized noise kernel for three different groups
of spacetimes:
1) ultrastatic metrics including the Einstein universe 
\cite{Phillips-Hu97}, hot flat space and
optical Schwarzschild spacetimes. 
2) Robertson-Walker universe and Schwarzschild black holes, from which 
structure formation from quantum fluctuations
\cite{strfor} and backreaction of Hawking radiation on 
the black hole spacetime \cite{HRS} can be studied. 
3) de Sitter and anti-de Sitter spacetimes:
The former is necessary for scrutinizing 
primordial fluctuations in the cosmic background radiation while
the latter is related to black hole phase transition and  AdS/CFT issues in 
string theory.

When the Green function is available in closed analytic
form, as is the case for optical metrics including hot flat space
and Einstein Universe, conformally static spaces such as Robertson-Walker Unvierses, 
and maximally symmetric spaces such as de Sitter and Anti de Sitter , 
one can carry out an  end point expansion according
to (\ref{general-endpt-series}),  displaying  the ultraviolet
divergence. Subtraction of the  Hadamard 
ansatz (\ref{ref-HadamardAnsatz}), expressed as a series 
expansion (\ref{ref-vwseries}), will render this Green function
finite in the coincident limit. With this, one can calculate the noise
kernel for a variety of spacetimes.

%%%%%%%%%%%%

\subsubsection{The Meaning of Regularization Revisited}

Thus far we have focussed on  isolating and removing the divergences
in the stress energy bitensor and the noise kernel in the coincident limit.
In  closing, we would like to redirect the reader's attention to the intrinsic 
values of the point-separated geometric quantities like the bitensor. 
As stressed earlier \cite{stogra}, in the stochastic
gravity program, point separated expression of stress energy
bi-tensor have fundamental physical meaning as it  contains information  on
the fluctuations and correlation of quantum fields,  and by consistency with
the gravity sector, can provide a probe into the coherent properties of quantum
spacetimes. Taking this view, we may also gain a new perspective on ordinary
quantum field theory defined on single points:
The coincidence limit depicts the low energy limit of the full quantum theory
of matter and spacetimes. Ordinary (pointwise) quantum field theory, classical
general relativity and  semiclassical gravity are the lowest levels of
approximations and should be viewed not as fundamental,  but only as effective
theories.  As such, even the  way how the conventional point-defined field 
theory 
emerges from the full theory when the two points (e.g., $x$ and $y$ in the noise 
kernel) are
brought together is interesting. For example, one can ask if there will also be
a quantum to classical transition in spacetime accompanying the  coincident
limit? Certain aspects like decoherence has been investigated before (see,
e.g., \cite{decQC}), but here the non-local  structure of spacetime and their
impact on quantum field theory become the central issue. (This may also be a
relevant issue in noncommutative geometry). The point-wise limit of field theory
of course has ultraviolet divergences and requires regularization. A new 
viewpoint
towards regularization evolved from this perspective of treating  conventional 
pointwise field theory as an effective theory in the coincident limit of the 
point-separated 
theory of extended spacetime. 

%%%%%%%%%%%%%%%
To end this discussion, we venture one philosophical point 
we find resounding in these investigations reported here.
It has to do  with the meaning of a point-defined versus
a point-separated field theory, the former we take as
an effective theory  coarse-grained from the latter, 
the point-separated theory  reflecting a finer level of
spacetime structure. 
It bears on the meaning of regularization, not
just at the level of technical procedures, but
related to finding  an effective description and
matching with physics observed at a coarser
scale or lower energy.

In particular, we feel that finding a  finite  energy momentum
tensor (and its fluctuations as we do here) which occupied
the center of attention in the research of quantum field theory 
in curved spacetime in the 70's is only a small part of a much 
larger and richer structure of theories of fields and spacetimes . 
We come to understand that whatever regularization method one 
uses to get these finite parts in a point-wise field theory should 
not be viewed as universally imparting meaning beyond its 
specified function, i.e., to identify the divergent pieces and 
provide a prescription for their removal. We believe the 
extended structure of spacetime (e.g., via point-separation or smearing)
and the field theory defined therein has its own much fuller
meaning beyond just reproducing the well-recognized result in
ordinary quantum field theory as we take the point-wise or coincident limit.
In this way of thinking, the divergence- causing terms are only
`bad' when they are forced to a point-wise limit, because of our
present inability to observe or resolve otherwise . If we accord them
with the full right of existence beyond this limit, and acknowledge
that their misbehavior is really due to our own  inability to cope,
we will be rewarded with the discovery of new physical phenomena 
and ideas of a more intricate world. (Maybe this is just another
way to appreciate the already well-heeded paths of string theory.)

\acknowledgements

BLH wishes to thank the organizers of this course for the invitation to lecture
and their warm hospitality at Erice. The materials presented here
originated from several research papers, three of BLH with Nicholas Phillips
and three of EV with Rosario Mart\'{\i}n. We thank Drs. Martin and Phillips 
as well as Antonio Campos, Andrew Matacz, Sukanya Sinha, Tom Shiokawa, Albert Roura and
Yuhong Zhang for fruitful collaboration and their cordial friendship since their Ph. D. 
days. We enjoy lively discussions with our long-time collaborator and dear friend
Esteban Calzetta, who contributed greatly to the establishment of this field.
We acknowledge useful discussions with Paul Anderson, Larry Ford, Ted Jacobson, 
Renaud Parentani and Raphael Sorkin.
This work is supported in part by NSF grant PHY98-00967
and the CICYT Reserach Project No. AEN98-0431. EV also
acknowledges support from the Spanish Ministery of Education under the
FPU grant PR2000-0181 and the University of Maryland for hospitality.

%\appendix

\newpage

%%%%%%%%%%%%%%%%%%%%%%%%%%%%%%%%%%%%%%%%%%
% @ do you want to put figures? I do not know how to do that
%%%%%%%%%%%%%%%%%%%%%%%%%%%%%%%%%%%%%%%%%%

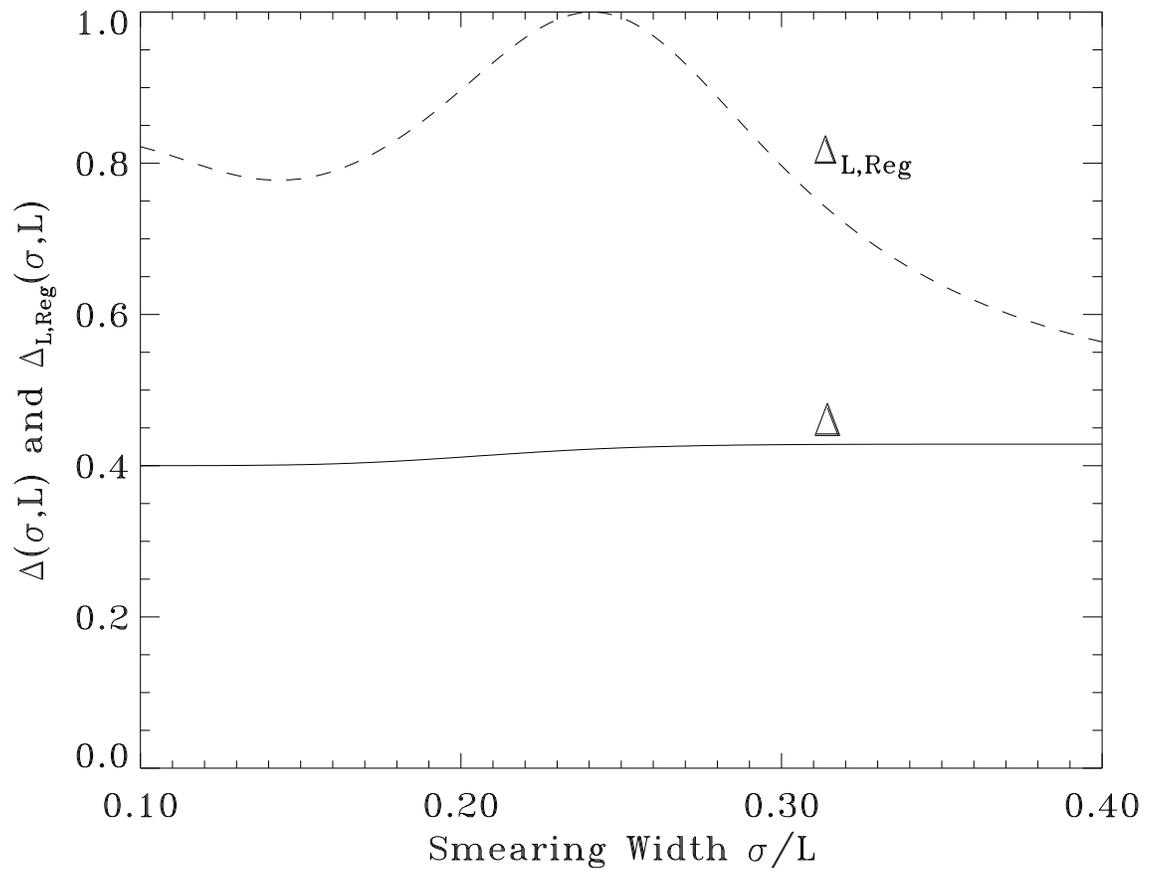
\begin{figure}
\caption{
  The dimensionless fluctuation measure
  $\Delta \equiv \left(\left<\hat\rho^2\right>-\left<\hat\rho\right>^2\right)/
                 \left<\hat\rho^2\right>$
  for the Casimir topology, along with $\Delta_{L,{\rm Reg}}$.
  The topology is that
  of a flat three spatial dimension manifold with one periodic
dimension of period
  $L=1$. The smearing width $\sigma$ represents the sampling width of the
  energy density operator $\hat\rho(\sigma)$. $\Delta$ is for the
complete fluctuations,
  including divergent and cross terms, while $\Delta_{L,{\rm Reg}}$
is just for the
  finite parts of the mean energy density and fluctuations. 
}
\end{figure}

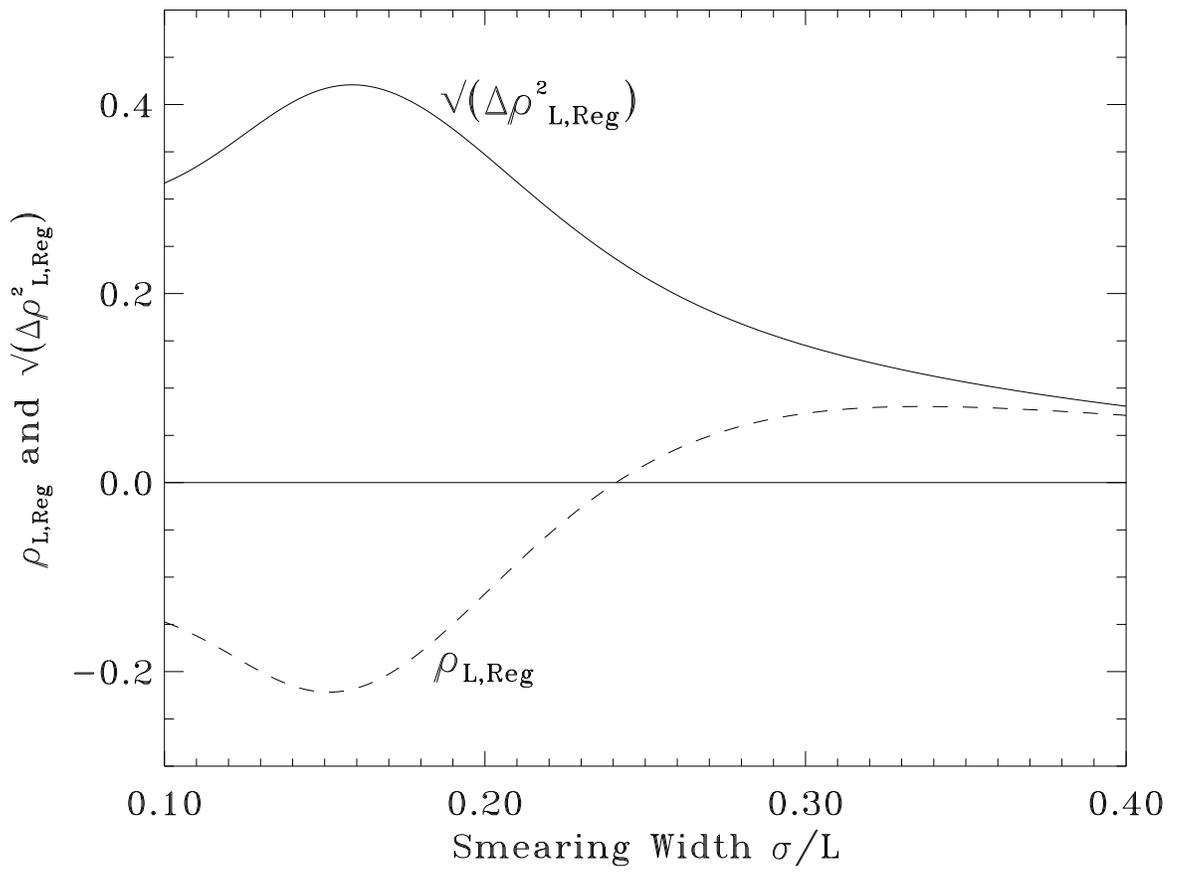
\begin{figure}
\caption{
  The finite parts of the mean energy density
$\rho^{\rm fin}(\sigma,L)$ and
  the fluctuations $\Delta\rho^{\rm fin}(\sigma,L)$
for the Casimir topology, as a
  function of the smearing width. 
}
\end{figure}


\newpage

\begin{references}

\bibitem{Physica} B. L. Hu, Physica A {\bf 158}, 399 (1989).

\bibitem{ELE}
  E. Calzetta and B.L. Hu,    Phys. Rev. D {\bf 49}, 6636 (1994);
  B. L. Hu and A. Matacz,    Phys. Rev. D {\bf 51}, 1577 (1995);
  B. L. Hu and S. Sinha,       Phys.  Rev. D {\bf 51}, 1587 (1995);
  A. Campos and E. Verdaguer,   Phys. Rev. D {\bf 53}, 1927 (1996);
  F. C. Lombardo and F. D. Mazzitelli,  Phys. Rev. D {\bf 55}, 3889 (1997).

\bibitem{stogra} 
B. L. Hu, Int. J. Theor. Phys. {\bf 38},  2987  (1999); gr-qc/9902064.

\bibitem{MV0} R. Mart\'{\i}n and E. Verdaguer,
Phys. Lett. B {\bf 465}, 113 (1999).

\bibitem{MV1}
R. Mart\'{\i}n and E. Verdaguer,  Phys. Rev. D {\bf 60}, 084008 (1999).

\bibitem{MV2}
R. Mart\'{\i}n and E. Verdaguer,  Phys. Rev. D {\bf 61}, 124024 (2000).

\bibitem{HP0} 
B. L. Hu and   N. G. Phillips,  Int. J. Theor. Phys.
{\bf 39}, 1817 (2000); gr-qc/0004006.

\bibitem{PH1}
N. G.  Phillips and  B. L. Hu, Phys. Rev. D {\bf 62}, 084017 (2000).

\bibitem{PH2}
N. G.  Phillips and  B. L. Hu, Phys. Rev. D {\bf 63},  104001 (2001).

\bibitem{qos}  See, e.g., E. B. Davies, {\it The Quantum Theory of Open
Systems} (Academic Press, London, 1976); K. Lindenberg and B. J. West,
{\it The Nonequilibrium Statistical Mechanics of Open and Closed Systems}
(VCH Press, New York, 1990); U. Weiss, {\it Quantum Dissipative Systems}
(World Scientific, Singapore, 1993)

\bibitem{if}
R. P. Feynman and F. L. Vernon, Ann. Phys. (NY) {\bf 24}, 118 (1963);
R. P. Feynman and A. R. Hibbs, {\it Quantum Mechanics and Path Integrals},
(McGraw-Hill, New York, 1965);
A. O. Caldeira and A. J. Leggett, Physica {\bf 121A}, 587 (1983);
Ann. Phys. (NY) {\bf 149}, 374 (1983);
H. Grabert, P. Schramm and G. L. Ingold, Phys. Rep. {\bf 168}, 115 (1988);
B. L. Hu, J. P. Paz and Y. Zhang, Phys. Rev. D {\bf 45}, 2843 (1992);
{\it ibid.} {\bf 47}, 1576 (1993).

\bibitem{ctp}  J. Schwinger, J. Math. Phys. {\bf 2} (1961) 407; P. M. Bakshi
and K. T. Mahanthappa, J. Math. Phys. {\bf 4}, 1 (1963);
{\bf 4}, 12 (1963); L. V.
Keldysh, Zh. Eksp. Teor. Fiz. {\bf 47 }, 1515 (1964) [Engl. trans. Sov.
Phys. JEPT {\bf 20}, 1018 (1965)]; K. Chou, Z. Su, B. Hao and L. Yu, Phys.
Rep. {\bf 118}, 1 (1985); Z. Su, L. Y. Chen, X. Yu and K. Chou, Phys. Rev.
B {\bf 37}, 9810 (1988). E. Calzetta and B. L. Hu, Phys. Rev. D
{\bf 40}, 656 (1989).


\bibitem{ctpcst} 
B. S. DeWitt, in {\it Quantum Concepts in Space and
Time} ed. R. Penrose and C. J. Isham (Claredon Press, Oxford, 1986).
R. D. Jordan, Phys. Rev. D {\bf 33}, 444 (1986);
E. Calzetta and B. L. Hu, Phys. Rev. D {\bf 35}, 495 (1987);
R. D. Jordan, Phys. Rev. D {\bf 36}, 3593 (1987);
J. P. Paz, Phys. Rev. D {\bf 41}, 1054 (1990);
A. Campos and E. Verdaguer, Phys. Rev. D {\bf 49}, 1861 (1994).


\bibitem{CH88}
E. Calzetta and B. L. Hu, Phys. Rev. D {\bf 37}, 2878 (1988).

\bibitem{dch}  E. Calzetta and B. L. Hu, ``Decoherence of Correlation
Histories'' in {\it Directions in General Relativity, Vol II: Brill
Festschrift}, eds B. L. Hu and T. A. Jacobson
(Cambridge University Press, Cambridge, 1993).

\bibitem{cddn}
E. Calzetta and B. L. Hu, ``Correlations, Decoherence,
Disspation and Noise in Quantum Field Theory'', in {\it Heat Kernel
Techniques and Quantum Gravity},
ed. S. Fulling (Texas A\& M Press, College Station 1995); hep-th/9501040.

\bibitem{stobol}  
E. Calzetta and B. L. Hu,  
Phys. Rev. D {\bf 61}, 025012 (2000).
%\bibitem{CH00} 
%" Stochastic dynamics of correlations in quantum field theory: 
% From Schwinger-Dyson  to Boltzmann-Langevin equation

\bibitem{MTW} C. W. Misner, K. S. Thorne and J. A. Wheeler,
{\it Gravitation} (Freeman, San Francisco, 1973).

\bibitem{Wald84} 
  R. M. Wald,       {\sl General Relativity}
      (The University of Chicago Press, Chicago, 1984).

\bibitem{Birrell-Davies82} N. D. Birrell and P. C. W. Davies, {\it
Quantum fields in curved space} (Cambridge University Press,
Cambridge, England, 1982).

\bibitem{Fulling89} S. A. Fulling, {\it
Aspects of quantum field theory in curved spacetime}
(Cambridge University Press, Cambridge, England, 1989).

\bibitem{Wald94} R. M. Wald, {\it
Quantum field theory in curved spacetime and black hole thermodynamics}
(Cambridge University Press, Cambridge, England, 1994).

\bibitem{mostepanenko}
  A. A. Grib, S. G. Mamayev and V. M. Mostepanenko,
      {\sl Vacuum Quantum Effects in Strong Fields} 
      (Friedmann Laboratory Publishing, St.~Petersburg, 1994).

\bibitem{scg}
%\bibitem{cpcbkr}
Ya. Zel'dovich and A. Starobinsky, Zh. Eksp. Teor. Fiz {\bf 61}, 2161 (1971)
[Sov. Phys.- JETP {\bf 34}, 1159 (1971)];
%\bibitem{Gri76}
L. Grishchuk, Ann. N. Y. Acad. Sci. 302, 439 (1976);
%\bibitem{HuPar77}
B. L. Hu and L. Parker, Phys. Lett. A {\bf 63}, 217 (1977);
%\bibitem {HuPar78}
B. L. Hu  and L. Parker, Phys. Rev. D {\bf 17}, 933 (1978);
%\bibitem {FHH}
F. V. Fischetti, J. B. Hartle and B. L. Hu, Phys. Rev. D
{\bf 20}, 1757 (1979);
%\bibitem{HarHu}
J. B. Hartle and B. L. Hu, Phys. Rev. D {\bf 20}, 1772 (1979);
{\it ibid.} {\bf 21}, 2756 (1980);
%\bibitem{Har81}
J. B. Hartle, Phys. Rev. D {\bf 23}, 2121 (1981);
P. A. Anderson, Phys. Rev. D {\bf 28}, 271 (1983);
{\it ibid.} {\bf 29}, 615 (1984).

\bibitem{envdec}
W. H. Zurek, Phys. Rev. D {\bf 24}, 1516 (1981); D26, 1862
(1982); in {\it Frontiers of Nonequilibrium Statistical Physics}, 
ed. G. T. Moore and M. O. Scully (Plenum, N. Y., 1986);
Physics Today {\bf 44}, 36 (1991);
E. Joos and H. D. Zeh, Z. Phys. B {\bf 59}, 223 (1985);
A. O. Caldeira and A. J. Leggett, Phys. Rev. A {\bf 31}, 1059 (1985);
W. G. Unruh and W. H. Zurek, Phys. Rev. D {\bf 40}, 1071 (1989);
B. L. Hu, J. P. Paz and Y. Zhang, Phys. Rev. D {\bf 45}, 2843 (1992);
W. H. Zurek, Prog. Theor. Phys. {\bf 89}, 281 (1993);   D. Giulini
{\it et al}, {\it Decoherence and the Appearance of a
Classical World in Quantum Theory}
(Springer Verlag, Berlin, 1996);
J. T. Whelan, Phys. Rev. D {\bf 57}, 768 (1998).

\bibitem{conhis}
R. B. Griffiths, J. Stat. Phys. {\bf 36}, 219 (1984);
R. Omn\'es, J. Stat Phys. {\bf 53}, 893 (1988);
{\it ibid.} {\bf 53} 933 (1988);
{\it ibid.} {\bf 53} 957 (1988);
{\it ibid.} {\bf 54}, 357 (1988); Ann. Phys. (NY)
{\bf 201}, 354 (1990); Rev. Mod. Phys. {\bf 64}, 339 (1992); {\it The
Interpretation of Quantum Mechanics} (Princeton UP, Princeton, 1994);
M. Gell-Mann and J. B. Hartle, in {\it Complexity, Entropy and the Physics
of Information}, ed. by W. H. Zurek (Addison-Wesley, Reading, 1990);
Phys. Rev. D {\bf 47}, 3345 (1993);
J. B. Hartle, ``Quantum Mechanics of Closed
Systems'' in {\it Directions in General Relativity} Vol. 1, eds B. L. Hu,
M. P. Ryan and C. V. Vishveswara (Cambridge Univ., Cambridge, 1993);
H. F. Dowker and J. J. Halliwell, Phys. Rev. D {\bf 46}, 1580 (1992);
J. J. Halliwell, Phys. Rev. D {\bf 48}, 4785 (1993);
{\it ibid.} {\bf 57}, 2337 (1998);
T. Brun, Phys. Rev. D {\bf 47}, 3383 (1993);
J. P. Paz and W. H. Zurek, Phys. Rev. D {\bf 48} 2728 (1993);
J. Twamley, Phys. Rev. D {\bf 48}, 5730 (1993);
C. J. Isham, J. Math. Phys. {\bf 35}, 2157 (1994);
C. J. Isham and N. Linden, {\it ibid.} {\bf 35}, 5452 (1994);
{\it ibid.} {\bf 36}, 5392 (1994);
J. J. Halliwell, Ann. N.Y. Acad. Sc. {\bf 755}, 726 (1995);
F. Dowker and A. Kent, Phys. Rev. Lett {\bf 75}, 3038
(1995); J. Stat. Phys. {\bf 82}, 1575 (1996);
A. Kent, Phys. Rev. A {\bf 54}, 4670 (1996);
Phys. Rev. Lett {\bf 78}, 2874 (1997);
{\it ibid.} {\bf 81}, 1982 (1998);
C. J. Isham, N. Linden, K. Savvidou and S. Schreckenberg,
J. Math. Phys. {\bf 39}, 1818 (1998).

\bibitem{decQC}
C. Kiefer, Clas. Quant. Grav. {\bf 4}, 1369 (1987);
 J. J. Halliwell, Phys. Rev. D {\bf 39}, 2912 (1989);
T. Padmanabhan, {\it ibid.} 2924 (1989);
 B. L. Hu ``Quantum and Statistical Effects in Superspace Cosmology'' in
{\it Quantum Mechanics in Curved Space time}, 
ed. J. Audretsch and V. de Sabbata (Plenum, London 1990);
E. Calzetta, Class. Quan. Grav. {\bf 6}, L227 (1989); Phys. Rev. D
{\bf 43}, 2498 (1991); J. P. Paz and S. Sinha, Phys. Rev. D {\bf44},
1038 (1991);
{\it ibid} {\bf 45}, 2823 (1992);
 B. L. Hu, J. P. Paz and S. Sinha,
``Minisuperspace as a Quantum Open System''
in {\it Directions in General Relativity}
Vol. 1, (Misner Festschrift) eds B. L. Hu,
M. P. Ryan and C.V. Vishveswara
(Cambridge University Press, Cambridge, England, 1993).

\bibitem{CH87}
E. Calzetta and B. L. Hu, Phys. Rev. D {\bf 35}, 495 (1987).

\bibitem{cv94}
A. Campos and E. Verdaguer, Phys. Rev. D {\bf 49}, 1861 (1994).

\bibitem{HuSin}
B.L. Hu and S. Sinha,   Phys.  Rev. D {\bf 51}, 1587 (1995).

\bibitem{CH94} 
  E. Calzetta and B.L. Hu,
       Phys. Rev. D {\bf 49}, 6636 (1994).

\bibitem{Banff}
B.L. Hu, in {\it Proceedings of the Third International
Workshop on Thermal 
Fields and its Applications}, CNRS Summer Institute,
Banff, August 1993, edited by R. Kobes and 
G. Kunstatter (World Scientific, Singapore, 1994), gr-qc/9403061.


\bibitem{HuShio} B. L. Hu and K. Shiokawa, Phys. Rev. D {\bf 57},
3474 (1998).

\bibitem{Ford} L. H. Ford and N. F. Svaiter, Phys. Rev. D. {\bf 56}, 2226
(1997).

\bibitem{WuFor} C.-H. Wu and L. H. Ford, Phys. Rev. D
{\bf 60}, 104013 (1999).

\bibitem{Sorkin} R. Sorkin, 
``How Wrinkled is the Surface of a Black Hole?'',
    in {\sl Proceedings of the First Australasian
Conference on General Relativity and Gravitation}
    February 1996, Adelaide, Australia, edited by
David Wiltshire, pp. 163-174 
   (University of Adelaide, 1996); gr-qc/9701056; 
R. D. Sorkin and D. Sudarsky, Class. Quantum Grav. {\bf 16}, 3835 (1999).

\bibitem{BFP00}
C. Barrab\`es, V. Frolov and R. Parentani, Phys. Rev. D
{\bf 62}, 044020 (2000).

\bibitem{MasPar00}
S. Massar and R. Parentani, Nucl. Phys. B {\bf 575}, 333 (2000).

\bibitem{calver99}
  E. Calzetta and E. Verdaguer,
      Phys. Rev. D {\bf 59}, 083513 (1999); 
 E. Calzetta, Int. J. Theor. Phys. {\bf 38}, 2755  (1999).

\bibitem{strfor}  
E. Calzetta and B. L. Hu, Phys. Rev. D {\bf 52}, 6770 (1995);
A. Matacz, Phys. Rev. D {\bf 55}, {1860} (1997);
E. Calzetta and S. Gonorazky,  Phys. Rev. D {\bf 55}, {1812}
(1997);  A. Roura and E. Verdaguer, Int. J. Theor. Phys.
{\bf 38}, 3123 (1999); {\it ibid.} {\bf 39}, 1831 (2000).

%\bibitem{CH95} E. Calzetta and B.L. Hu, Phys. Rev. D {\bf 52} 6770 (1995).

%\bibitem{Roura99} A. Roura and E. Verdaguer, Int. J. Theor. Phys. {\bf 38}, 3123 (1999).

%\bibitem{Roura00a} A. Roura and E. Verdaguer, Int. J. Theor. Phys. {\bf 39}, 1831 (2000).

\bibitem{Starobinsky80} A. A. Starobinsky, Phys. Lett. B {\bf 91}, 
99 (1980).

\bibitem{Vilenkin85} A. Vilenkin, Phys. Rev. D {\bf 32}, 2511 (1985).

\bibitem{Hawking01} S. W. Hawking, T. Hertog and H. S. Reall,
Phys. Rev. D {\bf 63}, 083504 (2001).

\bibitem{HuVer01} B. L. Hu and E. Verdaguer, in preparation (2001).

\bibitem{CanSci}
P. Candelas and D. W. Sciama, Phys. Rev. Lett. {\bf 38}, 1372 (1977).

\bibitem{Mottola}   E. Mottola,   Phys. Rev. D {\bf 33}, 2136 (1986).

\bibitem{HRS} B. L. Hu, A. Raval and S. Sinha,
``Notes on Black Hole Fluctuations  and Backreaction"
in {\it Black Holes, Gravitational Radiation and the Universe:
Essays in honor  of C. V. Vishveshwara}
eds. B. Iyer and B. Bhawal (Kluwer Academic Publishers, Dordrecht, 1998);  
gr-qc/9901010.

\bibitem{CamHu} A. Campos and B. L. Hu, Phys. Rev. D {\bf 58} (1998) 125021;
              Int. J. Theor. Phys. 38 (1999) 1253.

\bibitem{Par01}
R. Parentani, Phys. Rev. D {\bf 63}, 041503 (2001).

\bibitem{NiePar01}
J. C. Niemeyer and R. Parentani, astro-ph/0101451.

\bibitem{STFoam}
S. Carlip, Phys. Rev. Lett. 79, 4071 (1997);
Class. Quan. Grav. 15, 2629 (1998);
L. J. Garay, Phys. Rev. Lett. {\bf 80}, 2508 (1998);
Phys. Rev. D {\bf 58}, 124015 (1998);
Int. J. Mod. Phys. A {\bf 14}, 4079 (1999).

\bibitem{Shiokawa} K. Shiokawa, Phys. Rev. D {\bf 62}, 024002 (2000).

\bibitem {grhydro}
B. L. Hu, ``General Relativity as Geometro-Hydrodynamics"   
%Seminars at the University of Maryland,  Spring, 1994 (unpublished).
Invited talk at the Second Sakharov Conference,
Moscow, May, 1996; gr-qc/9607070.

\bibitem{meso}
B. L. Hu, ``Semiclassical Gravity and Mesoscopic Physics"
in {\it Quantum Classical  Correspondence} eds. D. S. Feng and B. L. Hu
(International Press, Boston, 1997);  gr-qc/9511077.

\bibitem{HuPeyr6} 
B. L.  Hu, Invited Talk  at Peyresq 6 (June 2001), 
Int. J. Theor. Phys. (2002).

\bibitem{Hartle-Horowitz81} J.B. Hartle and G.T. Horowitz, Phys. Rev.
D {\bf 24}, 257 (1981).

\bibitem{Wald77} R. M. Wald, Commun. Math. Phys. {\bf 54}, 1 (1977).

\bibitem{Christensen76} S. M. Christensen, Phys. Rev. D 
{\bf 14}, 2490 (1976); {\it ibid.} {\bf 17}, 946 (1978).

\bibitem{Bunch79} T. S. Bunch, J. Phys. A {\bf 12}, 517 (1979).

\bibitem{Ford82} L. H. Ford, Ann. Phys. (N.Y.) {\bf 144}, 238 (1982).

\bibitem{Kuo-Ford93} C.-I. Kuo and L. H. Ford, Phys. Rev. D
{\bf 47}, 4510 (1993).

\bibitem{Phillips-Hu97} N. G. Phillips and B. L. Hu, Phys. Rev. D
{\bf 55}, 6132 (1997).

\bibitem{Zurek91}  W. H. Zurek, Physics Today {\bf 44}, 36 (1991);
Prog. Theor. Phys. {\bf 81}, 281 (1993).

\bibitem{gell-mann-hartle} 
  M. Gell-Mann and J. B. Hartle,
      Phys. Rev. D {\bf 47}, 3345 (1993).

\bibitem{hartle}
  J. B. Hartle, 
      in {\sl Gravitation and Quantizations}, proceedings
      of the 1992 Les Houches Summer School, edited by B. Julia and 
      J. Zinn-Justin (North-Holland, Amsterdam, 1995);
      gr-qc/9304006, and references therein.

\bibitem{dowker}
  H. F. Dowker and J. J. Halliwell, 
      Phys. Rev. D {\bf 46}, 1580 (1992).

\bibitem{halliwell} 
  J. J. Halliwell, 
      Phys. Rev. D {\bf  48}, 4785 (1993);
      {\it ibid.} {\bf  57}, 2337 (1998).

\bibitem{whelan} 
  J. T. Whelan, 
      Phys. Rev. D {\bf  57}, 768 (1998).

\bibitem{MV3}
R. Mart\'{\i}n and E. Verdaguer, Int. J. Theor. Phys.
{\bf 38}, 3049 (1999).

\bibitem{Calzetta-Roura00a} E. Calzetta, A. Roura and E. Verdaguer,
quant-ph/0011097.

\bibitem{Hartle} J. B. Hartle, Phys. Rev. Lett. {\bf 39}, 1373 (1977); M.
Fischetti, J. B. Hartle and B. L. Hu, Phys. Rev. D {\bf 20}, 1757 (1979); 
J. B. Hartle and B. L. Hu, Phys. Rev. D {\bf 20}, 1772 (1979).


\bibitem{greiner} 
  C. Greiner and B. M\"{u}ller,
       Phys. Rev. D {\bf  55}, 1026 (1997).

\bibitem{cv96} 
  A. Campos and E. Verdaguer,
       Phys. Rev. D {\bf 53}, 1927 (1996).

\bibitem{campos-hu}
  A. Campos and B. L. Hu,
       Phys. Rev. D {\bf 58}, 125021 (1998).

\bibitem{morikawa}
  M. Morikawa,
     Phys. Rev. D {\bf 33}, 3607 (1986);
  D.-S. Lee and D. Boyanovsky,
     Nucl. Phys. B {\bf 406}, 631 (1993).

\bibitem{feynman-vernon}  
  R. P. Feynman and F. L. Vernon, 
      Ann.  Phys. {\bf 24}, 118 (1963).

\bibitem{feynman-hibbs}
  R. P. Feynman and A. R. Hibbs, 
      {\sl Quantum mechanics and path integrals}
      (McGraw-Hill, New York, 1965).

\bibitem{hu-paz-zhang} 
  B. L. Hu, J. P. Paz and Y. Zhang,
     Phys. Rev. D {\bf 45}, 2843 (1992).

\bibitem{hu-matacz94}
  B. L. Hu and A. Matacz,
     Phys. Rev. D {\bf 49}, 6612 (1994).

\bibitem{donoghue}
  J. F. Donoghue, 
      Phys. Rev. Lett. {\bf 72}, 2996 (1994);
      Phys. Rev. D {\bf 50}, 3874 (1994);
      Helv. Phys.Acta {\bf 69}, 269 (1996);
      in {\sl Advanced School on Effective Theories},
      edited by F. Cornet and M. J. Herrero (World Scientific,
      Singapore, 1996), gr-qc/9512024; gr-qc/9712070.

\bibitem{weinberg} 
  S. Weinberg,
     {\sl The Quantum Theory of Fields}, vols. I and II
     (Cambridge University Press, Cambridge, England, 1995 and 
     1996).

\bibitem{gleiser}
  M. Gleiser and R. O. Ramos,
     Phys.  Rev. D {\bf 50}, 2441 (1994);
  D. Boyanovsky, H. J. de Vega, R. Holman, D.S-Lee and A. Singh,
      Phys. Rev. D {\bf 51}, 4419 (1995);
  E. Calzetta and B. L. Hu,
     Phys. Rev. D {\bf 55}, 3536 (1997);
  M. Yamaguchi and J. Yokoyama,
     Phys. Rev. D {\bf 56}, 4544 (1997);
  S. A. Ramsey, B. L. Hu and A. M. Stylianopoulos,
     Phys. Rev. D {\bf 57}, 6003 (1998).

\bibitem{kubo} 
  R. Kubo, 
       J. Phys. Soc. (Japan) {\bf 12}, 570 (1957); 
  in {\sl Lectures in Theoretical Physics}, 
       vol.\ I, proceedings of the 1958 Summer Institute for
       Theoretical Physics, University of Colorado, Boulder,
       edited by W. E. Brittin and L. G. Dunham
       (Interscience, New York, 1959);
   Rep. Prog. Phys. {\bf 29}, 255 (1966).

\bibitem{martin}
  P. C. Martin and J. Schwinger,
       Phys.  Rev. {\bf 115}, 1342 (1959);
  L. P. Kadanoff and P. C. Martin,
       Ann. Phys. {\bf 24}, 419 (1963);
  M. Plischke and B. Bergersen, 
      {\sl Equilibrium Statistical Physics}, 
      2nd ed.\ (World Scientific, Singapore, 1994).


\bibitem{RouVer99}  A. Roura and E. Verdaguer, Phys. Rev. D {\bf 60},
107503 (1999).

\bibitem{ccv97}
  E. Calzetta, A. Campos and E. Verdaguer, 
       Phys. Rev. D {\bf 56}, 2163 (1997);
  A. Campos and E. Verdaguer,
       Int. J. Theor. Phys. {\bf 36}, 2525 (1997).

\bibitem{flanagan} 
  \'E. \'E. Flanagan and R. M. Wald, 
      Phys. Rev. D {\bf 54}, 6233 (1996).

\bibitem{lomb-mazz}
  F. C. Lombardo and F. D. Mazzitelli,
        Phys. Rev. D {\bf 55}, 3889 (1997).

\bibitem{jones}
  D. S. Jones, 
     {\sl Generalised functions}
     (McGraw-Hill, New York, 1966).

\bibitem{schwartz} 
   L. Schwartz,
     {\sl Th\'eorie des distributions}, Tomes I et II
     (Hermann, Paris, 1957 and 1959);
   A. H. Zemanian,
     {\sl Distribution Theory and Transform Analysis}
     (Dover, New York, 1987).

\bibitem{cmv95}
  A. Campos, R. Mart\'\i n and E. Verdaguer,
       Phys. Rev. D {\bf 52}, 4319 (1995).

\bibitem{horowitz}
  G. T. Horowitz,
      Phys. Rev. D {\bf 21}, 1445 (1980).

\bibitem{horowitz81}
  G. T. Horowitz,      
      in {\sl Quantum Gravity 2: A Second Oxford
      Symposium}, edited by C. J. Isham, R. Penrose
      and D. W. Sciama (Clarendon Press, Oxford, 1981).

\bibitem{horowitz_wald}
  G. T. Horowitz and R. M. Wald, 
      Phys. Rev. D {\bf 21}, 1462 (1980);
      {\it ibid.} {\bf 25}, 3408 (1982);
  A. A. Starobinsky, 
      Pis'ma Zh.\ Eksp.\ Teor.\ Fiz.\ {\bf 34}, 460 (1981)
      [JEPT Lett.\ {\bf 34}, 438, (1981)].

\bibitem{jordan} 
  R. D. Jordan,
       Phys. Rev. D {\bf 36}, 3593 (1987).

\bibitem{tichy}
  W. Tichy and \'E. \'E. Flanagan,
       Phys. Rev. D {\bf 58}, 124007 (1998).

\bibitem{horowitz-wald78}
  G. T. Horowitz and R. M. Wald, 
      Phys. Rev. D {\bf 17}, 414 (1978);
  S. Randjbar-Daemi,
      J. Phys. A {\bf 14}, L229 (1981);
      {\it ibid.~}{\bf 15}, 2209 (1982);
  W.-M. Suen,
      Phys. Rev. Lett.~{\bf 62}, 2217 (1989);
      Phys. Rev. D {\bf 40}, 315 (1989).

\bibitem{hartle_horowitz}
  J. B. Hartle and G. T. Horowitz,
      Phys. Rev. D {\bf 24}, 257 (1981).

\bibitem{simon}
   J. Z. Simon,
      Phys. Rev. D {\bf 43}, 3308 (1991); see also the
      discussion in W.-M. Suen, gr-qc/9210018.

\bibitem{Tomboulis77}
  E. Tomboulis,
      Phys. Lett. B {\bf 70}, 361 (1977).


\bibitem{Casimir} H. B. Casimir, Proc. Kon. Ned. Akad. Wet. {\bf 51},
793 (1948).

\bibitem{Barton} G. Barton, J. Phys. A. {\bf 24}, 991 (1991);
{\it ibid.} {\bf 24} 5533 (1991).

%\bibitem{valSCG} 

\bibitem{states}
See any textbook in quantum optics, e.g.,
D. F. Walls and G. J. Milburn, {\it  Quantum Optics}
(Springer-Verlag, Berlin, 1994);
M. O.  Scully and S. Zubairy, {\it Quantum Optics}
(Cambridge University Press, Cambridge 1997);
P. Meystre and M. Sargent III, {\it Elements of Quantum Optics}
(Springer-Verlag, Berlin, 1990).

\bibitem{NPhD}
N. G. Phillips, Ph. D. Thesis, University of Maryland (1999)

\bibitem{HuO'C}
B. L. Hu and D. J. O'Connor, Phys. Rev. D {\bf 36}, 1701 (1987).

\bibitem{Wheeler}
J. A. Wheeler,  Ann. Phys. (N. Y.) {\bf 2}, 604 (1957); 
{\it Geometrodynamics} (Academic Press, London, 1962);
in {Relativity, Groups and Topology}, eds B. DeWitt  and C. DeWitt
(Gordon and Breach, New York, 1964).

\bibitem{Hawking}
S. W. Hawking, Nucl. Phys. B {\bf 144}, 349 (1978).
S. W. Hawking, D. N. Page and C. N. Pope, Nucl. Phys. B
{\bf 170} 283 (1980).

\bibitem{Thorne}
K. S. Thorne, {\it Black Holes and Time Warps} (Norton Books, 1994).

\bibitem{Wald75} R. M. Wald, Commun. Math. Phys. {\bf 45}, 9 (1975).

\bibitem{WuFor2}
C. H. Wu and L. H.  Ford,  gr-qc/0102063.

\bibitem{ALN77} S. L. Adler, J. Lieberman and Y. J. Ng, Ann. Phys. (N.Y.)
                {\bf 106}, 279 (1977).

\bibitem{Wald78} R. M. Wald, Phys. Rev. D {\bf 17}, 1477 (1977).

\bibitem{DeWitt65}
 B. S. DeWitt,  {\it Dynamical Theory of Groups and Fields} 
 (Gordon and Breach, 1965).

\bibitem{DeWitt75} B. S. DeWitt, Phys. Rep. {\bf 19C} 297 (1975).

\bibitem{NPsc}
N. G. Phillips, ``Symbolic Computation of Higher Order Correlation Functions 
of Quantum Fields in Curved Spacetimes"  (in preparation).

\bibitem{Duff74} D. M. Capper and M. J. Duff, Nouvo Cimento {\bf 23A}, 
		173 (1974);
             M. J. Duff, {\it Quantum Gravity: An Oxford symposium}, ed.
             C. J. Isham, R. Penrose and D. W. Sciama
             (Oxford University Press, Oxford, 1975).

\bibitem{Hadamard} J. Hadamard, {\it Lectures on Cauchy's Problem in
	Linear Partial Differential Equations} (Yale University Press,
          New Haven, 1923).

%\bibitem{PazSin}  J. P. Paz and S. Sinha, Phys. Rev. D {\bf 44}, 1038 (1991).


\end{references}
\end{document}